\documentclass[a4paper,fleqn,usenatbib]{mnras}

\usepackage{newtxtext,newtxmath}

\usepackage[T1]{fontenc}
\usepackage{ae,aecompl}
\usepackage{hyperref}

\usepackage{graphicx}	
\usepackage{amsmath}	
\usepackage{amssymb}	
\usepackage{hyperref}	
\usepackage{xspace}
\hypersetup{colorlinks=true,linkcolor=blue,citecolor=blue,filecolor=blue,urlcolor=blue}

\newcommand{\mlr}{$\lambda(r)$\xspace}
\newcommand{\nh}{$N_{\mathrm{H}}$\xspace}

\title[Magnetic fields in star-forming systems (II)]{Magnetic fields in star-forming systems (II): examining dust polarization, the Zeeman effect, and the Faraday rotation measure as magnetic field tracers}

\author[S. Reissl]{Stefan Reissl$^{1}$\thanks{E-mail: reissl@uni-heidelberg.de},
Amelia M. Stutz$^{2,3}$,
Ralf S. Klessen$^{1,4}$,
\newauthor
Daniel Seifried$^{5}$,
and Stefanie Walch$^{5}$
\\
$^{1}$Universit\"{a}t  Heidelberg, Zentrum f\"{u}r Astronomie, Institut f\"{u}r theoretische Astrophysik,  Albert-Ueberle-Str. 2, 69120 Heidelberg, Germany\\
$^{2}$Departamento de Astronom\'{i}a, Universidad de Concepci\'{o}n,Casilla 160-C, Concepci\'{o}n, Chile\\
$^{3}$Max-Planck-Institute for Astronomy, K\"{o}nigstuhl 17, 69117 Heidelberg, Germany\\
$^{4}$Universit{\"a}t Heidelberg, Interdisziplin{\"a}res Zentrum f{\"u}r Wissenschaftliches Rechnen, Im Neuenheimer Feld 205, 69120 Heidelberg, Germany\\
$^{5}$Universit\"at zu K\"oln,I. Physikalisches Institut,  Z\"ulpicher Str. 77, 50937 K\"oln, Germany
}

\date{Accepted XXX. Received YYY; in original form ZZZ}

\pubyear{2015}

\begin{document}
\label{firstpage}
\pagerange{\pageref{firstpage}--\pageref{lastpage}}
\maketitle
\begin{abstract}
The degree to which the formation and evolution of clouds and filaments in the interstellar medium is regulated by magnetic fields remains an open question. Yet the fundamental properties of the fields (strength and 3D morphology) are not readily observable.  We investigate the potential for recovering magnetic field information from dust polarization, the Zeeman effect, and the Faraday rotation measure ($RM$) in a SILCC-Zoom magnetohydrodynamic (MHD) filament simulation. The object is analyzed at the onset of star formation, and it is characterised by a line-mass of about $\mathrm{\left(M/L\right) \sim 63\ \mathrm{M}_{\odot}\ pc^{-1}}$ out to a radius of $1\,$pc and a kinked 3D magnetic field morphology. We generate synthetic observations via {\sc POLARIS} radiative transfer (RT)  post-processing, and compare with an analytical model of helical or kinked field morphology to help interpreting the inferred observational signatures.  We show that the tracer signals originate close to the filament spine. We find regions along the filament where the angular-dependency with the line-of-sight (LOS) is the dominant factor and dust polarization may trace the underlying kinked magnetic field morphology.  We also find that reversals in the recovered magnetic field direction are not unambiguously associated to any particular morphology. Other physical parameters, such as density or temperature, are relevant and sometimes dominant compared to the magnetic field structure in modulating the observed signal. We demonstrate that the Zeeman effect and the $RM$ recover the line-of-sight magnetic field strength to within a factor {2.1 - 3.4}. We conclude that the magnetic field morphology may not be unambiguously determined in low-mass systems by observations of dust polarization, Zeeman effect, or $RM$, whereas the field strengths can be reliably recovered.
\end{abstract}

\begin{keywords}
ISM: magnetic fields -- radiative transfer -- methods: numerical -- techniques: polarimetric --  dust, extinction -- line: profiles -- radio continuum: ISM
\end{keywords}

\section{Introduction}
One of the most fundamental question in astrophysics is how stars are formed in their natal molecular clouds. Processes on multiple scales such as cloud-cloud collisions \citep[][]{Kitsionas2008,Fukui2014}, gravitational cloud collapse  \citep[][]{McKee2002}, turbulent motion \citep[][]{MacLow2004,Wollenberg2020}, and gas accretion  \citep[][]{Bonnell2004,Krumholz2009,Mejia2017} have been identified as key ingredients of star-formation. Such collapsing clouds seem to be commonly associated with a large-scale filamentary gas structure \citep[][]{Gutermuth2009,Schneider2012,Konyves2015,Stutz2018,Arzoumanian2019}.

Additionally, magnetic fields can regulate the star-formation efficiency by providing support against cloud collapse \citep[][]{Mouschovias1999,LiGoodman2014} or by compressing the gas due to pinch instabilities \citep[][]{Fiege2000A,Fiege2000A}. However, the role of magnetic fields involved in such processes is still a field of ongoing research \citep[][]{Mouschovias2009,Stutz2016,Koertgen2018,Reissl2018,Girichidis2018,Seifried2019,Seifried2020}.  Charged gas can exert a considerable amount of drag on the magnetic field, warping the field lines into distinct characteristic configurations. This ambipolar diffusion process has been observed e.g. in the low-mass star-forming cloud {NGC 1333 IRAS 4A} ending up in an hourglass-like field morphology\citep[][]{Girart2006}. 
While a hourglass field seems to be a common scenario on cloud and core scales as shown by numerical modelling efforts \citep[][]{Goncalves2008,Frau2011}, such a field structure seems to be unlikely for large scale filaments. It can be shown from a stability argument that a filament may be wrapped by a helical field \citep[][]{Fiege2000A,Fiege2000B}. In an alternative scenario a filament may form when a moving cloud runs into a high-density e.g. a much larger cloud \citep[][]{Inoue2013,Inoue2018}. In this so called "sushi" model a filament would form perpendicular to the plane of the shock front and the magnetic field would be bent with a characteristic kink all along the filament. Similar configurations of bent field lines in filaments are suggested by models \citep[][]{Reissl2018,LiKlein2019,Tahani2019} and observations \citep[e.g.][]{Pattle2017} alike. It comes naturally that this collision scenario results in shocks, which  then  fragment  further into  dense  cores which subsequently form stars. This scenario indicates the significance of filamentary structures as a critical step in the star-formation process. However, \citep[][]{Stutz2016} favor a helical field morphology in Orion as a natural field configuration in high-mass filaments.
Altogether, both magnetic field strength and morphology provide clues about the particular star-formation scenarios and the relevant physical processes. Both quantities can be tested observationally by exploring the numerous physical processes related to light polarization \citep[][]{Reissl2018}. 

A well-established practice of probing the magnetic field directions is by means of dust polarimetry \citep[][]{Hiltner1949,Hall1949,Jones1967}. As indicated by observations the alignment of non-spherical dust grains with the magnetic field seems to be common from molecular clouds \citep{Girart2006,Sadavoy2018} up to galactic scales \citep[][]{PlanckXIX2015}. However, such observations remain inconclusive. In particular, infrared (IR) dust polarimetry is utilised in magnetic field measurements of filaments as e.g. in \cite{Hildebrand2000}, \cite{Matthews2000}, \cite{Pillai2015}, or \cite{Pillai2017}. While field directions perpendicular as well as parallel to filaments are reported in \cite{Pillai2015}, the observations of \cite{Matthews2000} in Orion suggest a field direction being mainly perpendicular to the filament. Furthermore, a significant complication is the complexity of the grain alignment physics \citep[see][for review]{Andersson2015}. Depending on the local conditions this efficiency may be highly variable along the line-of-sight (LOS) and regions without aligned grains cannot contribute to the polarization signal at all \citep[see e.g.][]{Seifried2019}.

The Zeeman effect is a key probe \citep[][]{Crutcher1993} in measuring the LOS field strength. Here, the magnetic field splits  certain transition lines into sub-level. Reliable magnetic field measurements toward dense cores, molecular clouds, as well as filaments have been successfully performed with HI, OH, CSS, CN and maser lines \citep[e.g.][]{Fiebig1989,Crutcher1993,Crutcher1996a,Crutcher1996b,Caswell1995,Heiles1997,Crutcher1999,Fish2003,Falgarone2008,Crutcher2012,Pillai2016,Nakamura2019}. The critical aspect is a line separation detectable by instrumentation. Hence, a successful field measurement depends on the density regime and tracer \citep[see e.g.][]{Crutcher2012,Brauer2017A,Nakamura2019}. Because of the comparatively high technical and precision requirements, Zeeman splitting measurements, and therefore the more direct inference of the magnetic field strengths and LOS direction in astrophysical systems, are not yet as common as we would like \citep[][]{Crutcher2012}.  However, because of their critical importance they will very likely become routine in the near future. A study exploring synthetic dust polarization and Zeeman observations is presented in \cite{Reissl2018} (see below). 

Magnetised plasmas tend to rotate the polarization vectors along the LOS. This effect, called Faraday rotation, is most prominent in radio observations and depends on magnetic field strength and electron fraction (density). Consequently, with a known or assumed electron density, the rotation measurement ($RM$) traces the LOS magnetic field strength \citep[][]{Rybicki1979,Huang2011}.  
Radio observations of the $RM$ were used for detecting the magnetic field in the galaxies {NGC 253} and M51, respectively \citep[][]{Heesen2011,Fletcher2011,Beck2015}. A first all sky map of the Milky Way was derived in \cite{Oppermann2012} on the basis of polarised background sources and later improved by \cite{Hutschenreuter2019}.  More recently, $RM$ measurements where applied to derive the magnetic field strength of filaments in the nearby regions Orion, Perseus, and California based on a chemical network model to derive the electron fractions \citep[][]{Tahani2018} and they have been employed to study the spiral arm structure in the Milky Way \citep{Shanahan2019,ReisslRM2020}.

From the observational side, the $RM$ may become unreliable when a significant amount of Faraday rotation takes place within a single frequency bin. Furthermore, the complex ionization processes and the subsequent electron fractions are hard to constrain along the LOS. For instance, protostars are factories of cosmic rays (CR) which may feed back into the electron fraction \cite{Padovani2016}. Modeling efforts have been made to overcome these uncertainties \citep[see e.g.][]{Pakmor2018,Tahani2018,Reissl2019,Pellegrini2019}. Hence, the $RM$ can be considered as a reliable tool to study the magnetic field strength across a wide range of conditions. 



Altogether, the interpretation of line polarization, dust polarization observations as well as Faraday measurements should be treated with care because projecting intrinsically 3D fields onto the LOS perpendicular to the plane-of-the-sky produces the usual loss of information and accompanying degeneracies that hinder the inference of full 3D-space morphology. Different magnetic field configurations may even  produce similar measured  polarization signals being indistinguishable by observations \citep[see e.g.][]{Reissl2017}. Efforts were made by \cite{Reissl2014} to tackle this problem by analyzing a combination of synthetic linear and circular dust polarization maps. The origin along a particular LOS was traced in \cite{Reissl2017}.

All these techniques presented above (i.e. dust polarization, Zeemann effect, and Faraday rotation) have an intrinsic angular-dependency between the LOS and magnetic field direction. In principle, it should be possible to detect a unique field morphology from observations and modeling by focusing on these dependencies. However, systematic studies concerning that particular topic are rare. \cite{Reissl2018} modeled such angular signatures and show that dust polarimetry combined with complementary Zeeman observations would allow to identify distinct field morphologies by their characteristic radial profiles of linear and circular dust polarization as well as Zeeman observations. A strength of this study is that it presents a simplified analysis based on the observations of \cite{Stutz2016} for filamentary cloud structures. However, analytical models alone cannot fully account for the complexity arising for example from velocity variations and small-scale structures such as density condensations (e.g. "cores").  


In this paper we expand on the modeling framework established by \cite{Reissl2014} and \cite{Reissl2018} with the goal of illuminating the usefulness of the above magnetic field tracers in a controlled experiment where the underlying 3D shape of the magnetic field is known and the action of gravity and supernova (SN) driving in the galactic disk is forming filaments \citep[][]{Walch2015}. The predictions of \cite{Reissl2018} are tested with mock observations based on a MHD simulation and subsequent radiative transfer (RT) post-processing with {\sc POLARIS} (see below). The advantage of such mock observations from MHD simulations is that the origin of the signal remains accessible. In particular, we focus on a case study of a MHD simulation performed with the  {\sc FLASH} code within the SILCC-Zoom project \citep[][]{Seifried2017,Haid2019}. The simulation provides an isolated low-mass filament that is surrounded by a kinked magnetic field structure. Such a kinked structure is similar to the one suggested in \cite{Inoue2013} and modeled in \cite{Reissl2018}. In addition to the tracers of dust polarization and Zeeman effect considered in \cite{Reissl2018}, we include synthetic Frarady $RM$ in order to trace characteristic signatures of the magnetic field morphology with observationally accessible measures.

Synthetic images are created with the versatile and publicly available RT code  {\sc POLARIS}\footnote{\href{http://www1.astrophysik.uni-kiel.de/~polaris/}{http://www1.astrophysik.uni-kiel.de/${\sim}$polaris/}} \citep[][]{Reissl2016}. The advantage of this code is that it unites the physics of polarised radiation of dust emission \citep[][]{Reissl2016,Reissl2017}, grain alignment \citep[][]{Reissl2016,Reissl2018}, Zeeman effect \citep[][]{Brauer2017B,Brauer2017A}, and synchrotron emission \citep[][]{Reissl2019} into a single framework. {\sc POLARIS} runs RT simulations on an octree grid\footnote{Indeed, {\sc POLARIS} provides also spherical, cylindrical, and Voronoi grid geometries suitable for post-processing various analytical models and MHD simulations.} allowing to maintain the native grid structure of the {\sc FLASH} code. 

The paper is structured as follows: In \S~\ref{sect:MHDSimulation} we describe the setup of the SILCC-Zoom simulation we use for RT post-processing. We construction an analytical model for reference in \S~\ref{sect:ToyModel}. In \S~\ref{sect:Obs} we introduce the RT post-processing pipeline for the creation of synthetic observations. Here, we outline the physical principles of different magnetic field tracers. We present the synthetic observations and derived data products in \S~\ref{sect:FilProp}. In \S~\ref{sect:Origin} we discuss the origin of the polarization signals. We discuss our results in \S~\ref{sect:Discussion} in the context of real observations of filaments. Finally, we summarise our findings and conclusions in \S~\ref{sect:Summary}.

\begin{figure}
\hspace{-5mm}
\begin{minipage}[c]{1.0\linewidth}
      \includegraphics[width=1.1\textwidth]{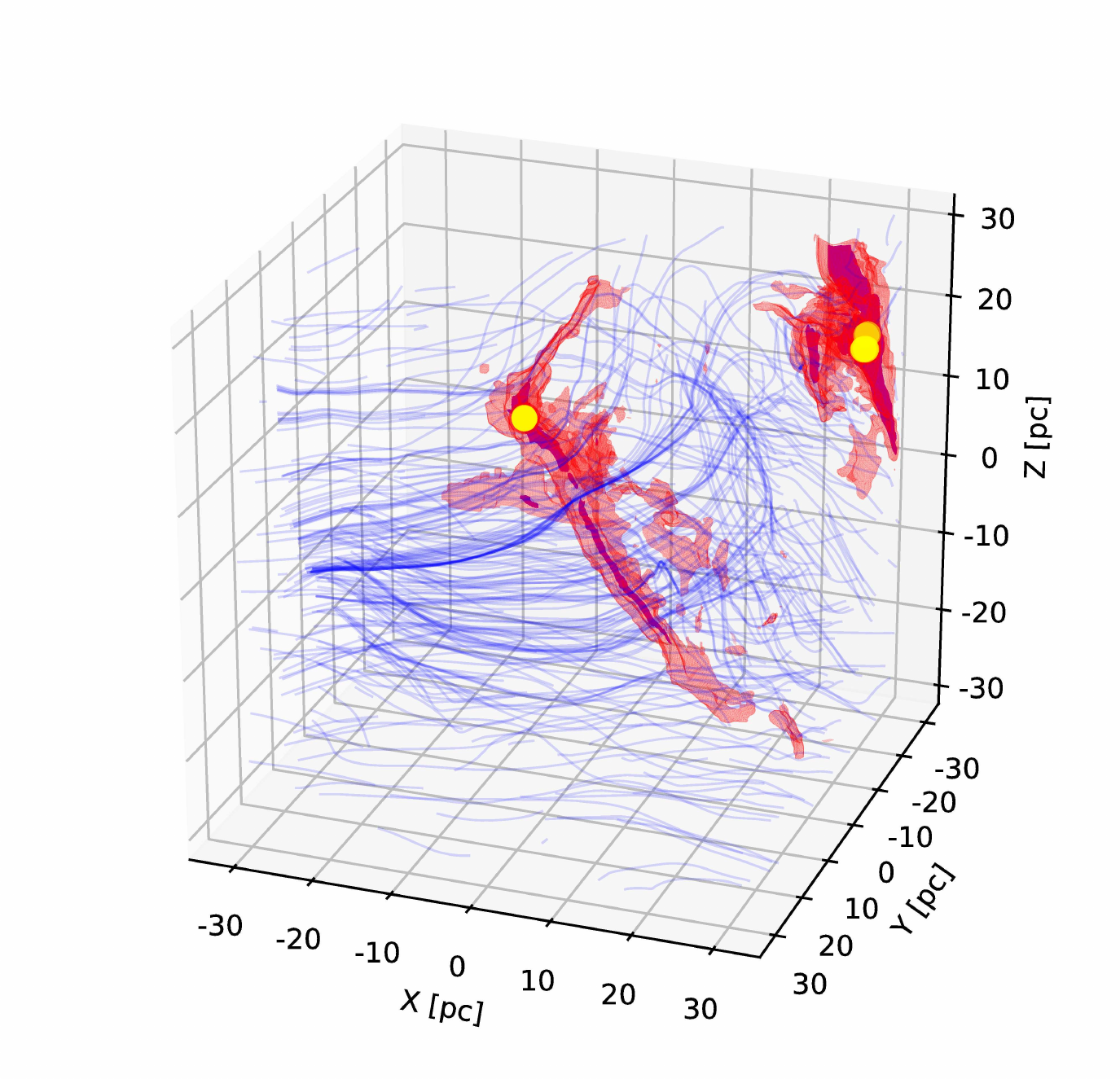}
\end{minipage}

\caption{Magnetic field lines (blue) and the corresponding isosurface of the gas distribution of the SILCC-Zoom cutout at $n_{\mathrm{gas}}=100\ \mathrm{cm}^{-3}$ (red) and $n_{\mathrm{gas}}=800\ \mathrm{cm}^{-3}$ (purple), respectively. Star-forming regions are marked by yellow spheres. The filament moves predominantly towards the positive $Y-$direction. The LOS for the synthetic observations is defined to be parallel to the $Y-$axes.}
\label{fig:3DMHD}
\end{figure}

\begin{figure*}
\centering
\begin{minipage}[c]{1.0\linewidth}
      \includegraphics[width=0.49\textwidth]{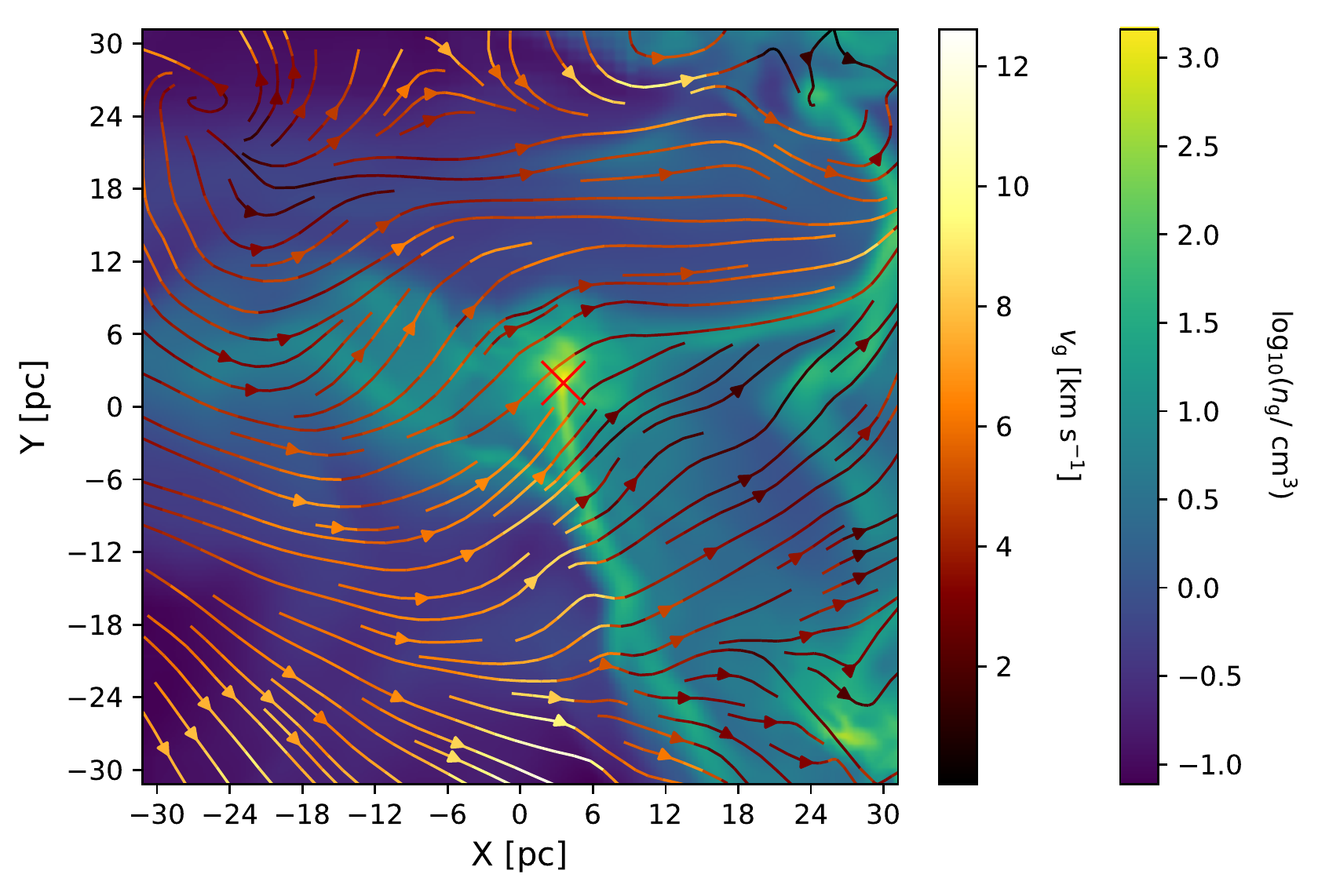}
      \includegraphics[width=0.49\textwidth]{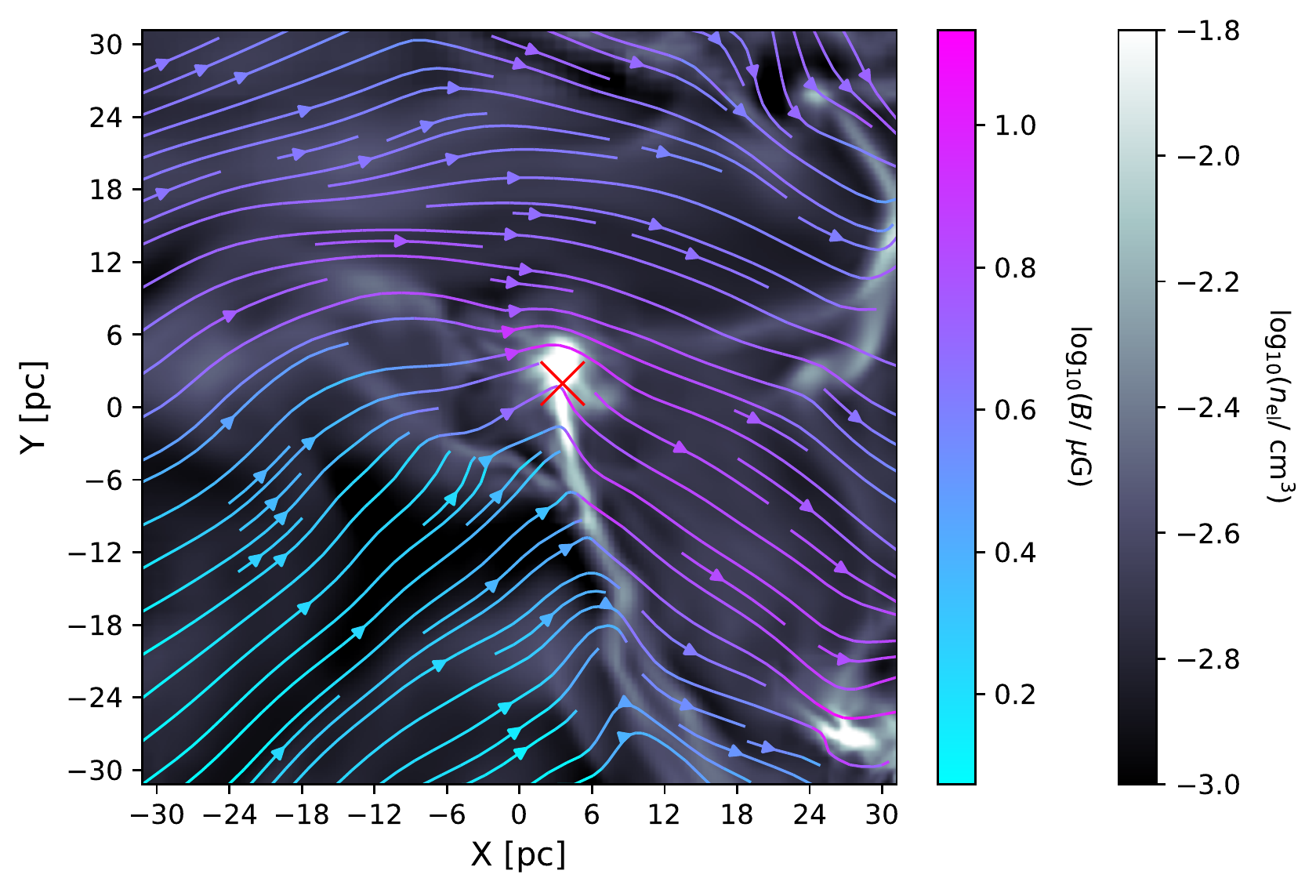}
\end{minipage}
\caption{ Cut through the $XY$ plane at $Z=0\ \mathrm{pc}$. The filament spine (red cross) runs roughly perpendicular through the image. Left panel: The gas density distribution $n_{\mathrm{g}}$ overlaid with the vector field of the gas velocity $v_{\mathrm{g}}$ of the SILCC-Zoom cube. Right panel: The same as the left panel for  electron density $n_{\mathrm{el}}$ and the streamlines are the direction and magnitude of the magnetic field B. }
\label{fig:Midplanes}
\end{figure*}



\section{The SILCC-Zoom MHD simulation}
\label{sect:MHDSimulation}
The ideal MHD simulation setup used here was originally presented in \citet{Seifried2019} with the later addition of radiative feedback \citep{Haid2019}. The dust polarization in this simulations without feedback were already investigated in \citet{Seifried2019}.  
 
The simulation is part of the SILCC-Zoom project \citep{Seifried2017}, where we model the evolution of molecular clouds embedded in their galactic environment with the {\sc FLASH} code version 4.3 \citep{Fryxell00,Dubey08} and the entropy-stable MHD solver developed in  \citet{Derigs2016,Derigs2018}. The zoom-in region has a side-length of about $100\ \mathrm{pc}$ in which we resolve the cloud structure with a resolution of $0.12\ \mathrm{pc}$. In particular, we model the chemical and thermodynamical evolution of the gas using a chemical network for H$^+$, H, H$_2$, C$^+$, CO, and O \citep{Nelson97,Glover07,Glover10}. The shielding of the additional interstellar radiation field (ISRF, G$_0$ = 1.7) as well as the effect of self-gravity are solved with a tree-based approach \citep{clark2012,Wunsch18}. For the density distribution of free electrons we evaluate the ionization state of hydrogen and carbon and assume ${ n_{\mathrm{el}}= n_{\mathrm{H^+}} + n_{\mathrm{C^+}} }$.
 
The initial gas distribution follows a Gaussian profile with a scale height of $30\ \mathrm{pc}$ and a total surface density of $10\ \mathrm{M}_{\odot}\ \mathrm{pc}^{-2}$ reminiscent of the solar neighborhood. We include an external gravitational potential due to a pre-existing stellar component \citep{Spitzer42}. We initialise the magnetic field along the $X$-direction as ${ B_{\mathrm{X}} = B_{\mathrm{X,0}} \left(\rho(Z)/\rho_0\right)^{1/2} }$ where we set the magnetic field in the midplane to ${ B_{\mathrm{X,0}} = 3\ \mu\mathrm{G} }$ in accordance with recent observations \citep[e.g.][]{Beck13}.
The initial grid resolution is $3.9\ \mathrm{pc}$ and we inject supernovae (SNe) with a rate of $15\ \mathrm{SNe/Myr}$ to drive turbulence and form a self-consistent 3-phase ISM discussed in \citet{Walch2015} and \citet{Girichidis16}. At $t = 16\ \mathrm{Myr}$, we stop the injection of SNe and increase the resolution in the aforementioned zoom-in region to a $0.12\ \mathrm{pc}$ refinement based on the Jeans length and variations in the gas density.
In the surroundings of the zoom-in region we keep the lower resolution in its surroundings. In this way we can follow the formation of the molecular cloud located inside the Zoom-in region. Furthermore, throughout the simulation, the magnetic field evolution is followed self-consistently with the gas flow.
 
In order to investigate the effect of stellar radiative feedback, we include sink particles, which form once the gas density exceeds a threshold of $1.1 \times 10^{-20}\ \mathrm{g\ cm}^{-3}$. Every $120\ \mathrm{M}_{\odot}$ of accreted mass, one massive star between $9\ \mathrm{M}_{\odot}$ and $120\ \mathrm{M}_{\odot}$ is randomly sampled from an initial mass function assuming a slope of $-2.3$ \citep{Salpeter55}. We follow the evolution of each massive star using stellar evolutionary tracks \citep{Ekstrom12,Gatto17,Peters17}. In addition, we calculate the amount of photoionizing radiation released by each star and its effect on the chemical state of the gas \citep{Haid18,Haid19} with the backwards ray-tracing algorithm {\sc TreeRay} \citep[][W\"unsch et al., in prep.]{Wunsch18}. For further details on the simulation and radiative feedback, we refer to \citet{Seifried2017,Seifried2019} and \citet{Haid2019}.
 
From this high-resolution zoom-in region we select a cutout with a side length $62.5\ \mathrm{pc}$ that contains a relatively isolated filament. In this cutout we find twelve massive stars that have formed in three tight star-forming regions. Here, a star-forming region is loosely defined by all massive stars that fall within the same sphere with a radius of $0.2\ \mathrm{pc}$. This SILCC-Zoom cutout allows us to create mock observations of dust emission, molecular line polarization and Faraday effect in a self-consistent manner, as it provides us with the densities, velocity field, molecular abundances, electron fractions, and magnetic fields at any given point in the simulation domain.

In Fig. \ref{fig:3DMHD} we show a 3D representation of the gas density distribution and the magnetic field configuration in the SILCC-Zoom cutout as well as the location of the star-forming regions. The coordinate system is defined by the $X$, $Y$, and $Z$ axis as given by the SILCC-Zoom setup but the origin of the coordinate system is defined to be in the center of the SILCC-Zoom cutout. We emphasise that this coordinate system is strictly applied in all the following sections and figures of this paper.

The gas mass is mostly accumulated in two distinct regions, an irregularly shaped diffuse cloud-like structure at one of the corners of the cutout and a relatively isolated filamentary structure in the center with one star-forming region. We refer to this filamentary structure as the main filament hereafter. The cloud harbours the other two of the three star-forming regions within the cutout and is penetrated by an unordered magnetic field with many twisted field lines. In contrast, the main filament is embedded in a more regular field with a direction mostly parallel to the $X-$axis and perpendicular to the spine of the main filament. 

In Fig. \ref{fig:Midplanes} we present cuts through the MHD grid of gas and electron densities, the velocity field, and the magnetic field directions in the $XY$ plane at $Z=0\ \mathrm{pc}$. Here, we define the spine to be at the density maximum of the main filament in each cut. 

Fig. \ref{fig:Midplanes} reveals that the spine is close to the center of the plane and that the main filament has an extended tail along $Y<0\ \mathrm{pc}$. The gas shocks at the main filament and the tail. Thus, the gas is dragging the magnetic field lines mostly along the $Y-$direction. The situation of this kinked magnetic field configuration is similar to the analytical model $flow$ presented in \cite{Reissl2018}, the "sushi" model of \cite{Inoue2013,Inoue2018}, and the simulations presented in \cite{LiKlein2019}.

\begin{figure*}
\centering
\begin{minipage}[c]{1.0\linewidth}

\begin{minipage}[c]{1.0\linewidth}
      \includegraphics[width=0.49\textwidth]{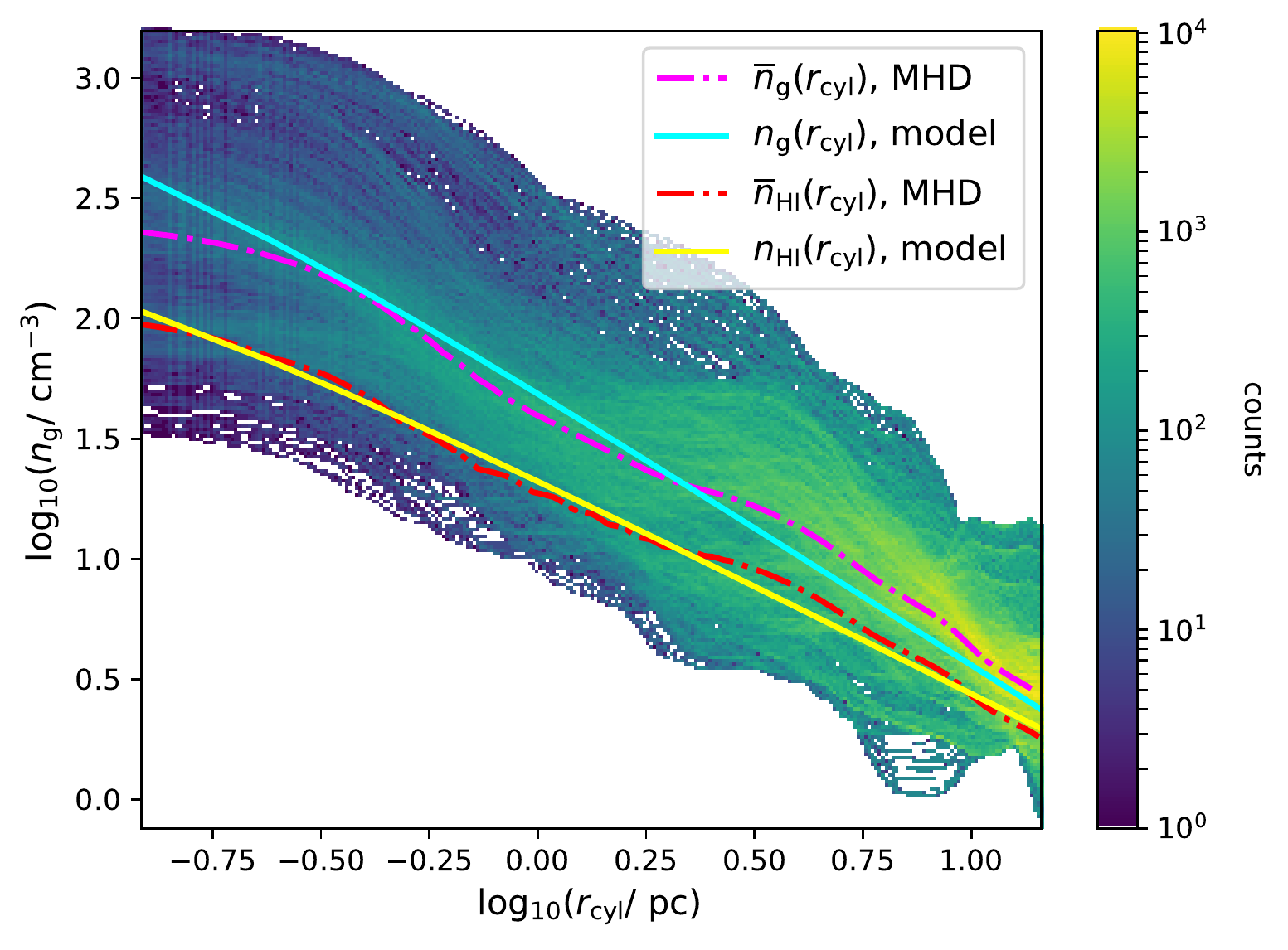}
      \includegraphics[width=0.49\textwidth]{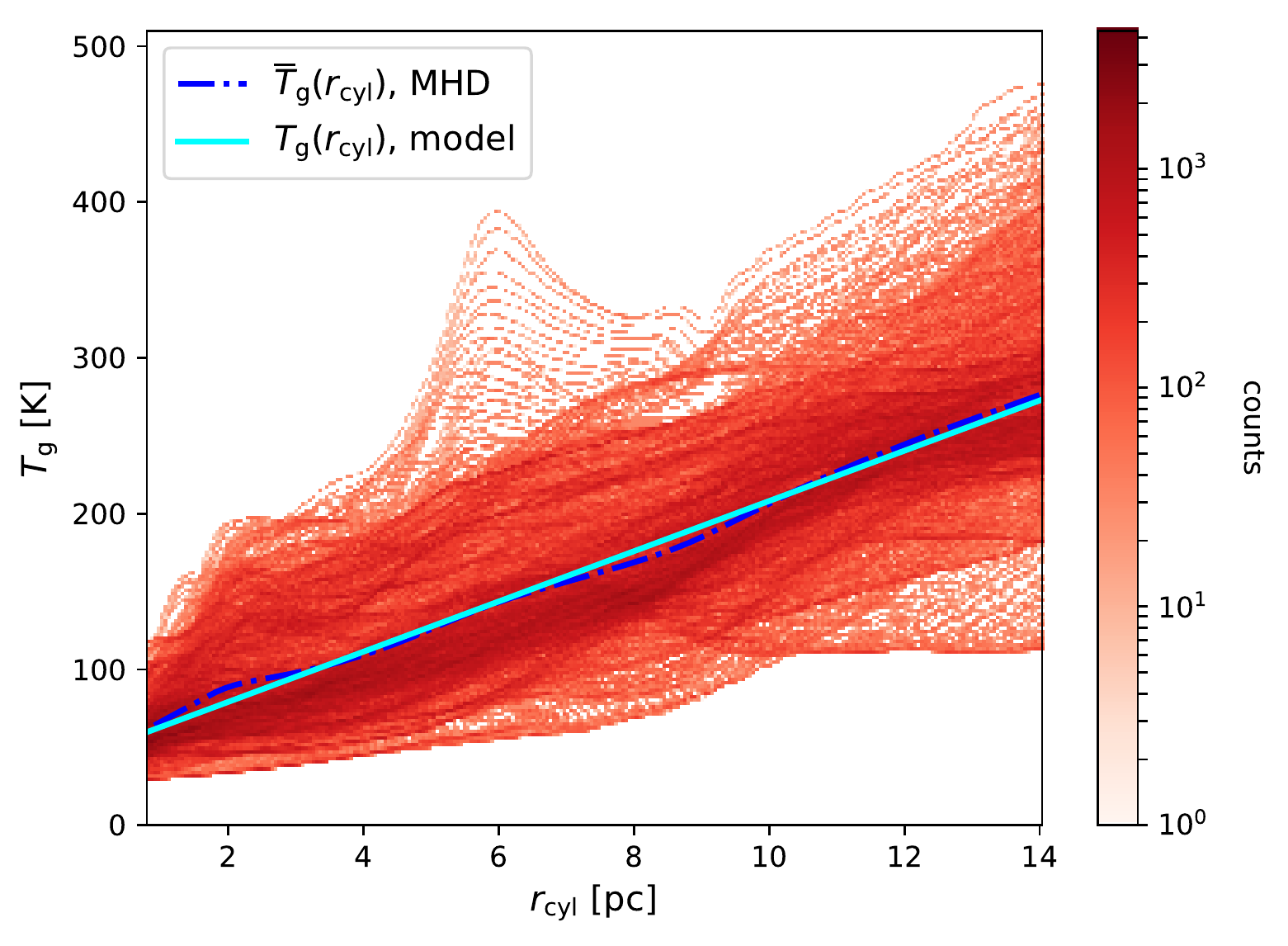}
\end{minipage}
      
\begin{minipage}[c]{1.0\linewidth}
      \includegraphics[width=0.49\textwidth]{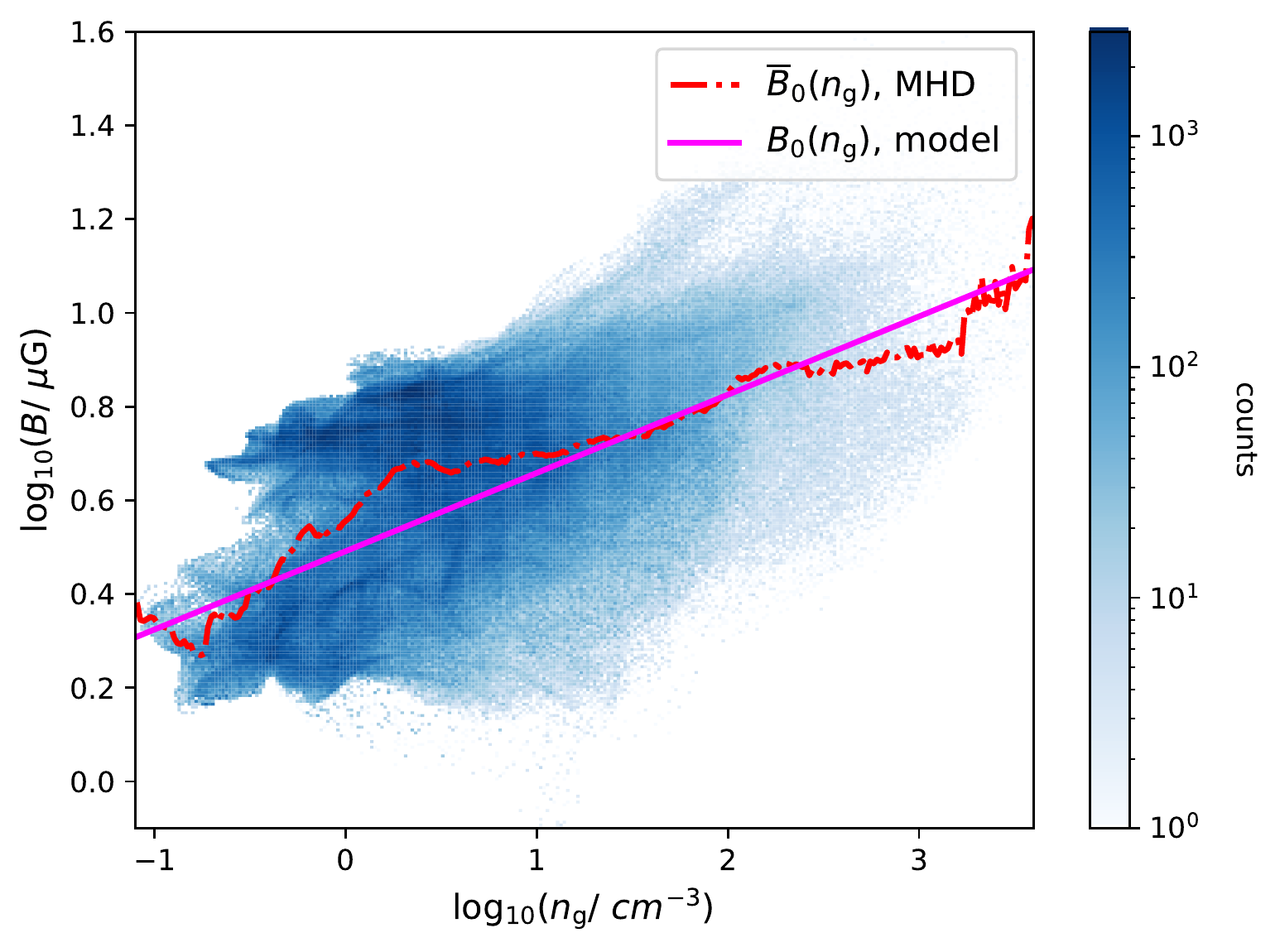}
      \includegraphics[width=0.49\textwidth]{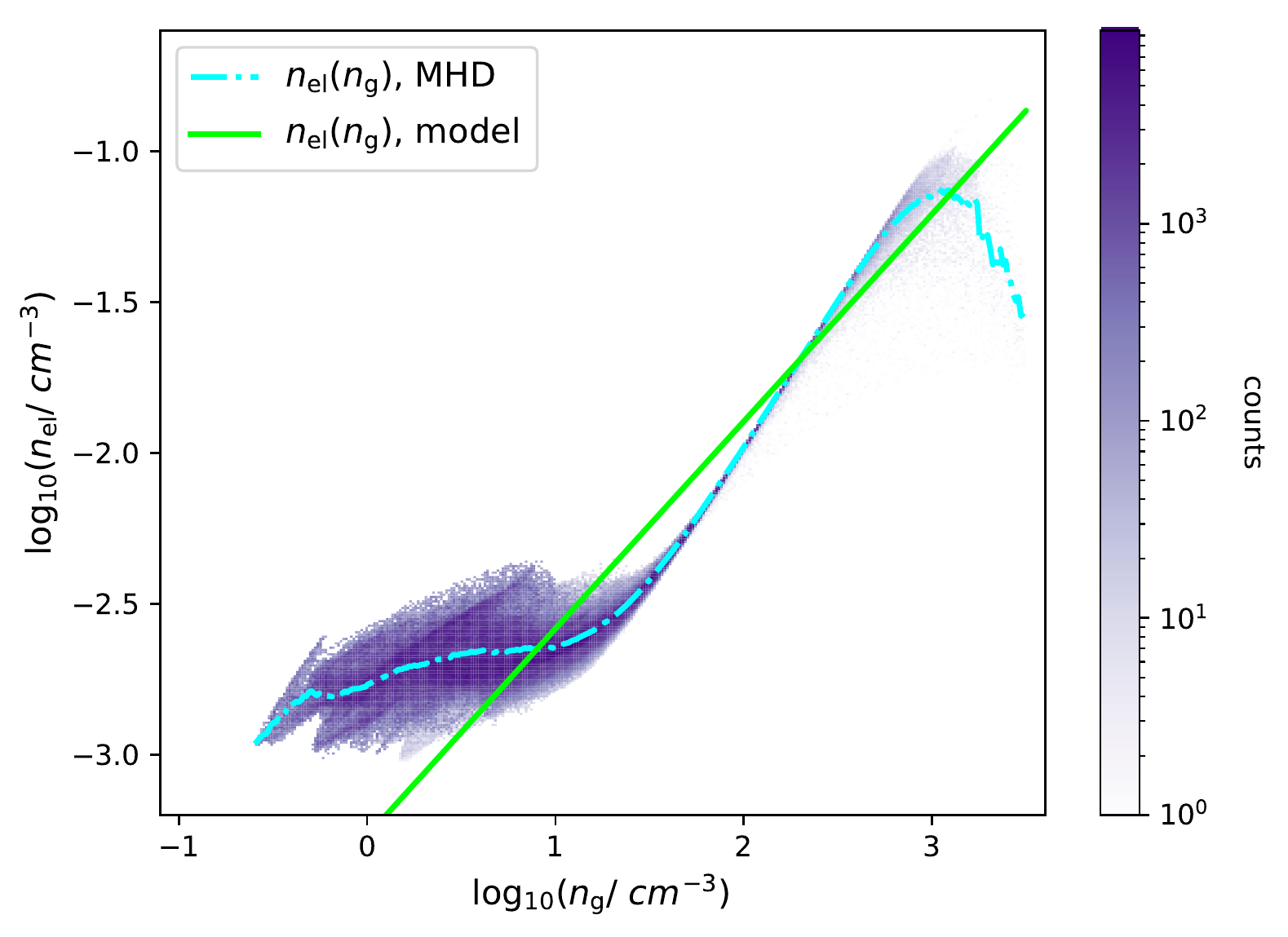}
\end{minipage}

\end{minipage}
\caption{The panels show the distribution of physical quantities taken from the SILCC-Zoom simulation. The best fit relations of these quantities are used to construct the analytical model.  Upper left panel: Distribution of radial gas density profiles $n_{\mathrm{g}}(r_{\mathrm{cyl}})$ from spine outward to the edge of the SILCC-Zoom cutout. The dash dotted line is the average $\overline{n}_{\mathrm{g}}(r_{\mathrm{cyl}})$  over all SILCC-Zoom profiles and the solid lines represents the best fit of the analytical model. The panel shows also the related fit data of the radial profile of HI for comparison. Upper right panel: The same as the upper left panel for the radial gas temperature $T_{\mathrm{g}}(r_{\mathrm{cyl}})$. Bottom left panel: The same as the upper panels for the gas density dependent scaling parameter  $B_{\rm 0}\left( n_{\mathrm{g}} \right)$ of the magnetic field strength. Bottom right panel: The same as the bottom left panel for the electron density  $n_{\rm el}\left( n_{\mathrm{g}} \right)$  dependent on gas density $n_{\mathrm{g}}$. }
\label{fig:ModProf}
\end{figure*}

\section{The analytical filament model}
\label{sect:ToyModel}
%

The SILCC-Zoom simulation exhibits a high degree of complexity due to the large number of physical processes at act as described in the previous Section. In order to facilitate the astrophysical interpretation of the synthetic observations, we seek help from simple analytic models of magnetized filaments which are constructed to have similar bulk properties. For that we follow the approach that was introduced by \cite{Reissl2018} to match the observations of {Orion A} as presented by \cite{Stutz2016}. 

The analytic model presented here, consists of a filamentary density distribution in cylindrical symmetry on a regular Cartesian $256^3$ grid with a side length of $62.5\ \mathrm{pc}$, which makes the results directly comparable to SILCC-Zoom simulations and eliminates potential grid artefacts in the RT post-processing. The gas distribution of the analytical model follows a Plummer power-law profile \citep[][]{Plummer1911} with:
\begin{equation}
n_{\rm g}(r_{\mathrm{cyl}}) = n_{\rm 0} \left[ 1 + \left( \frac{r_{\mathrm{cyl}}}{r_{\rm flat}}  \right)^2 \right]^{-\beta}\, .
\label{eq:DensDist}
\end{equation}
Here, the radius ${ r_{\mathrm{cyl}}=\left(X^2+Y^2\right)^{1/2} }$ is the cylindrical radius i.e. distance from the spine of the filament and $n_{\rm 0}$ is its central density. The characteristic radius $r_{\rm flat}$ and the power-law index $\beta$ control the shape in the center and the slope in the outskirts of the model filament, respectively. A Plummer profile with $r_{\rm flat}=0.05\ \mathrm{pc}$ is consistent with observations \citep[][]{Palmeirim2013,Stutz2016}. 

In order to determine the remaining parameters of the analytical model, we take the SILCC-Zoom filament and select $120$ radial profiles of density and temperature in $256$ midplane cuts going from the spine of the SILCC-Zoom main filament outwards. With these profiles we perform a least-squares fit for the  parameters $n_{\rm 0}$ and $\beta$. We find a central density of $n_{\rm 0} = 1650\  \mathrm{cm^{-3}}$ and a power-law index of $\beta=1.15$ for the analytical model. Compared to the model of \cite{Reissl2018}, the total gas mass is about a factor of 10 smaller. Furthermore, the power-law index $\beta$ in \cite{Reissl2018} is taken to be $1.6 - 2.0$ in concordance with \cite{Stutz2016}. Consequently, the density slope of the analytical model falls more slowly as observed in {Orion A}. 

For the RT calculations including the Zeeman effect discussed below we utilise the HI $1420\ \mathrm{MHz}$ ($21\ \mathrm{cm}$) line as a tracer of the magnetic field properties in the analytical model and the SILCC-Zoom cutout. Hence, for the radial HI density of the analytical model we apply a similar profile as Eq. \ref{eq:DensDist} and find $n_{\rm 0} = 324\ \mathrm{cm^{-3}}$ and $\beta=0.9$ (see Fig. \ref{fig:ModProf}).


For the radial gas temperature distribution  $T_{\rm g}(r)$ of the analytical model we find a linear correlation with the radius $r_{\mathrm{cyl}}$ and get 
\begin{equation}
T_{\rm g}(r_{\mathrm{cyl}}) =  16.1\ \mathrm{K} \left( \frac{r_{\mathrm{cyl}}}{\mathrm{pc}} \right) + 46.8 \ \mathrm{K}\, .
\label{eq:TempDist}
\end{equation}
A central gas temperature of about $47\ \mathrm{K}$ is higher than the $10-20\ \mathrm{K}$ derived from observations \citep[][]{Genzel1989,Stutz2013}. The gas temperature $T_{\rm g}$ plays a role in calculating the grain alignment efficiency, line broadening and level population of our RT simulations. For grain alignment, a factor of $10$ results only in an increase in dust polarization of about $2\ \%$ as outlined in \cite{Reissl2020} in great detail.
Judging from the distribution of energy levels and the line broadening profile (see \S~\ref{sect:ObsZeeman} for details) of the HI $21\ \mathrm{cm}$ line we estimate that the emission may be higher by a factor $1-7$ using a gas temperature of $47\ \mathrm{K}$ instead of $10\ \mathrm{K}$.  However, this would only affect the comparison of the main filament with actual observations. For the comparison of the main filament of with the analytical model we have roughly the same temperature range. Hence, the higher gas temperature does only marginally effect our results and comparisons. The same holds for the assumptions about electron fractions and velocity field: The magnitude of Zeeman observations and $RM$ may be slightly different for a more sophisticated modeling but barely effect the characteristic angular-dependencies of the pattern associated with  different magnetic field structures. 

In Fig. \ref{fig:ModProf} we show the selected SILCC-Zoom profiles as well as their average value in comparison with the  density and temperature profiles of the analytical model. The Plummer profile of the density as well as the linear temperature profiles of our analytical model matches well with the average trends of the SILCC-Zoom main filament. 

We note that the electron fraction in the SILCC-Zoom simulation is calculated self-consistently as outlined in Sect. \ref{sect:MHDSimulation}. However, for the analytical model we take a simpler approach. We assume that the ionization is CR dominated and that the ionization is in equilibrium with the recombination. Hence, the electron density should depend on gas density via $n_{\mathrm{el}} \left( n_{\mathrm{g}}  \right) \propto n_{\mathrm{g}}^{1/2}$ \citep[see e.g.][chapter 7]{Tielens2005}. However, fitting the exact parameters gives a relation of $n_{\rm el}(n_{\rm g}) = 5.3\times10^{-4}\ n_{\rm g}^{0.69}$ for the SILCC-Zoom filament where the exponent is a little larger than $1/2$ (see Fig. \ref{fig:ModProf}). We note that the average ratio of electrons to gas for all cells of the SILCC-Zoom filament is ${ n_{\mathrm{el}} / n_{\mathrm{g}} = 8.1\times 10^{-4} }$. Such an electron fraction is of the same order as the one derived with a chemical model in \citet{Tahani2018}. For the gas velocity of the analytical model we take an average vector field over all velocity vectors per cell of the SILCC-Zoom grid.

For the synthetic dust polarization and Zeeman observations of the analytical model we consider two distinct magnetic field morphologies. For later comparison we use the helical field parametrization,
\begin{equation}
\begin{split}
{\bf B}_{\mathrm{heli}}(X,Y,Z) = \qquad\qquad\qquad\qquad\qquad\qquad\qquad\qquad\qquad\qquad \\  B_{\rm 0}\left( n_{\mathrm{g}} \right) \left(-Z\cos \alpha /( X^2+Y^2)^{1/2},X\cos \alpha /( X^2+Y^2)^{1/2}, \sin \alpha \right)^T\; ,
\end{split}
\label{eq:MagHeli}
\end{equation}
presented in \cite{Reissl2018} written in the coordinate system of the SILCC-Zoom simulation. $T$ denotes the transposed vector, the quantity $\alpha$ is the pitch angle of the helical field, and $B_{\rm 0}\left( n_{\mathrm{g}} \right)$ is a scaling factor dependent on gas density $n_{\mathrm{g}}$. Furthermore, we limit our investigation to a representative case of $\alpha=30^\circ$ named model $heli_{\rm 30}$ in the nomenclature of \cite{Reissl2018}. For the scaling of the magnetic field strength of the analytical model we follow the relation $B_{\rm 0}\left( n_{\mathrm{g}} \right) \propto n_{\rm g}^{\kappa}$ reported in \cite{Crutcher2010}. Fitting this relation to the SILCC-Zoom data of the main filament, we find $B_{\rm 0}\left( n_{\mathrm{g}} \right) = 3.1\ \mu \mathrm{G}\ \times (n_{\rm g}/ 1\ \mathrm{cm}^{-3} )     ^{0.17}$. However, in \cite{Crutcher2010} they derive a power law index of $\kappa=0.65$ for $n_{\rm g}  \gtrsim 500\ \mathrm{cm^{-3}}$ and a constant tail for smaller densities. The selected SILCC-Zoom sub-region with a maximal density of $n_{\rm g}  \lesssim 3000\ \mathrm{cm^{-3}}$ falls exactly in this transition region and hence $\kappa=0.17$ seems to be justifiable. The magnetic field - density relation is plotted in Fig. \ref{fig:ModProf} showing a good match between the analytical-model and the average  magnetic field of the SILCC-Zoom main filament. 

A kinked magnetic field is reported by \cite{Inoue2013} and \cite{LiKlein2019}. Their field possesses a similar configuration as the SILCC-Zoom simulation introduced in \S~\ref{sect:MHDSimulation}. A comparable magnetic field parametrization is presented in \cite{Reissl2018}. For our analytical model we use their magnetic field parameterization:
\begin{equation}
{\bf B}_{\mathrm{flow}}(X,Y,Z) = B_{\rm 0}\left( n_{\mathrm{g}} \right)\frac{\left(1,-5\zeta X(2-Y)^2\exp(-8\zeta^2 X^2), 0\right)^T}{1+25\zeta^2 X^2 (2-Y)^4\exp(-16 \zeta^2 X^2)}\, .   
\label{eq:MagFlow}
\end{equation}
For this parametrization \citep[named model $flow$ in][see also their Figure 1 for detailed 3D illustrations of the field morphology]{Reissl2018} we introduce the additional parameter $\zeta$ for our analytical model. In order to control the width of the kink for a better comparison with the SILCC-Zoom results we use a range of $0.15 - 0.25$ for $\zeta$ in the following sections. 


\section{Synthetic observations}
\label{sect:Obs}
Determining polarization requires measurements of the different orientations of the electric field vector ${\bf E}$ and is typically characterised by the Stokes vector ${\bf S}=(I,Q,U,V)^T$ where the parameters describe the total intensity $I$, the linear polarization represented by $Q$ and $U$, and the circular polarization $V$. The polarization state of radiation is then determined by the degree of linear polarization
\begin{equation}
P_{\mathrm{l}}=\frac{\sqrt{Q^2+U^2}}{I} = \frac{I_{\mathrm{p}}}{I},
\label{eq:PlDefinition}
\end{equation}
where the quantity $I_{\mathrm{p}}$ is the linearly polarised part of the total intensity. The orientation angle of linear polarization is give by
\begin{equation}
\chi=\frac{1}{2} \tan^{-1}\left(\frac{U}{Q} \right)\, ,
\label{eq:LinearOrientation}
\end{equation}
and the degree of circular polarization by
\begin{equation}
P_{\mathrm{c}}= \frac{I_{\mathrm{R}} - I_{\mathrm{L}}}{I} =  \frac{V}{I}\, ,
\label{eq:PcDefinition}
\end{equation}
where $I_{\mathrm{R}}$ and $I_{\mathrm{L}}$ are the  left-hand and right-hand, respectively, components of the circularly polarised light. We note that $P_{\mathrm{l}} \in [0,1]$, whereas $P_{\mathrm{c}} \in [-1,1]$. {\sc POLARIS} solves the RT problem in all four Stokes parameters simultaneously for dust polarization, line RT with Zeeman effect, and synchrotron polarization alike by a Runge-Kutta-Fehlberg (RFK45) solver with an inbuilt error correction. For details we refer the reader to Appendix B in \cite{Reissl2019}.


\subsection{Dust heating and grain alignment}
\label{sect:ObsDust}
{\sc POLARIS} allows the user to calculate the radiation field considering a number of distinct photon emitting sources by means of a Monte-Carlo (MC) RT scheme \cite[][]{Reissl2016}. It includes scattering, absorption, and re-emission on dust grains.
The life cycle of each photon package is governed by random sampling of physical quantities such as the path length between scattering events and the change in light propagation direction after scattering. Consequently, the radiation field contains an inherent amount of noise. Once the radiation field is known, {\sc POLARIS} calculates the dust temperature assuming an energy balance between absorbed and emitted radiation \cite[][]{Lucy1999,Bjorkman2001,Reissl2016} as well as the grain alignment efficiency (see below). 

Linear polarization as well as circular polarization of grains arises from elongated dust grains. In general, polarization on dust grains may also arise due to scattering processes but such effects can be neglected in the IR and sub-mm regime of wavelength. Elongated grains emit light preferentially along their majors axis while having the tendency to align with their minor axis with the magnetic field direction. Hence, dust emission traces the magnetic field orientation rotated by $90^\circ$.

The total dust mass is determined by applying a constant dust mass to gas mass ratio of $m_{\mathrm{dust}}/m_{\mathrm{gas}}=0.01$ in each grid cell. The dust itself is a mixture of materials consisting of $62.5\ \%$ astronomical silicate and $37.5\ \%$ graphite. The size distribution follows the canonical power-law ${ N(a)\propto a^{-3.5} }$ where $a$ is the effective radius of a {non-spherical} dust grain equivalent to the volume of sphere of the same mass \citep[][]{Mathis1977}. We apply a sharp cut off for the lower size distribution at $a_{\mathrm{min}}=5\ \mathrm{nm}$ as well as the upper grain size at $a_{\mathrm{max}}=250\ \mathrm{nm}$. We note that the grain distribution is most certainly different within the filament with grain sizes up to $~1\ \mu\mathrm{m}$ \citep{Rouan1984,Hankins2017}. However, the grain growth processes  \citep[][]{Ossenkopf1994} as well as the size redistribution e.g. by rotational disruption \citep[][]{Hoang2019} is still a field of ongoing research. Using a different grain model would influence the magnitude of dust emission and polarization degree but has little effect on the angular-dependencies of the dust polarization. For instance, in \cite{Seifried2020} the entire SILCC-Zoom region is post-processed with {\sc POLARIS} using a similar dust setup but with a maximal grain size of $a_{\mathrm{max}}=2\ \mu\mathrm{m}$. They report a peak dust polarization that is about a factor of $1.2$ higher than ours while the polarization pattern itself is basically the same.

Assuming an energy equilibrium between absorption and emission, the linearly polarised intensity in emission depends on the wavelength $\lambda$ and follows a modified black body spectrum. For simplicity, we make use of a dust temperature $T_{\mathrm{d}}$ averaged over all grain sizes and dust materials within the scope of this paper. Hence, the contribution per path element $\mathrm{d}\ell$ to the total dust intensity is
\begin{equation}
\mathrm{d}I  \propto n_{\mathrm{g}} C_{\mathrm{abs}} B_\lambda \left( T_{\mathrm{d}} \right) \mathrm{d}\ell\, ,
\label{eq:Idust}  
\end{equation}
and the contribution to linear polarization may be calculated by
\begin{equation}
\mathrm{d} I _{\mathrm{p}}  \propto n_{\mathrm{g}} \Delta C_{\mathrm{abs}} \times B_\lambda \left( T_{\mathrm{d}} \right) \mathrm{d}\ell\, 
\label{eq:Ipdust}  
\end{equation}
where $B_\lambda \left( T_{\mathrm{d}} \right)$ is the Planck function. The quantities  $C_{\mathrm{abs}}$ and $\Delta C_{\mathrm{abs}}$ are the cross sections of total absorption and polarised absorption, respectively, size-averaged over the distribution function $N(a)$. For IR and sub-mm emission these cross sections can be calculated as
\begin{equation}
C_{\mathrm{abs}} = \int_{a_{\mathrm{min}}}^{a_{\mathrm{max}}}N(a)\left[C_{\bot}(a) + C_{||}(a) \right]\ \mathrm{d}a
\label{eq:Cpol}  
\end{equation}
and
\begin{equation}
\Delta C_{\mathrm{abs}}\cong \sin^2 \vartheta  \int_{a_{\mathrm{alg}}}^{a_{\mathrm{max}}}R(a)N(a)\times  \left[ C_{\bot}(a) - C_{||}(a) \right]\ \mathrm{d}a\, ,
\label{eq:DCpol}  
\end{equation}
where $\vartheta$ is the angle between the LOS and the magnetic field direction, $C_{\bot}(a)$ and $C_{||}(a)$ are the pre-calculated cross sections perpendicular and parallel, respectively, to the grain's minor principal axis \citep[see][for details]{Reissl2017}, and $R(a)$ is the Rayleigh reduction factor accounting for the grain alignment efficiency.


The key aspect in modelling accurate synthetic dust polarization is calculating the grain alignment efficiency. The reduction in polarization is handled by {\sc POLARIS} following the physics of the radiative torque (RAT) alignment theory \citep[][]{Lazarian2007}. The quintessential part of the RAT theory entails a spin-up process of irregular grains by an anisotropic radiation field and the subsequent alignment of paramagnetic grains with the magnetic field direction. This effect is opposed by random bombardment of gas particles. Thus, the spin-up process of grains can be quantified by the ratio \citep[see e.g.][]{Lazarian2007,Hoang2014}
\begin{equation}
\left(\frac{J_{\rm{RAT}}}{J_{\rm{g}}}\right)^2 \propto a \left(\frac{1}{ n_{\rm g} T_{\rm g}} \int \lambda Q_{\mathrm{RAT}} \gamma_\lambda u_\lambda \mathrm{d}\lambda \right)^2\, ,
\label{eq:OmegaRatio}
\end{equation}
where $J_{\rm{RAT}}$ is the angular momentum of the dust grain by radiation and $J_{\rm{g}}$ is the angular momentum induced by random gas collisions. The 
anisotropy factor, $\gamma_\lambda$, and the energy density, $u_\lambda$, of the radiation field are calculated by {\sc POLARIS } with a MC approach\footnote{For a detailed description of the latest implementation of the  {\sc POLARIS} RT in combination with RAT physics we refer to \cite{Reissl2020}.}. For the grain alignment efficiency we use the parametrization 
\begin{equation}
 Q_{\mathrm{RAT}} = \cos \Psi  \begin{cases} 0.4   &\mbox{if } \lambda \leq 1.8\ a \\ 
		    0.4 \times\left(\dfrac{\lambda}{1.8\ a} \right)^{-2.6}  & \mbox{otherwise} \end{cases}\, .
 \label{eq:QGamma}
\end{equation}
suggested in \cite{Herranen2018} based on the average of a large ensemble of distinct grain shapes. Here, $\Psi$ is the angle of the anisotropic radiation field and the magnetic field direction \citep[see][for details]{Reissl2016,Reissl2020}. 

The range of dust grain sizes $a$ that have a stable alignment i.e. $R(a) = 1$ is usually parameterised by $J_{\rm{RAT}}/J_{\rm{g}} > 3$  \citep[see e.g.][for details]{Hoang2014}. This allows to calculate the characteristic grain size $a_{\mathrm{alg}}$ where $J_{\rm{RAT}}/J_{\rm{g}} = 3$. Hence, all dust grain sizes $a>a_{\mathrm{alg}}$ can contribute to the dust polarization. We emphasise that the product $n_{\rm g} T_{\rm g}$ in Eq. \ref{eq:OmegaRatio} acts as a pressure term. Consequently, the dust polarization signal increases for a stronger radiation field and decreases in dense regions or regions of higher gas temperature.

A second criterion for grain alignment is related to the precession
timescales of the grains and the local magnetic field strength \citep[][]{Lazarian2007}. However, this criterion is always fulfilled for the parameters provided by the SILCC-Zoom simulation, the dust model, and the resulting radiation field \citep[for more details we refer to Appendix C in][]{Reissl2020}. Furthermore, we allow only silicate grains to align while graphite grains cannot due to their vastly different paramagnetic properties \citep[][]{Hunt1995,Draine1996,Hoang2014A}.

A small amount of the linearly polarised radiation can also be transferred into the Stokes $V$ component, i.e. circular polarization, by means of dichroism of non-spherical grains \citep[see][]{Martin1974,Whitney2002,Reissl2016}. This process is amplified by the amount of twisted field lines along the LOS. Thus, circular dust polarization profiles observed perpendicular to the filament may also help to probe a particular  magnetic field morphology. Indeed, this effect has already been explored as a technique complementary to linear dust polarization profiles and Zeeman measurements in \cite{Reissl2014} and \cite{Reissl2018}, respectively.

\subsection{Line transfer and Zeeman effect}
\label{sect:ObsZeeman}
In this paper we utilise the HI $21\ \mathrm{cm}$ line for the RT with Zeeman effect. This line is a good tracer for gas densities of the order $10-100\ \mathrm{cm^{-3}}$ because of its low critical density \citep[][]{Crutcher1999} matching well with the spine and the diffuse medium surrounding the filament.


In the presence of a magnetic field, characteristic line transitions split into well defined sub-levels. 
The lower and upper sub-levels are defined by the magnetic quantum numbers $M'$ and $M''$, respectively. Selection rules for these quantum states allow only for so called $\pi$ transitions with $\Delta M = 0$ and $\sigma_\pm$ transitions $\Delta M = \pm 1$. Linear polarization is associated with a $\pi$  transition, while circular polarization arises from the phase shift between the $\sigma_\pm$ transitions. The equation for line RT including Zeeman effect takes the following form: {$\mathrm{d} S_\nu/\mathrm{d} \ell =J_\nu - \hat{K}_\nu S_\nu$}. The propagation matrix 
\begin{equation}
\hat{K}_\nu = n_{\rm HI} \frac{h \nu}{8\pi} n' A_{\rm ij} \sum_{\rm M',M''} S_{\rm M',M''} F_{\rm B}(\nu,\Delta \nu_{\rm z})\hat{A} 
\label{eq:ZeemanK}
\end{equation}
and the emissivity 
\begin{equation}
J_\nu = n_{\rm HI} \frac{h \nu}{8\pi} \left(n' B_{\rm ji} - n'' B_{\rm ij}\right) \sum_{\rm M',M''}  S_{\rm M',M''}F_{\rm B}(\nu,\Delta \nu_{\rm z}) \left( \hat{A}S_{\rm 0}\right)
\label{eq:ZeemanJ}
\end{equation}
depend on the HI number density $n_{\rm HI}$, the characteristic line strengths $S_{\rm M',M''}$ between the lower level and upper level (indicated by $'$ and $''$, respectively) as well as the Einstein coefficient for spontaneous emission $A_{\rm ij}$, photon absorption $B_{\rm ji}$ and induced emission  $B_{\rm ij}$, and $h$ is the Planck constant. The matrix $\hat{A}$ describes the angular-dependencies of the $\pi$ and $\sigma_\pm$ transitions with respect to the LOS, the unit vector is defined to be ${{\bf S}_{\rm 0}=(1,0,0,0)^T }$, and $F_{\rm B}(\nu,\Delta \nu_{\rm z})$ is the line broadening profile \citep[see][]{Larsson2014,Brauer2017A}.  Furthermore, 
\begin{equation}
\Delta \nu_{\rm z}=B \frac{\mu_{\rm B}}{h}\left(g'M'-g''M''\right)
\end{equation}
is the Zeeman shift between the $\sigma_\pm$ transitions \citep[][]{Brauer2017A} where as the constant $\mu_{\rm B}$ is the Bohr magneton, and the quantities $g'$ and $g''$ are statistical weights. For simplicity we assume local thermodynamic equilibrium for the upper and lower population $n'$ and $n''$ of the energy level of HI, respectively, with
${ n'/n'' \propto \exp( 0.07\ \mathrm{K} / T_{\rm g} )}$. As discussed in \cite{Reissl2018} other level population schemes do not significantly change the line RT including Zeeman effect. This gives the spectrum of the observed total intensity
\begin{equation}
I_\nu = I_{\mathrm{L}} + I_{\mathrm{R}}\, ,
\label{eq:ZeemanI}
\end{equation}
as a function of the frequency $\nu$. For Zeeman split lines, the shape of the circular component is the first derivative of $I_\nu$,
\begin{equation}
V_\nu = \frac{\mathrm{d} I_\nu}{\mathrm{d} \nu}\Delta \nu_{\rm z} \cos \vartheta\
\label{eq:ZeemanV}
\end{equation}
Hence, the Zeeman effect is primarily dependent on the gas density $n_{\rm HI}$, the gas temperature $T_{\rm g}$, and the LOS magnetic field strength ${B}_{||}=B\cos \vartheta$. Consequently, observing Zeeman split line emission profiles in $I$ and $V$ in multiple velocity channels gives an estimate of the magnetic field strength and the field orientation with respect to the LOS.

\subsection{Faraday rotation}
\label{sect:ObsSync}
The {\sc POLARIS} code is capable of solving the full synchrotron RT problem including Faraday rotation and Faraday conversion \citep[see][for details]{Reissl2019}. However, in this paper we only focus on the particular implications of Faraday rotation. When a plasma is exposed to an external magnetic field, the polarised light passing though the medium experiences a rotation in the orientation angle $\chi$ of the linear polarization. Hence, the observed orientation of polarization, $\chi_{\mathrm{obs}}$, becomes
\begin{equation}
\chi_{\mathrm{obs}}=\chi_{\mathrm{source}}+\lambda^2 \times RM\,  ,
\end{equation}
where $\chi_{\mathrm{source}}$ is the polarization angle of the light when it gets emitted from some source and $\lambda$ is the wavelength of the radiation. Hence, the rotation measurement 
\begin{equation}
RM = \frac{1}{2\pi}\frac{e^3}{m_{\rm e}^2c^4}\  \int_{\ell_{\mathrm{source}}}^{\ell_{\mathrm{obs}}}\ n_{\rm el}(\ell) \times B_{||}(\ell)\ \mathrm{d} \ell
\label{eq:DefinitionRM}
\end{equation}
 is a wavelength independent quantity along the LOS. Here, $e$ is the electron charge, $m_{\rm e}$ is the mass of an electron, and $c$ is the speed of light. The source and observer are positioned at $\ell_{\mathrm{source}}$ and $\ell_{\mathrm{obs}}$, respectively. Since $\chi_{\mathrm{obs}}$ depends on $\lambda^2$, the $RM$ can be determined by multi-wavelength radio observations. 
Consequently, the $RM$ traces the LOS magnetic field strength, $B_{||}=B\cos \vartheta$,  modulated  by the density $n_{\rm el}$ of free electrons along each path element $\mathrm{d} \ell$. We note that $RM$ and the Zeeman effect have the same angular-dependency concerning LOS and magnetic field direction. For details about the implementation of synchrotron RT including the Faraday $RM$ refer to \cite{Reissl2019}. 

\subsection{Post-processing and image analysis}
As a first step, we calculate the dust temperature distribution and grain alignment efficiency for the SILCC-Zoom cutout as well the analytical model with its two different magnetic field parameterizations. For the  SILCC-Zoom cutout, we utilise the twelve massive stars located in three star-forming regions of the SILCC-Zoom cutout (see Fig. \ref{fig:3DMHD}) and an ISRF with $G_{\mathrm{0}}=1.7$ \citep[see][for details]{Mathis1983} as photon emitting sources. The spectral energy distribution (SED) of each individual star and its total luminosity $L$ is calculated from the corresponding stellar parameters, i.e. effective temperature and radius provided by the SILCC-Zoom simulation itself. Assuming a black body SED this results in a range of  luminosities of $11.2\ \mathrm{L}_{\odot}$ - $1.62\times 10^4\ \mathrm{L}_{\odot}$. The SED of the ISRF is taken from \cite{Mathis1983} and accounts for the typical stellar and thermal radiation of the ISM in our Milky Way as well as the contribution of the cosmic-microwave-background. For the analytical model we apply only the ISRF as a source since we expect the stars only locally to dominate the overall radiation field. In our particular runs we use $5\times 10^6$ photons per wavelength with $100$ bins logarithmically distributed between $91\ \mathrm{nm}$ and $2\ \mathrm{mm}$ for each of the stars of the SILCC-Zoom data. For the ISRF component we apply $1\times 10^9$ photons representing the optimal compromise between run-time and noise.

In a second step, we perform a ray-tracing scheme \citep[see][]{Reissl2016,Reissl2019}. Here, the grain alignment and the polarization properties of each of the grain size bins is taken individually into account. The LOS is designated to run along the $Y$ axis from $-31.21\ \mathrm{pc}$ to $31.21\ \mathrm{pc}$ (compare Fig. \ref{fig:3DMHD}). Consequently, the detector is parallel to the $\mathrm{XZ}$ plane. For the ray-tracing run, we use a plane-parallel detector with $512^2$ pixel and assume a distance of $400\ \mathrm{pc}$ for the filament from the observer. For the analytical model, we simply look perpendicular to the filament spine. The same detector configuration is applied for the ray-tracing considering the physics of the Zeeman and the Faraday $RM$, respectively, in the following sections. 

For the ray-tracing with dust we create synthetic intensity maps for the visual band ($0.55\ \mu m$) and in the sub-mm regime ($850\ \mu m$). For the visual band, we use the dust extinction to determine the optical depth, $\tau_{\mathrm{V}}$, assuming a constant background radiation $I_{\mathrm{V,back}}$. For convenience, we simply apply $I_{\mathrm{V,back}}=1\ \mathrm{a.u.}$. The observed extincted intensity $I_{\mathrm{V,obs}}$ is related to the optical depth by ${A_{\mathrm{V}}= 2.5\times\log_{\mathrm{10}}\left( I_{\mathrm{V,obs}} / I_{\mathrm{V,back}} \right)\ \mathrm{mag} \approx 1.086\ 
\tau_{\mathrm{V}}\ \mathrm{mag} }$ since ${ I_{\mathrm{V,obs}} = I_{\mathrm{V,back}}\exp(-\tau_{\mathrm{V}}) }$. We use the observed relation ${ N_{\mathrm{H}}=1.59 \times 10^{21} \mathrm{cm^{-2} / mag}\ \times  A_{\mathrm{V}} }$ of visual extincting \citep[][]{Savage1977} in order to get a synthetic map of the column density $N_{\mathrm{H}}$ from our mock observations. 

The spine of the SILCC-Zoom main filament is determined from the $N_{\mathrm{H}}$ map
within the narrow range of $X$ between $-6\ \mathrm{pc}$ and $18\ \mathrm{pc}$. Here, we use the publicly available tool {\sc FilFinder}\footnote{https://fil-finder.readthedocs.io/en/latest/} \citep[][]{Koch2015}.  Finally, we perform a RT ray-tracing simulations with {\sc POLARIS} at a wavelength $850\ \mu m$ in order to calculate the full Stokes vectors and subsequently the maps of linear and circular dust polarization $P_{\mathrm{l}}$ and $P_{\mathrm{c}}$, respectively. 

For the line ray-tracing with {\sc POLARIS } including the Zeeman feature we utilise pre-calculated transition parameters as outlined in \cite{Brauer2017A}. The HI $1420\ \mathrm{MHz}$ (21 cm) line has a characteristic frequency shift of $\Delta\nu / B = 1.4$ for the Zeeman measurements. Here, we apply $221$ velocity channels\footnote{Test runs revealed that the magnetic field strength cannot be reliably recovered for our RT simulations with a number of velocity channels smaller than $71$.} equally distributed between $\pm 13\ \mathrm{km\ s^{-1}}$ (for further details we refer to \citealt{Crutcher1999} and for the implementation of the Zeeman effect in  {\sc POLARIS} to \citealt{Larsson2014} and \citealt{Brauer2017A}). 

The ray-tracing of the Faraday $RM$ is completely defined by the electron distribution $n_{\mathrm{el}}$ as well as the magnetic field strength and direction ${\bf B}$ (see Eq. \ref{eq:DefinitionRM}). Here, we simply utilise the values of  $n_{\mathrm{el}}$ and ${\bf B}$ per cell as they are provided by the SILCC-Zoom cutout and the analytical model, respectively.


\begin{figure}
\centering
\begin{minipage}[c]{0.975\linewidth}
      \includegraphics[width=1.0\textwidth]{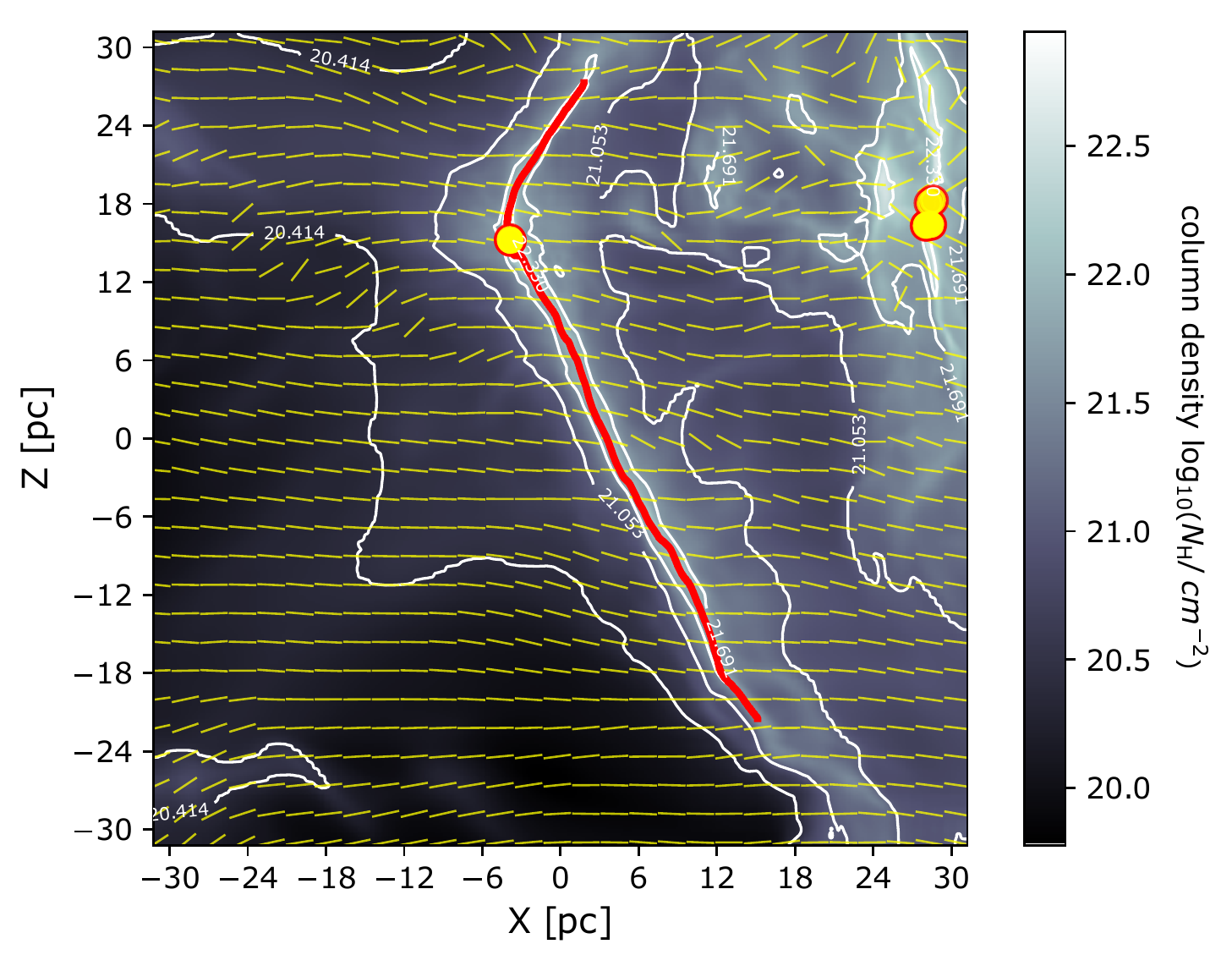}\\
      \vspace{-1mm}
      \includegraphics[width=1.0\textwidth]{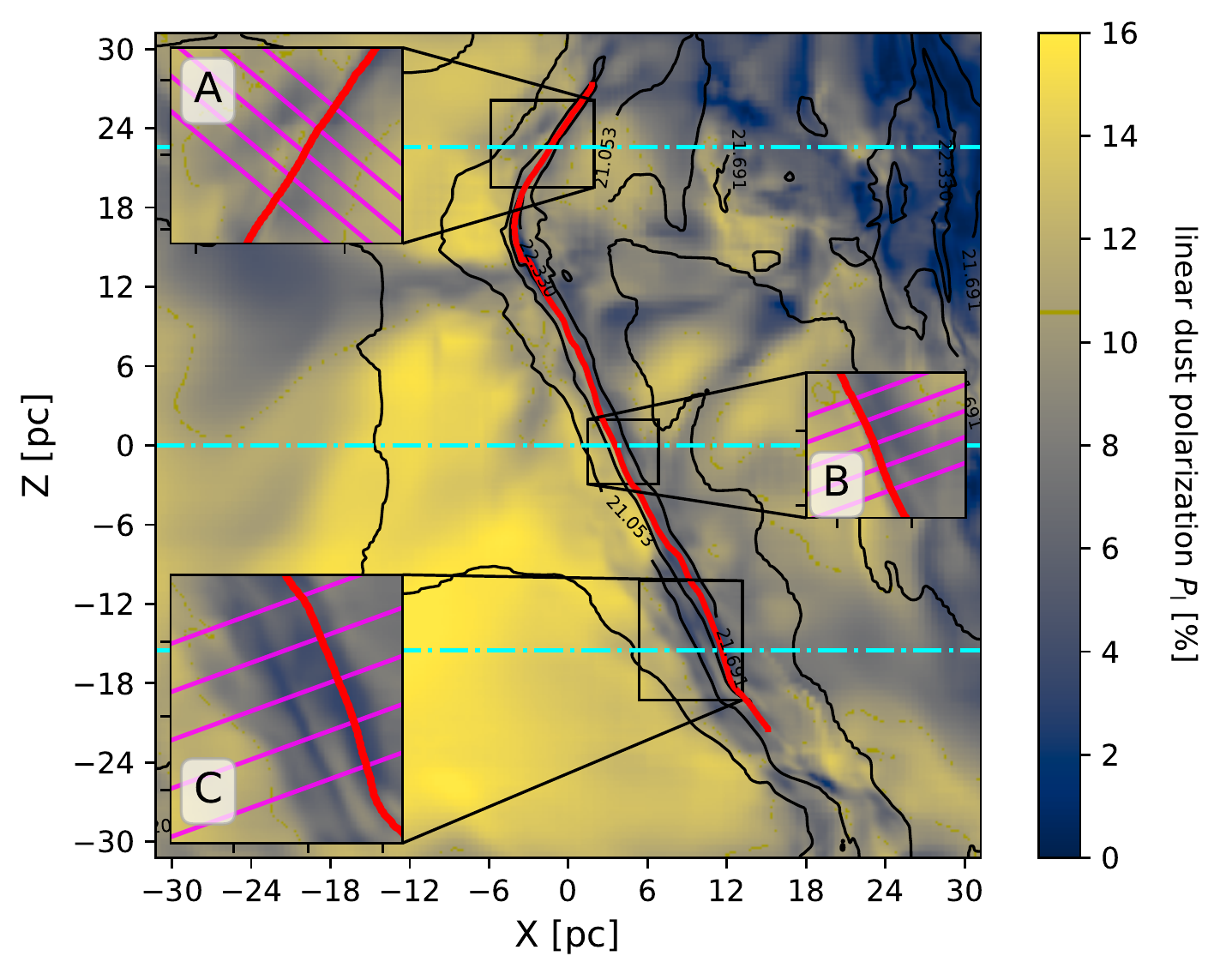}\\
      \vspace{-1mm}
      \includegraphics[width=1.0\textwidth]{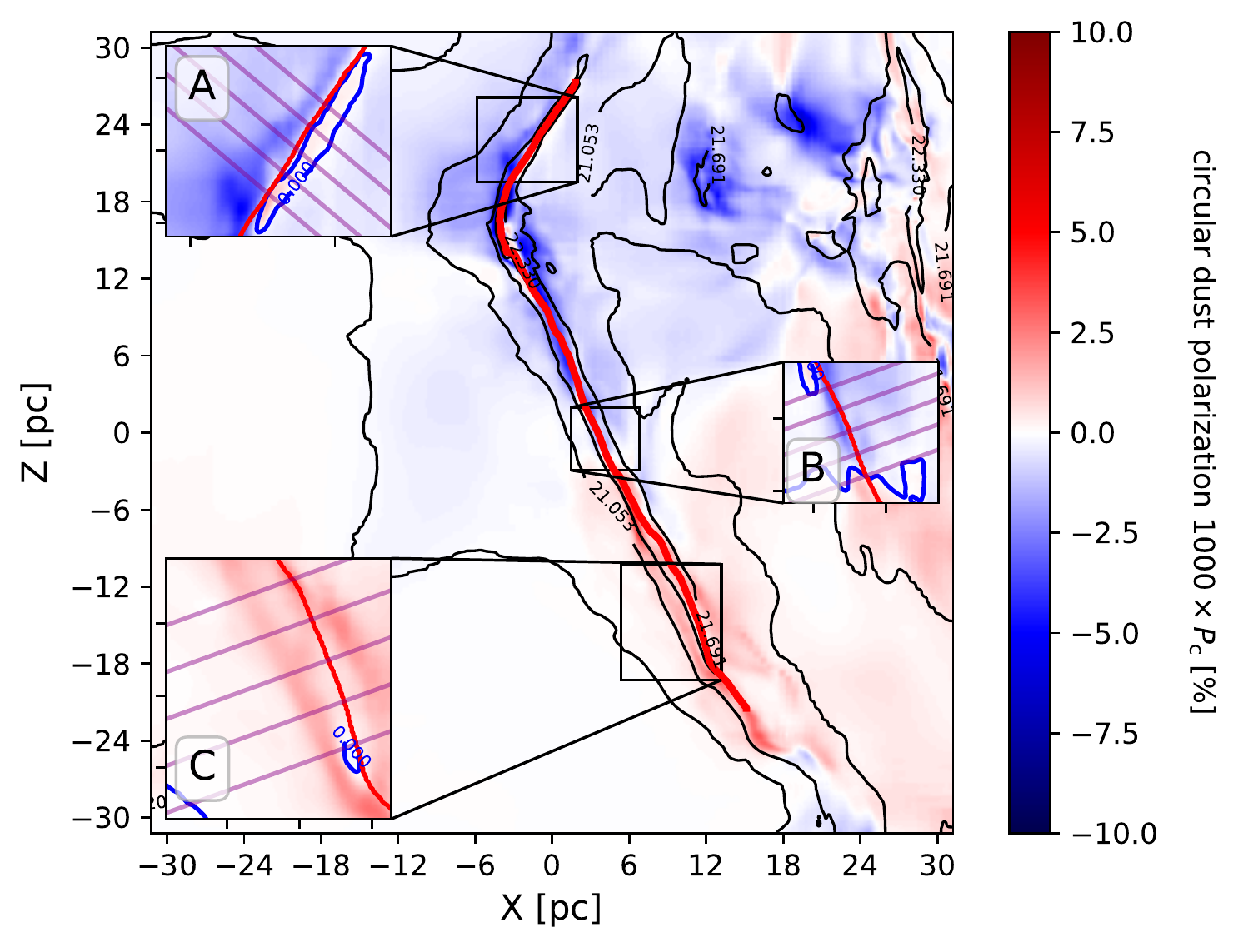}
\end{minipage}
\vspace{-1mm}
\caption{Maps of gas column density $N_{\mathrm{H}}$ (top), linear polarization $P_{\mathrm{l}}$ (middle), and circular dust polarization $P_{\mathrm{c}}$ (bottom), respectively. White and  black lines indicate the $N_{\mathrm{H}}$ contours. The filament's spine is depicted by a red line. Yellow dots correspond to the star-forming regions while yellow lines show the polarization vectors. The polarization vectors are rotated by $90^\circ$ to match the projected magnetic field morphology. Zoom-in panels are labeled A, B, and C and correspond to the regions of possible detection of the underlying magnetic field morphology. We emphasise that these regions are the only ones that show a characteristic dust polarization pattern that may be associated with a kinked field morphology. Other areas along the spine  remain inconclusive. The purple lines in the panels A - C correspond to the profiles shown in Fig. \ref{fig:DustProfiles}. Blue lines in the zoom-in regions of $P_{\mathrm{c}}$ are the contour of $P_{\mathrm{c}}=0\ \%$. Horizontal cyan dash-dotted lines correspond to the $XY-$planes shown in figures \ref{fig:AccDust} and \ref{fig:AccAppDust}.}
\label{fig:DustEmission}
\end{figure}

\begin{figure}
\centering
\begin{minipage}[c]{0.975\linewidth}
      \includegraphics[width=1.0\textwidth]{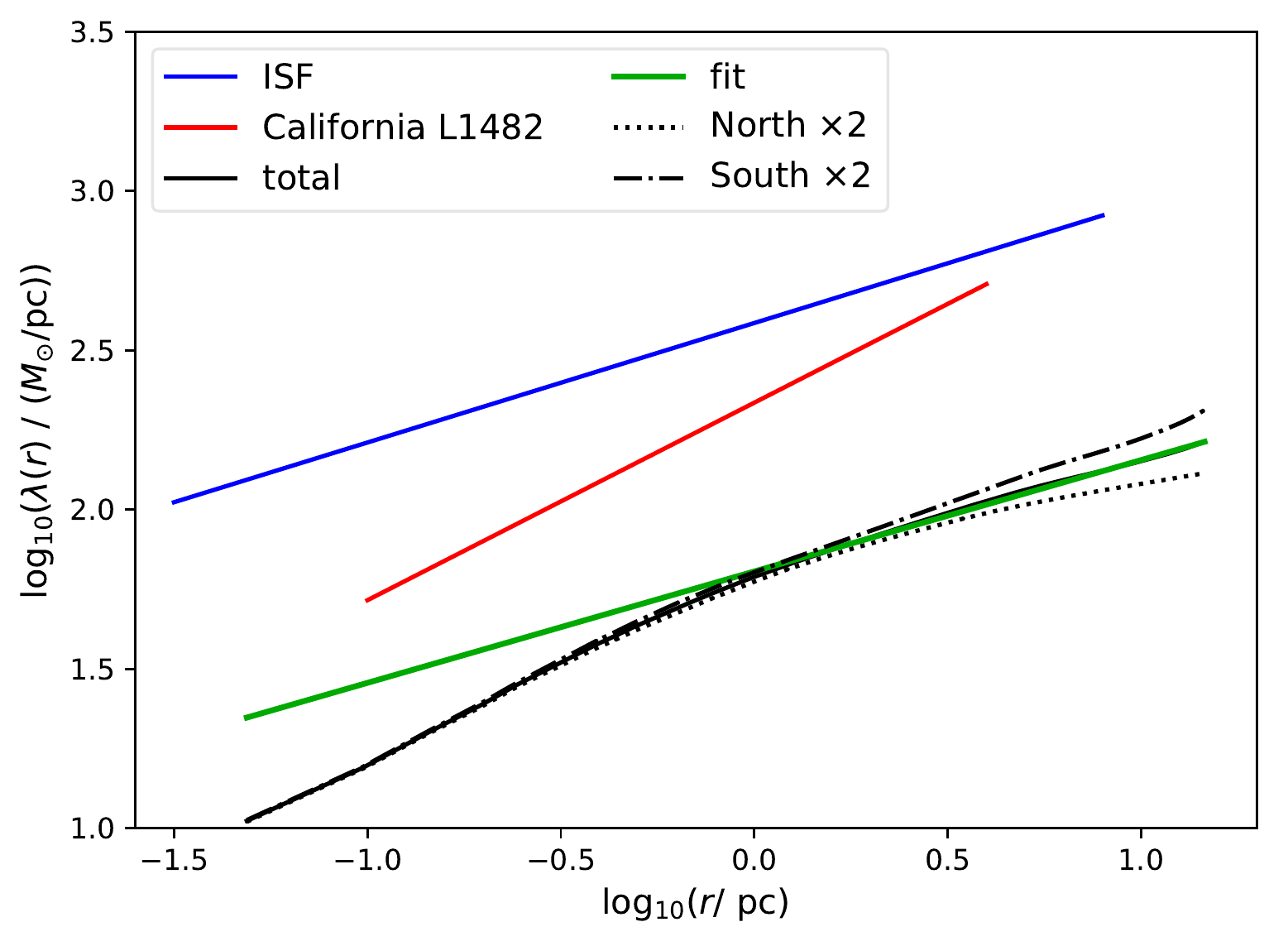}
\end{minipage}
\caption{Enclosed projected mass per unit length M/L for the ISF and California L1641 filament. The M/l profile follows a power law of ${\lambda(r) = K\times( r / \mathrm{pc} )^\gamma}$. The north and south section of the SILCC-Zoom main filament are shown separately (after multiplying by 2). The ISF, L1482 as well as SILCC-Zoom main filament  show roughly the same power-law dependency. For larger radii the slope of the main filament agrees more with that of the ISF where as towards smaller radii the slope is more comparable to that of L1641.}
\label{fig:DiagML}
\end{figure}

\begin{figure*}
\centering
\begin{minipage}[c]{1.0\linewidth}
      \includegraphics[width=0.49\textwidth]{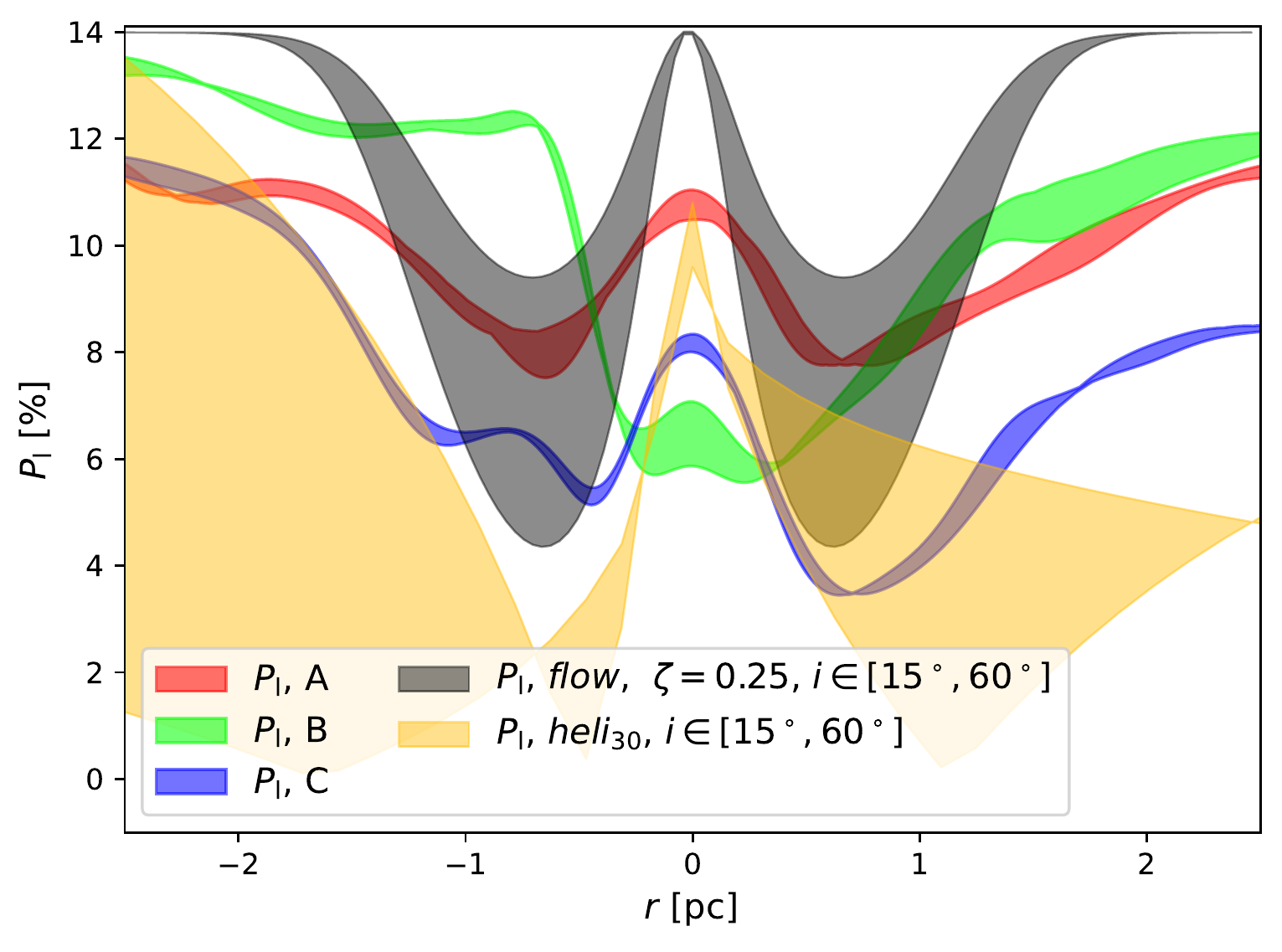}
      \includegraphics[width=0.49\textwidth]{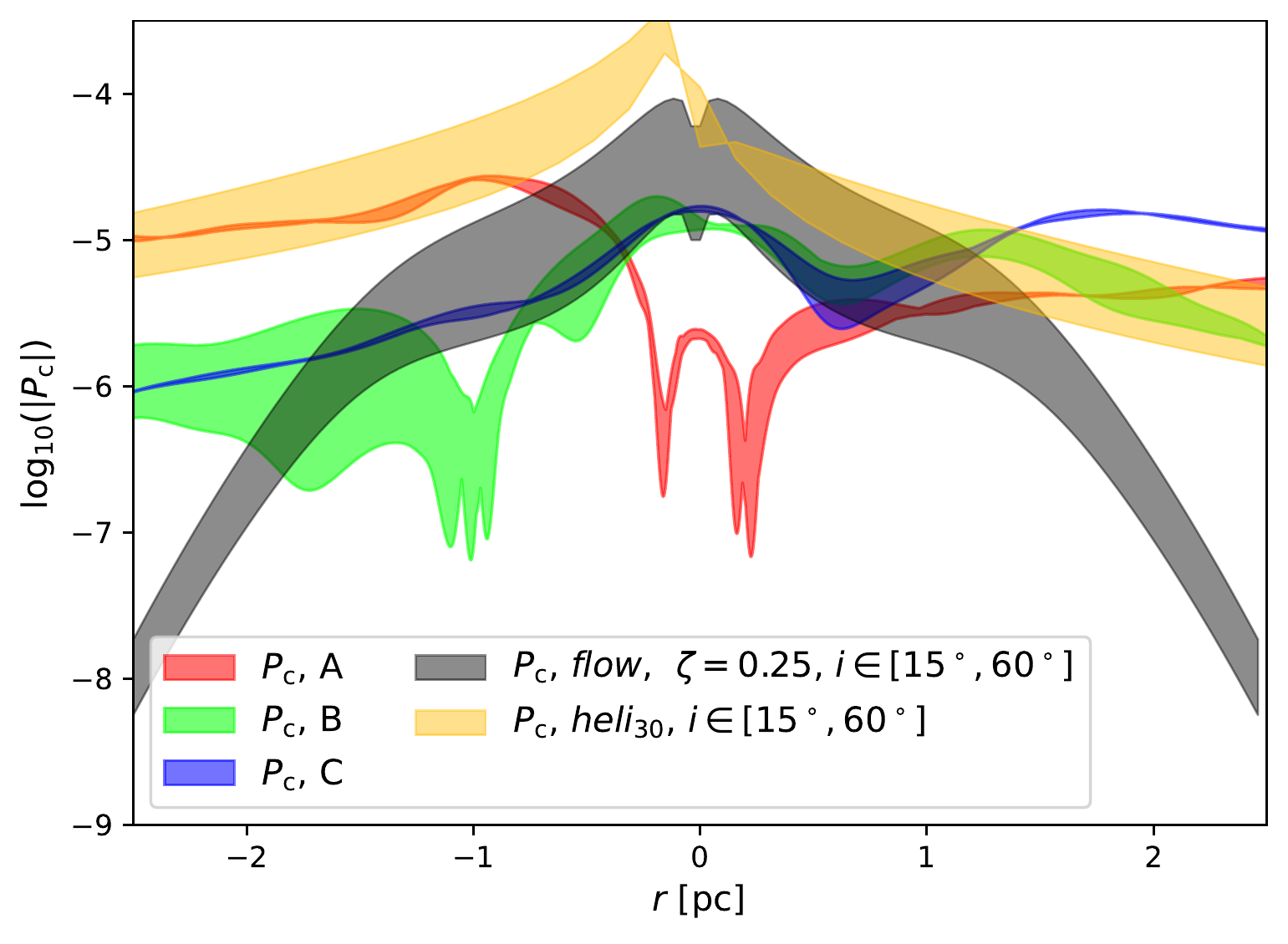}
\end{minipage}
\caption{Profiles of linear dust polarization (left) and circular dust polarization (right). 
The shaded areas in red, green, and blue represent the range of profiles of the regions A,  B, and C, respectively, as depicted in the zoom-in panels of {Fig.~\ref{fig:DustEmission}}. The shaded areas in yellow and gray correspond to the results of the analytical model with the magnetic field parameterizations $heli_{\mathrm{30}}$ and $flow$, respectively, for different inclination angles $i$ and scaling parameters $\zeta$. All central peaks of the profiles are shifted to $r=0\ \mathrm{pc}$ for a better compression. Only the SILCC-Zoom profiles of the linear dust polarization, especially of region A and roughly of region C agree with the analytical model $flow$ where as the circular polarization profiles cannot be assigned to any of the model predictions. }
\label{fig:DustProfiles}
\end{figure*}

\begin{figure*}
\centering
\begin{minipage}[c]{1.0\linewidth}
      \includegraphics[width=0.48\textwidth]{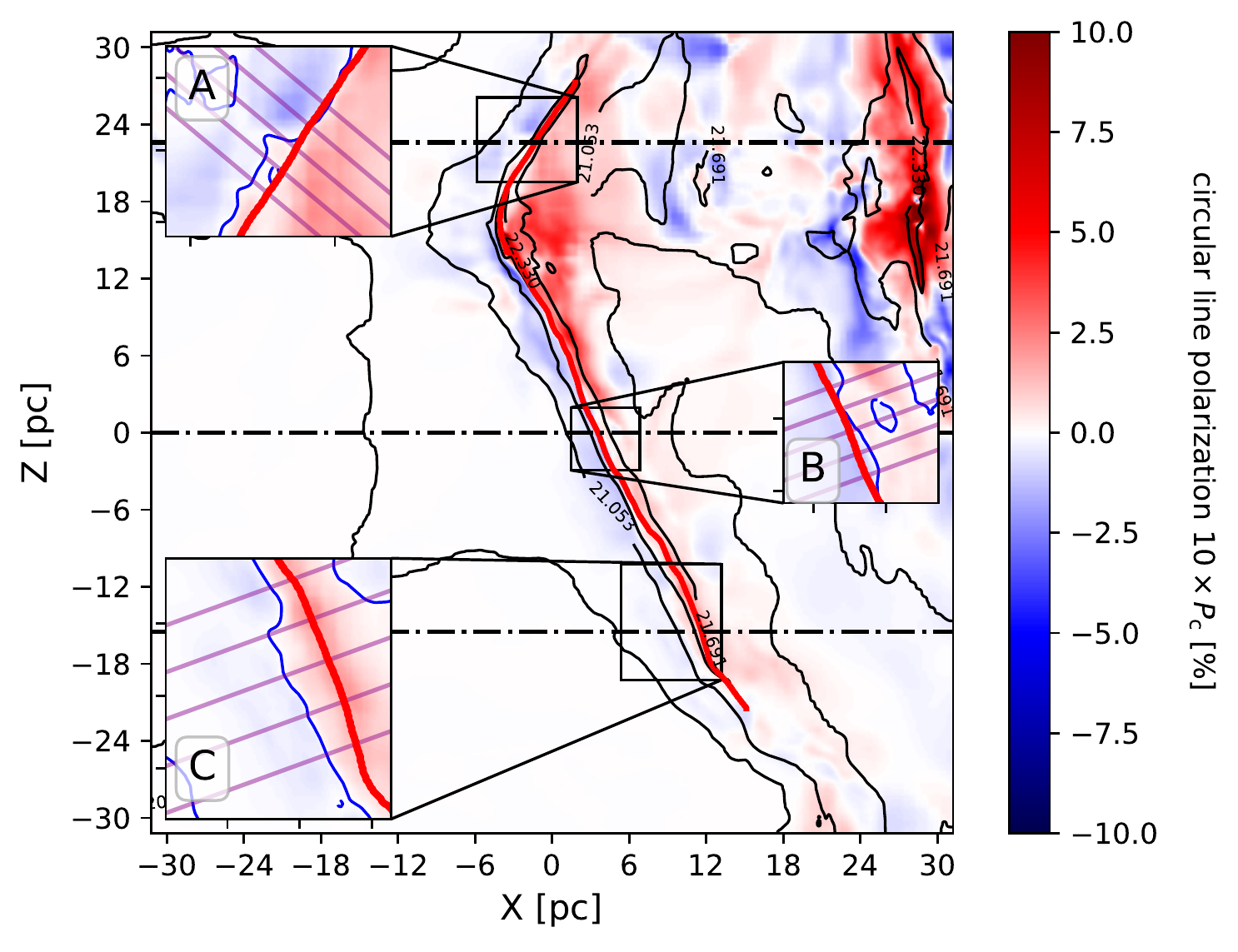}
      \includegraphics[width=0.50\textwidth]{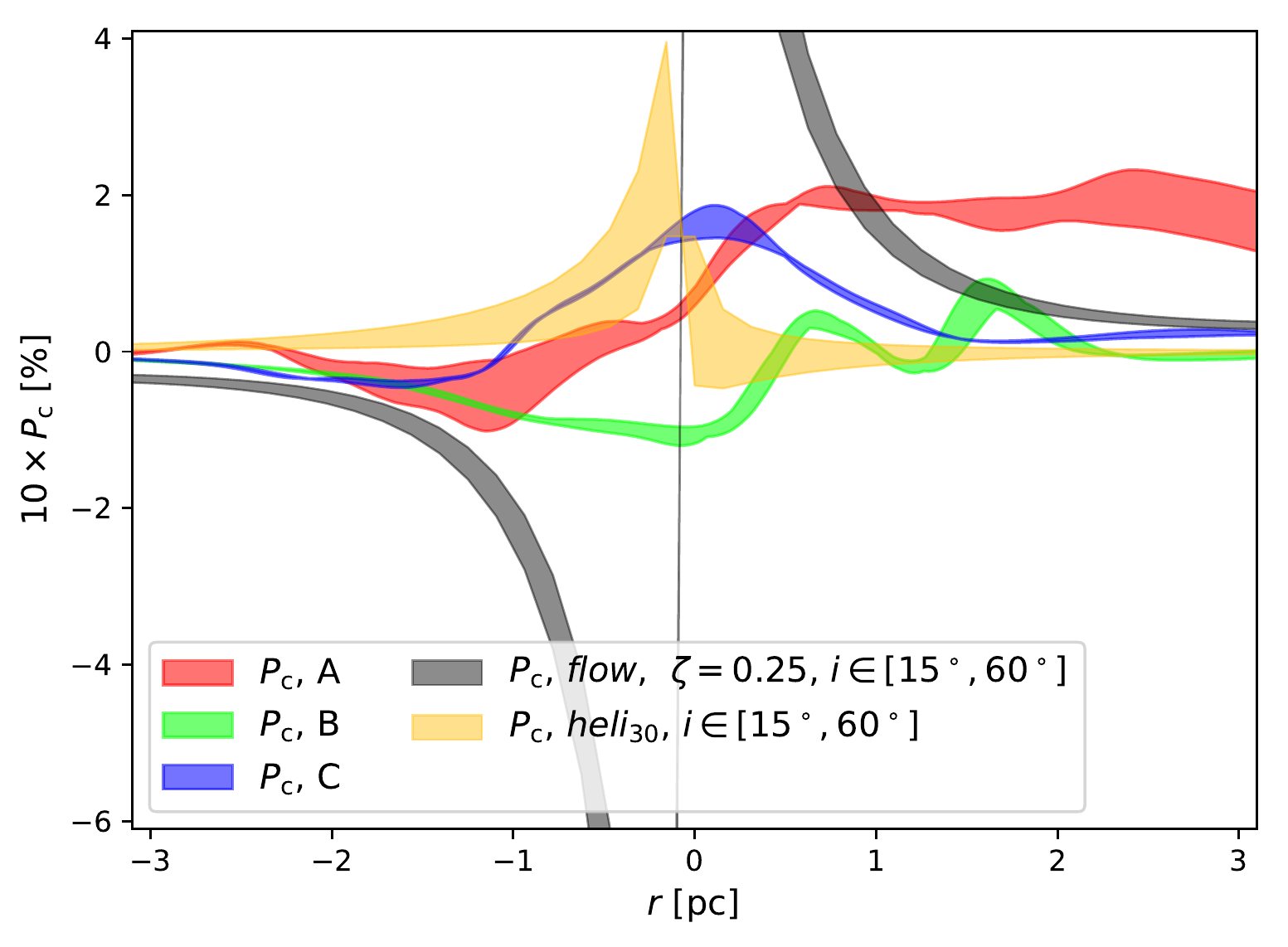}

\caption{Left: Map of the circular polarization $P_{\mathrm{c}}$ of the HI line with contours of $N_{\mathrm{H}}$ in black. Zoom-in regions A, B, and C correspond to the ones in Fig. \ref{fig:DustEmission}. Red lines depict the spine of the main filament, while the purple lines in the zoom-in panels indicate the selected profiles. Horizontal black dash dotted lines correspond to the $XY-$planes shown in figures \ref{fig:AccLine} and \ref{fig:AccAppLine}. Blue contours in are the zero crossings of the circular polarization i.e. $P_{\mathrm{c}} = 0\ \%$.
We  note that the zero crossings of $P_{\mathrm{c}}$ roughly coincide with the spine of the filament. Right: Range of $P_{\mathrm{c}}$  for the five selected profiles of the zoom-in regions A (red), B (green), and C (blue) in comparison with the analytical model with parameterization $flow$ (grey) and $heli_{\mathrm{30}}$ (yellow) for different inclination angles $i$ and scale parameters $\zeta$. All profiles are shifted so that the spine coincides with $r=0\ \mathrm{pc}$ for better compression. The zero crossings and overall shapes of the profiles in the regions A-C do not show any resemblance with the predictions of the analytical model with its magnetic field parameterizations $flow$ and $heli_{\mathrm{30}}$, respectively. }
\label{fig:HIMapProf}
\end{minipage}
\centering
\begin{minipage}[c]{1.0\linewidth}
      \includegraphics[width=0.48\textwidth]{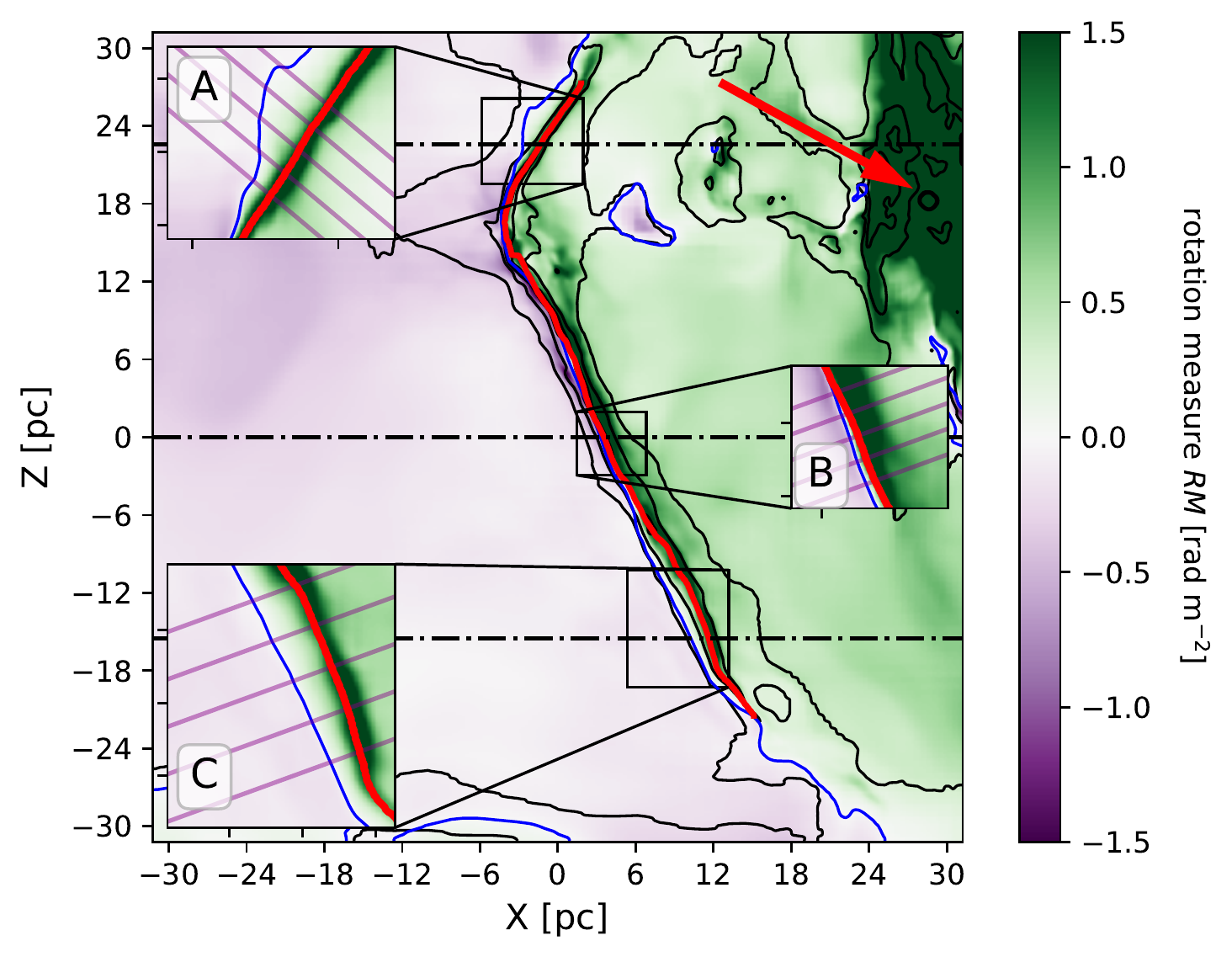}
      \includegraphics[width=0.52\textwidth]{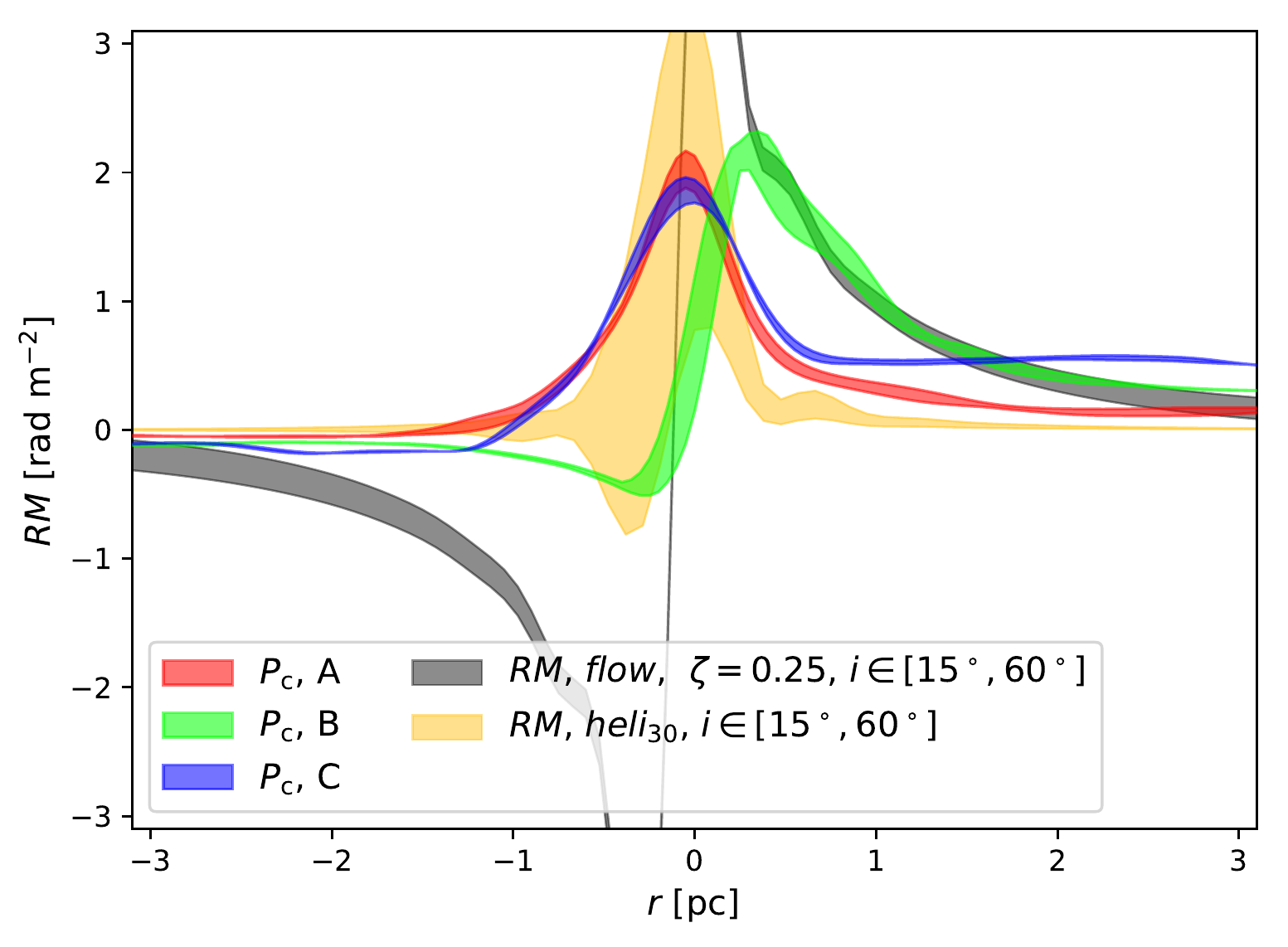}

\caption{ The same as Fig. \ref{fig:HIMapProf} but for the Faraday $RM$. The black contours show the electron dispersion measure $DM = \int n_{\mathrm{el}}(\ell) \mathrm{d}\ell$. Horizontal black dash dotted lines correspond to the $XY-$planes shown in figures \ref{fig:AccRM} and \ref{fig:AccAppRM}. The red arrow points toward the spot of a highly ionizing star-forming region where we report values up to $RM = 10^5\ \mathrm{rad\ m^{-2}}$. The zero crossings of the $RM$ in the map almost coincide with the spine of the filament in regions B and C while A has a larger offset. As fore the $RM$ profiles, only the profiles of region B somewhat agrees with the model $heli_{\mathrm{30}}$ with respect to the shape. Overall, the profile shapes of the main filament do not match the predictions of the analytical model $flow$. }
\label{fig:RMMapProf}
\end{minipage}
\end{figure*}

\subsection{The line of sight analysis technique}
\label{sect:LOSAnalysis}
The transport of polarised radiation is inherently a 3D problem. However, any astronomical observation is merely a projection and one of the spatial dimensions gets lost. Compared to actual observations, the advantage of synthetic observations by means of RT post-processing is that the full 3D information along the LOS remains accessible in each cell of our SILCC-Zoom grid.  Consequently, one can explore in detail the actual origin of any polarization signal.

We emphasise once again that any observation presented in this paper is in the $XZ-$plane while the LOS is along the $Y-$axis from $-31.21\ \mathrm{pc}$ to $31.21\ \mathrm{pc}$ i.e. towards the observer (see Fig. \ref{fig:3DMHD}). Consequently, the polarization signal accumulates in the $XY-$plane. In order to track the polarization signal, we perform RT simulations with {\sc POLARIS} for three distinct regions in the $XZ-$plane, which are selected because they exhibit the characteristic polarization pattern that may be associated with a kinked magnetic field morphology.

In detail, for exploring the origin of polarization we keep track of the polarization signal at each point of these exemplary $XZ-$planes and evaluate the
the relative change along each path element $\mathrm{d} \ell$ of the signal and define
\begin{equation}
\Delta x_{\mathrm{i}} =  \frac{x_{\mathrm{i}}(\ell + \mathrm{d} \ell) - x_{\mathrm{i}}(\mathrm{d} \ell) }{\mathrm{d} \ell}\, .
\label{eq:DeltaX}
\end{equation}
Here, the index $\mathrm{i}$ stands for any position within a certain $XY-$plane of the SILCC-Zoom simulation, the quantity $x$ may represent the linear dust polarization $P_{\mathrm{l}}$, the circular line polarization $P_{\mathrm{c}}$, and the $RM$, respectively. Consequently, the accumulated value $\Sigma x$ of the polarization signal onto the point $Y'$ along the LOS is defined by
\begin{equation}
\Sigma x_{\mathrm{i}} =  \int_{-31.21\ \mathrm{pc}}^{Y'} \Delta x_{\mathrm{i}} \mathrm{d}\ell\, .
\label{eq:SigmaX}
\end{equation}
We estimate the dependency of the polarization signal on the local conditions (represented by $y$) with the help of the Pearson correlation coefficient
\begin{equation}
\mathcal{R}=\frac{\sum_{\mathrm{i}=0}^N  \left( x_{\mathrm{i}}-\overline{x} \right)\times\left( y_{\mathrm{i}}-\overline{y} \right)  }{\left[ \sum_{\mathrm{i}=0}^N  \left( x_{\mathrm{i}}-\overline{x} \right)^2 \times \sum_{\mathrm{i}=0}^N\left( y_{\mathrm{i}}-\overline{y} \right)^2\right]^{1/2}}\, .
\label{eq:DefPearson}
\end{equation}
Here,  $\overline{x}$ and $\overline{y}$, respectively, are the arithmetic means over the entire sample set. Whereas $x$ represents the synthetic polarization signal, the quantity $y$ may serve as a placeholder for the gas pressure $p_{\mathrm{g}}$, gas temperature $T_{\mathrm{g}}$, the radiation field $\zeta$ (see below), magnetic field strength B, electron density $n_{\mathrm{el}}$, or the angular-dependencies $\cos \vartheta$ and $\sin^2 \vartheta$, respectively. Consequently, the possible range of ${-1 \leq \mathcal{R} }$ ( ${ \mathcal{R} \leq 1}$) allows to quantify the negative (positive) correlation of $x$ with $y$. 

In summary this technique allows to explore the origin of the polarization signal and the behaviour of the different tracers along the LOS dependent on the local gas and magnetic field properties. This kind of LOS analysis technique is similar to the one introduced in \cite{Reissl2017}\footnote{The ray-tracing within a certain plane is now a standard feature of {\sc POLARIS}.}. 

\section{Filamentary properties and magnetic field signatures}
\label{sect:FilProp}

\subsection{Mass to length ratio}
We begin our analysis of results by characterizing the mass distribution in our simulated filament, based on the \nh map discussed in \S~4.  One of the most fundamental properties of filaments is the line-mass (M/L) profile \mlr: the ratio of the total mass M of the filament at a given radius to its length L. \cite{Stutz2016} characterise the Integral Shaped Filament (ISF) in Orion~A using the \mlr profile derived from \nh maps of \cite{Stutz2015}, calculated from Herschel dust emission maps. Specifically, they measure the cumulative \mlr profile as a function of the projected radius $r$ (or equivalently, impact parameter) from the filament \nh ridgeline i.e. the spine. Here the \nh spine traces the maximum \nh along the filament. \cite{Stutz2016} and show that the \mlr profile of the ISF is well-approximated by a simple power-law:
\begin{equation}
\lambda(r) = K\times\left( \frac{r}{\mathrm{pc}} \right)^\gamma\, ,
\label{eq:MLRatio}
\end{equation}
where $K$ is a scaling parameter and $\gamma$ is the power-law exponent (see below for the numerical values of these parameters corresponding to the filaments considered in this work). Thus, \mlr is a cumulative measure of the mass of the filament per unit length for a radius $r$ from the spine.

In Fig. \ref{fig:DiagML} we show the \mlr profile of our synthetic \nh map (as seen in Fig.~\ref{fig:DustEmission}), following \cite{Stutz2016}. For comparison, we also present a measure of the \mlr profile of the Orion/ISF \citep{Stutz2016} and California/L1482 \citep{Alvarez2020} filaments, for comparison to observed structures. The ISF has a best fit profile of \mlr${= 385\,\mathrm{M_{\odot}\ pc^{-1}} (r/\mathrm{pc})^{0.38}}$. The California/L1482 filament is less massive and steeper, with \mlr${= 217\ \mathrm{M_{\odot}\ pc^{-1}} (r/\mathrm{pc})^{0.62}}$. We analyze the SILCC-Zoom main filament in the northern ($Z>14\ \mathrm{pc}$) and southern part ($Z<14\ \mathrm{pc}$) separately and find an average profile with \mlr${= 63.2\ \mathrm{M_{\odot}\ pc^{-1}} (r/\mathrm{pc})^{0.34}}$. 

Inspection of Fig. \ref{fig:DiagML} immediately reveals that the \mlr profile of the simulated filament is considerably below Orion/ISF and California, with a \mlr more similar to those of filaments in lower mass regions (although this statement remains to be investigated in detail).  In general, all detailed simulations like ones we use here do not yet capture the higher masses and \mlr values found in regions like the observations explored above. Reaching higher \mlr values is thus one of the next steps in simulation work aimed at understanding star and cluster formation.  That said, the particular SILCC-Zoom simulation we use is unique in that it provides viable means to evaluate the robustness of the inferred magnetic field morphology, given the large number of physical processes included and the high spatial resolution achieved. 



\subsection{Dust polarization and extinction}
\label{sect:DustAndExtionction}

Our RT simulations as outlined in \S~\ref{sect:ObsDust}
result in dust temperatures $T_{\mathrm{d}}$ of $8\ \mathrm{K} - 18\ \mathrm{K}$ for the SILCC-Zoom simulation as well as the analytical model. For the cells of the SILCC-Zoom harboring the massive stars we report temperatures of roughly $50\ \mathrm{K} - 100\ \mathrm{K}$. 

In Fig. \ref{fig:DustEmission} we present the corresponding synthetic maps of column density and dust polarization as well as the polarization vectors resulting from the SILCC-Zoom RT post-processing. 
We emphasise, that the polarization vectors are rotated by $90^\circ$ to match the projected magnetic field morphology. We find that the polarization vectors are mostly parallel to each other following the $X$ axis. Such a polarization pattern perpendicular to the spine of a filament is also reported e.g. by \cite{Pillai2017}. Only in the upper right corner of the column density map ($X>18\ \mathrm{pc}$ and $Z>6\ \mathrm{pc}$) the polarization vectors become increasingly unordered. This effect is mostly due to the twisted field geometry along the LOS through the diffuse cloud in that region  (compare Fig. \ref{fig:3DMHD}).

As we show in Fig. \ref{fig:Midplanes}, the characteristic kink of the dragged magnetic field correlates with the spine of the main filament. Indeed, evaluating the magnetic field structure in the MHD grid along the spine reveals that the kink is always correlated with the main filament. By design, this situation is similar to the analytical model introduced in \S~\ref{sect:ToyModel} with the magnetic field parameterization $flow$. Hence, one may expect to find similarities in the dust polarization features of the main filament and the analytical model. Since the main filament of the SILCC-Zoom simulation is not alwasy exactly parallel to one of the axes of the simulation domain, we vary the inclination angle $i$ of the analytical model between $15^\circ$ and $60^\circ$ for a better comparison. Here, the inclination $i$ is defined with respect to the normal vector of the spine of the filament. In the dust polarization maps of the SILCC-Zoom and analytical model we take radial profiles centered on the observed spines.


In Fig. \ref{fig:DustProfiles} we present radial profiles in comparison with the profiles calculated from the analytical model considering the magnetic field parameterizations $flow$ as well a the helical field $heli_{\mathrm{30}}$ (see Eq. \ref{eq:MagFlow} and Eq. \ref{eq:MagHeli}). In these plots, the radius $r$ is the distance from the spine of the filament in the $XZ-$plane.

 We analyze the polarization profiles perpendicular to the spine along the entire SILCC-Zoom main filament. Here, characteristic signatures predicted by the analytical model with parameterization $flow$ can only be found in three distinct regions that are labeled A, B, and C, hereafter. The region A is located in the northern part of the main filament where as regions B and C are in the southern part (see Fig. \ref{fig:DustEmission} and  following figures). 

On average, the magnitude of linear dust polarization $P_{\mathrm{l}}$ roughly agrees for the model $flow$ and the SILCC-Zoom data where as model $heli_{\mathrm{30}}$ is about $2\%$ lower. This is because of the twisted field lines of the parameterization $heli_{\mathrm{30}}$ leading to some depolarization along the LOS. As in \cite{Reissl2018}, our results of  $heli_{\mathrm{30}}$ shows an asymmetry of the profile on either side of the filament. The $P_{\mathrm{l}}$ profiles of the regions A and C qualitatively resemble well the profile of model $flow$. However, we had to scale model $flow$ by $\zeta=0.25$ ($\zeta=1.0$ corresponds to the model of \citet{Reissl2018}) to agree. The profiles of region B possesses less obvious features, the minimums are more narrow and would fit better to a analytical model with $\zeta=0.15$. Altogether, all profiles with their broad and smooth central maximums show some resemblance with parameterization $flow$ but none with the parameterization $heli_{\mathrm{30}}$. 

The circular dust polarization $P_{\mathrm{c}}$ of the analytical models result in profiles with a maximum near the center, i.e. $r=0\ \mathrm{pc}$. Yet again, the model $heli_{\mathrm{30}}$ is not symmetric. Comparing the profiles A, B, and C with the analytical model profiles reveals no clear match. We note that circular dust polarization is dependent on the amount of twisted field lines along the LOS. Hence, we see much more features in the $P_{\mathrm{c}}$ profiles coming from the main filament because of its more turbulent field. Consequently, such variations in the field render the circular dust polarization useless to detect a certain magnetic field morphology.


\begin{figure}
\centering
\begin{minipage}[c]{0.975 \linewidth}
      \includegraphics[width=1\textwidth]{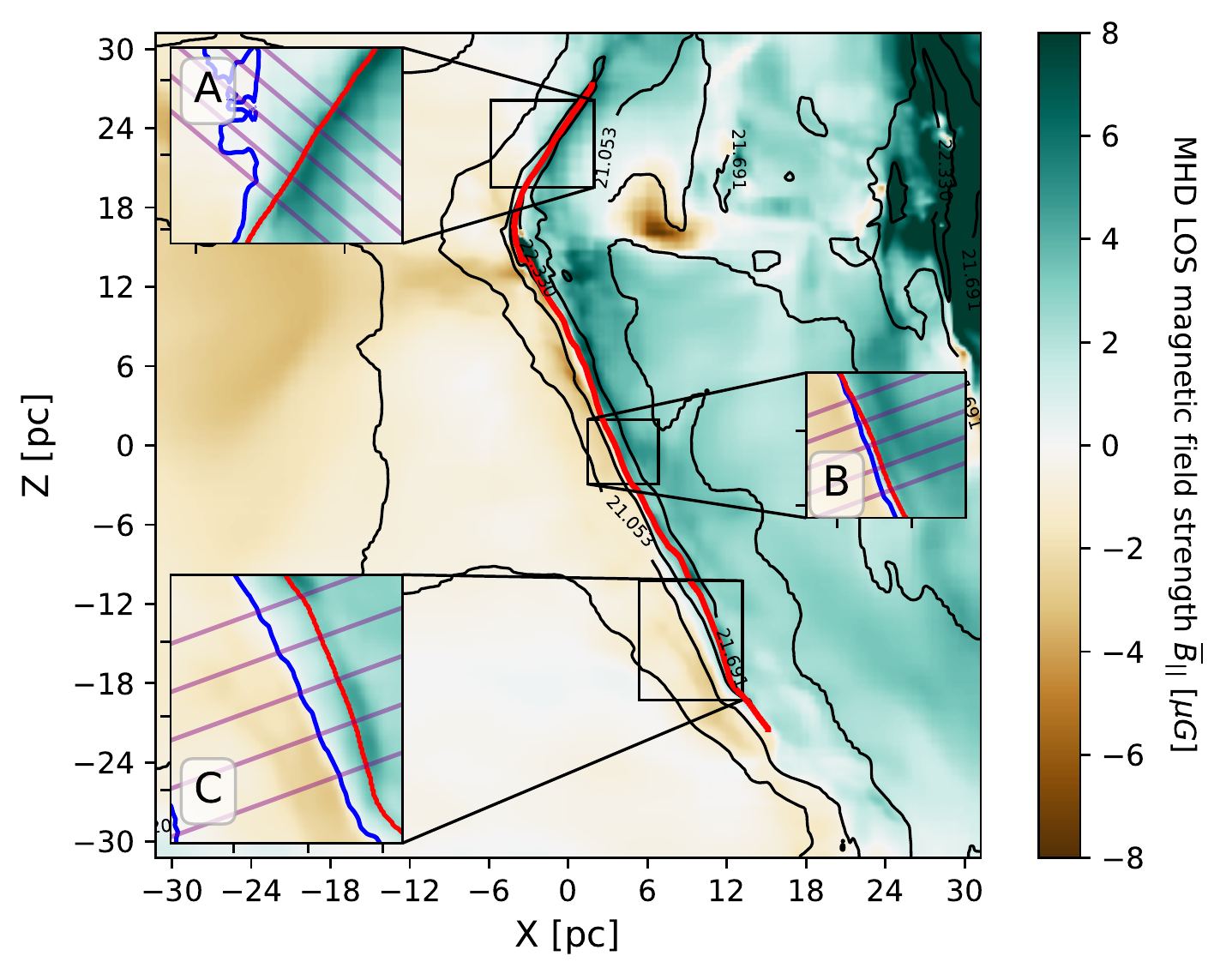}\\
      \includegraphics[width=1\textwidth]{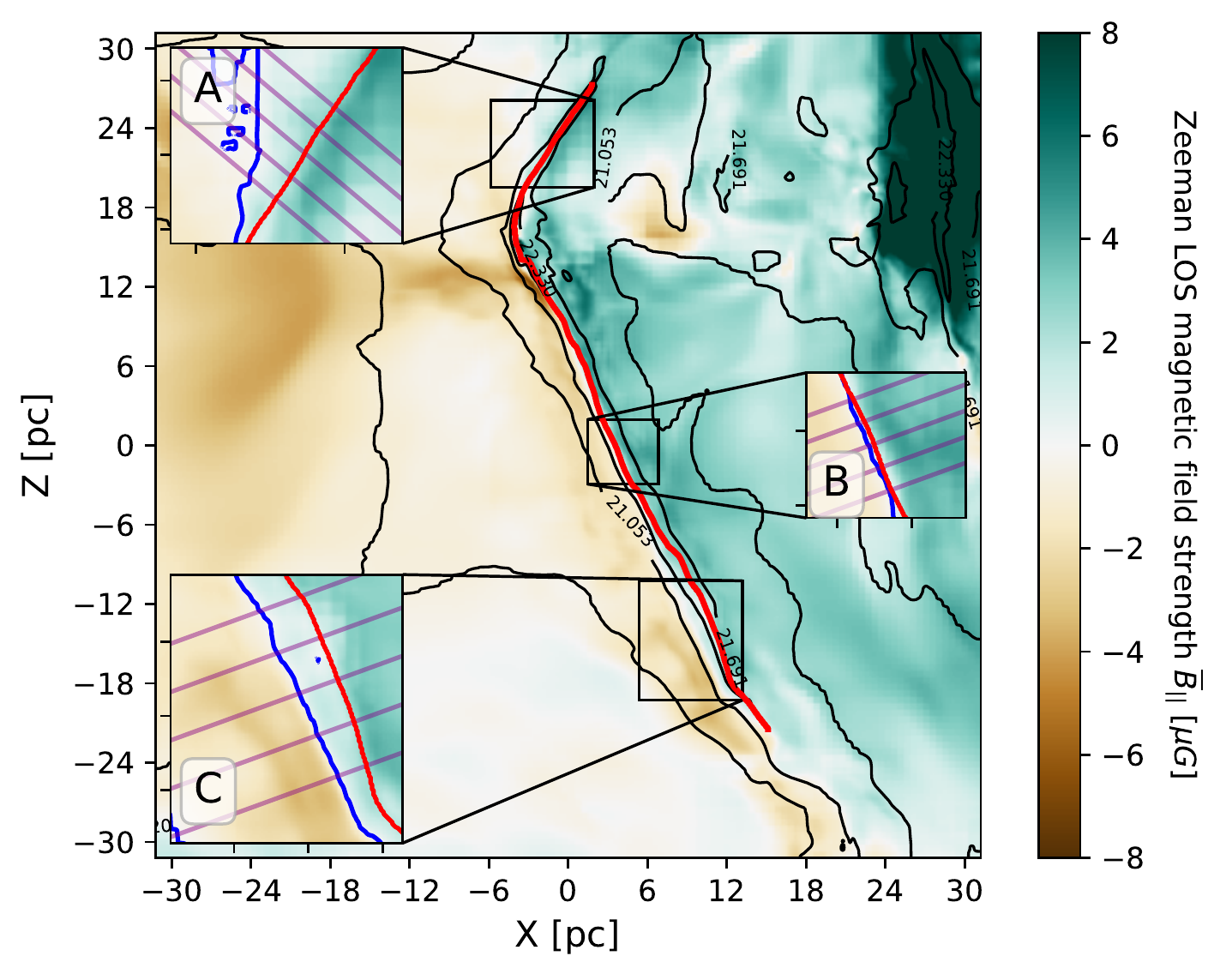}\\
      \includegraphics[width=1\textwidth]{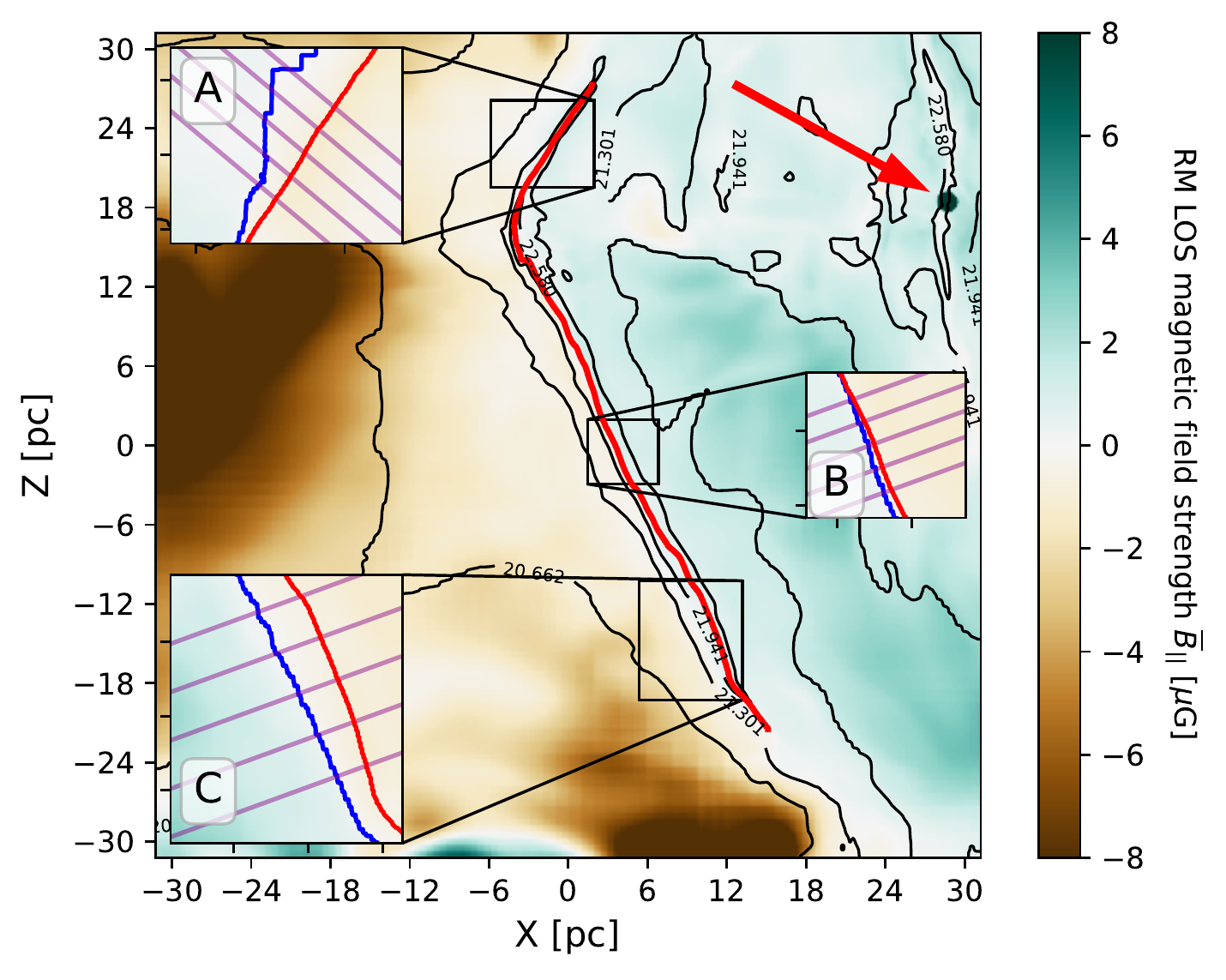}
\end{minipage}
\caption{Maps of magnetic field strength $\overline{B}_{||}$ density weighted along the LOS of the SILCC-Zoom cutout (top) in  comparison with the fields derived by Zeeman effect (middle) and Faraday $RM$ (bottom). Black contour lines show the column density $N_{\mathrm{H}}$, red lines are the spine of the main filament while blue lines are the contour of $\overline{B}_{||}=0\ \mu\mathrm{G}$. Zeeman and $RM$ basically recover the same magnetic field structure. We emphasise that the zero crossings of the magnetic field with respect to the spine of the main filament appear to be almost identical in all three maps. In the bottom panel a red arrow marks the position of a highly ionized bubble surrounding one of the star-forming regions. For this bubble we report a $\overline{B}_{||}$ of the order of $10^4\ \mu\mathrm{G}$ recovered by the $RM$.}
\label{fig:MagFields}
\end{figure}

\begin{figure*}
\centering
\begin{minipage}[c]{1.0\linewidth}
      \includegraphics[width=0.49\textwidth]{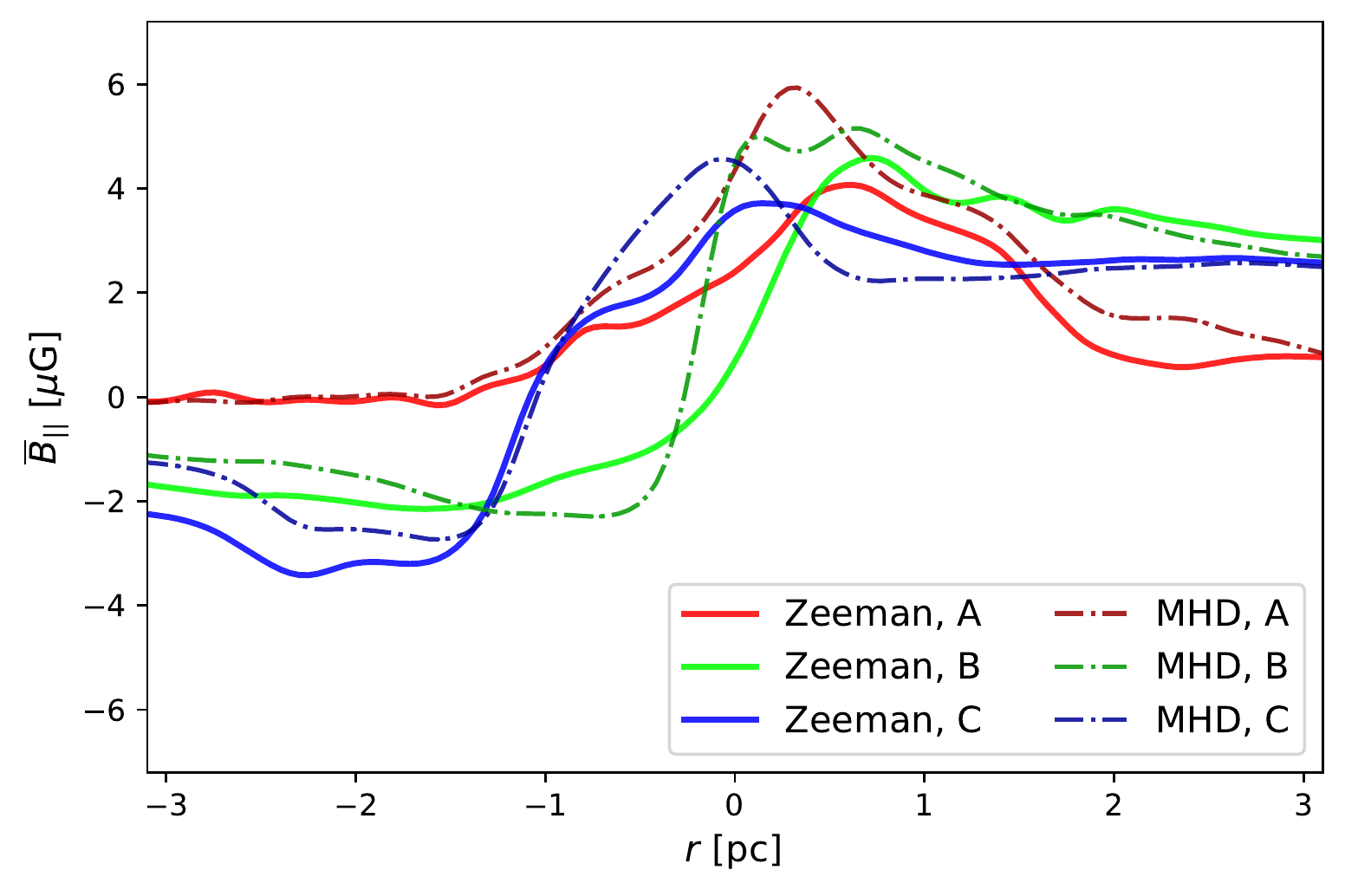}      \includegraphics[width=0.49\textwidth]{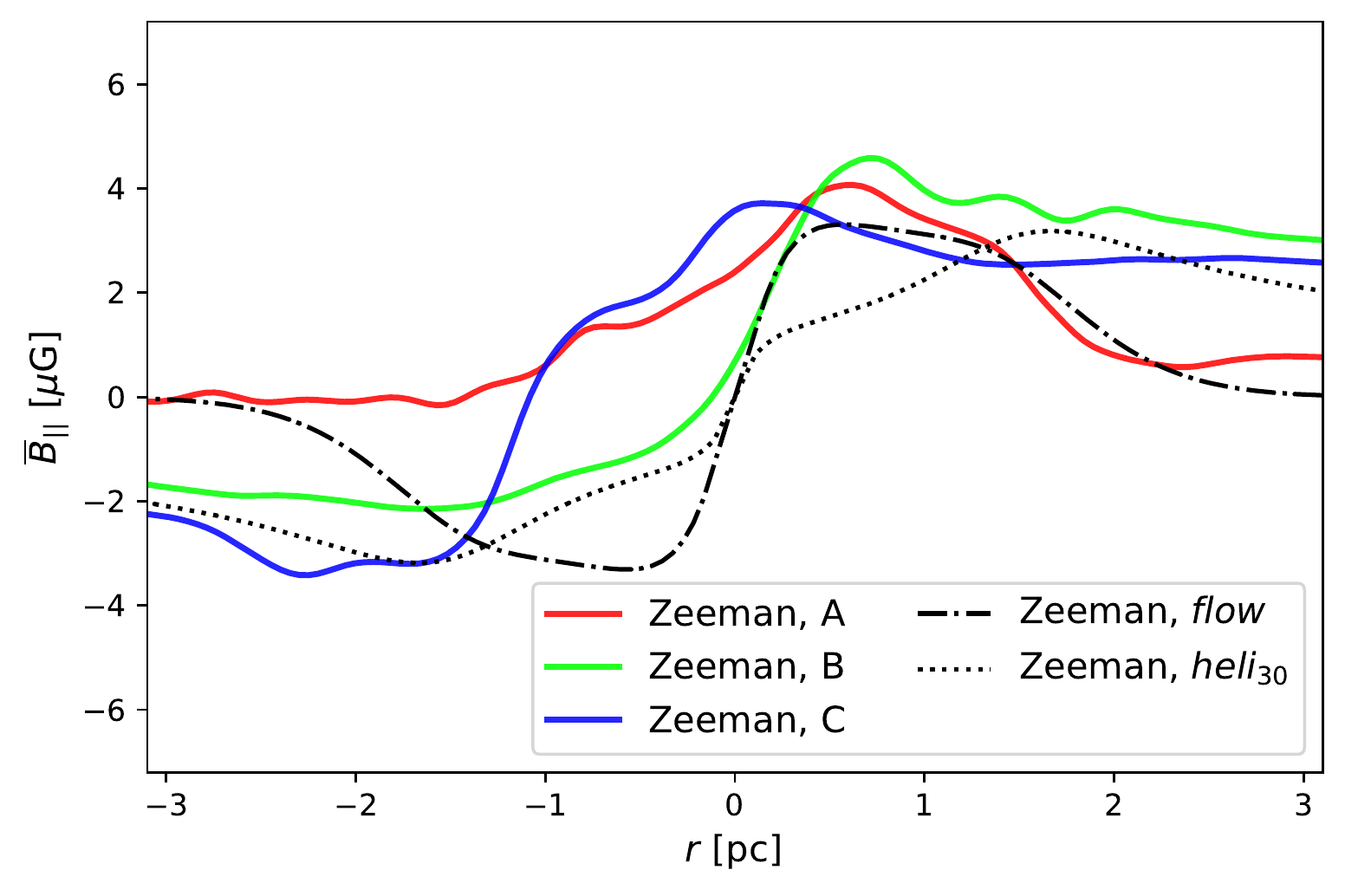}
\end{minipage}
\begin{minipage}[c]{1.0\linewidth}
      \includegraphics[width=0.49\textwidth]{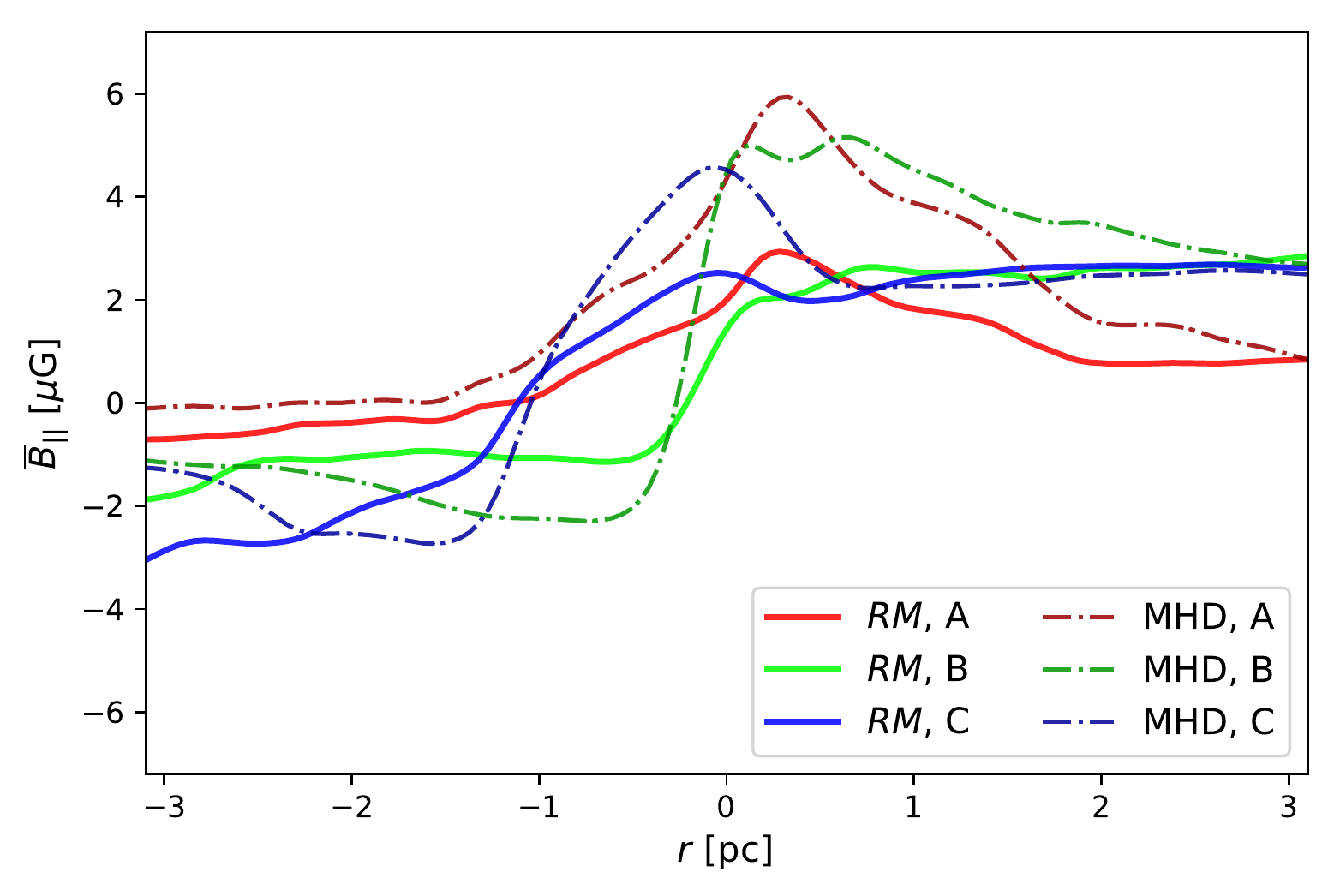}      \includegraphics[width=0.49\textwidth]{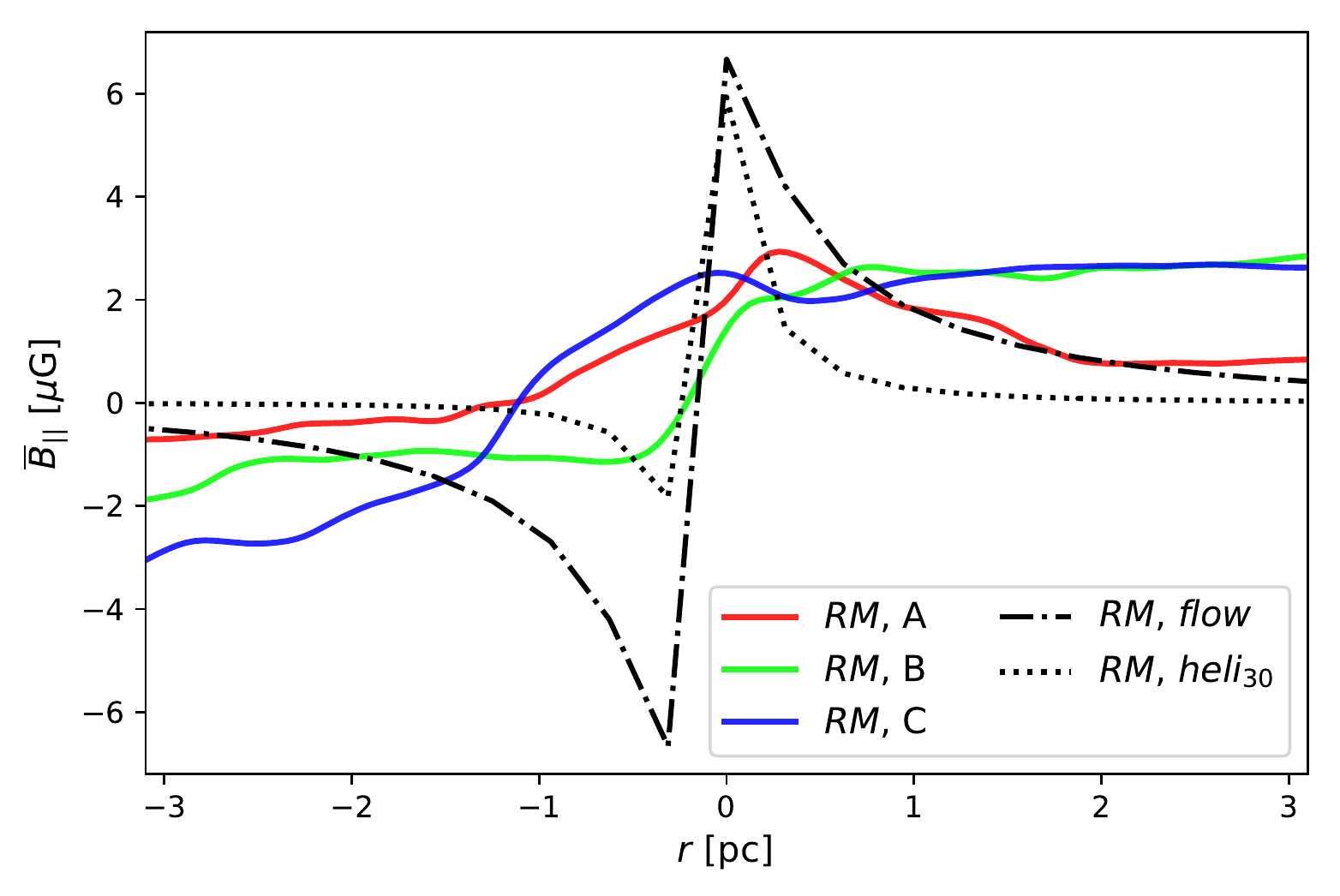}
\end{minipage}

\caption{Selected radial profiles of the average LOS magnetic field strength $\overline{B}_{||}$ in the regions A-C (see Fig. \ref{fig:MagFields}). Upper left panel: The original SILCC-Zoom MHD magnetic field $\overline{B}_{||}$ profiles (dash dotted lines) of the main filament in the regions A (red), B (green), and C (blue) in comparison with the magnetic field strength derived from Zeeman observations (solid lines). All profiles are shifted so that the spine coincides with $r=0\ \mathrm{pc}$ for better comparison. Upper right panel: The same as the upper left panel but for the comparison between Zeeman derived $\overline{B}_{||}$ field in the main filament and the analytical models $flow$ (black dash dotted line) and $heli_{\mathrm{30}}$ (black dotted line). Bottom panels: The same as the upper panels but for  $\overline{B}_{||}$ derived by the Faraday $RM$. We note that the $RM$ predicts slightly lower values of the magnetic field strength close to the spine as the Zeeman observations. None of the derived profiles for the main filament seem to resemble the predicted profiles of the analytical models.}
\label{fig:ProfAllMag}
\end{figure*}

\begin{figure}
\centering
\begin{minipage}[c]{1.0\linewidth}
      \includegraphics[width=1.0\textwidth]{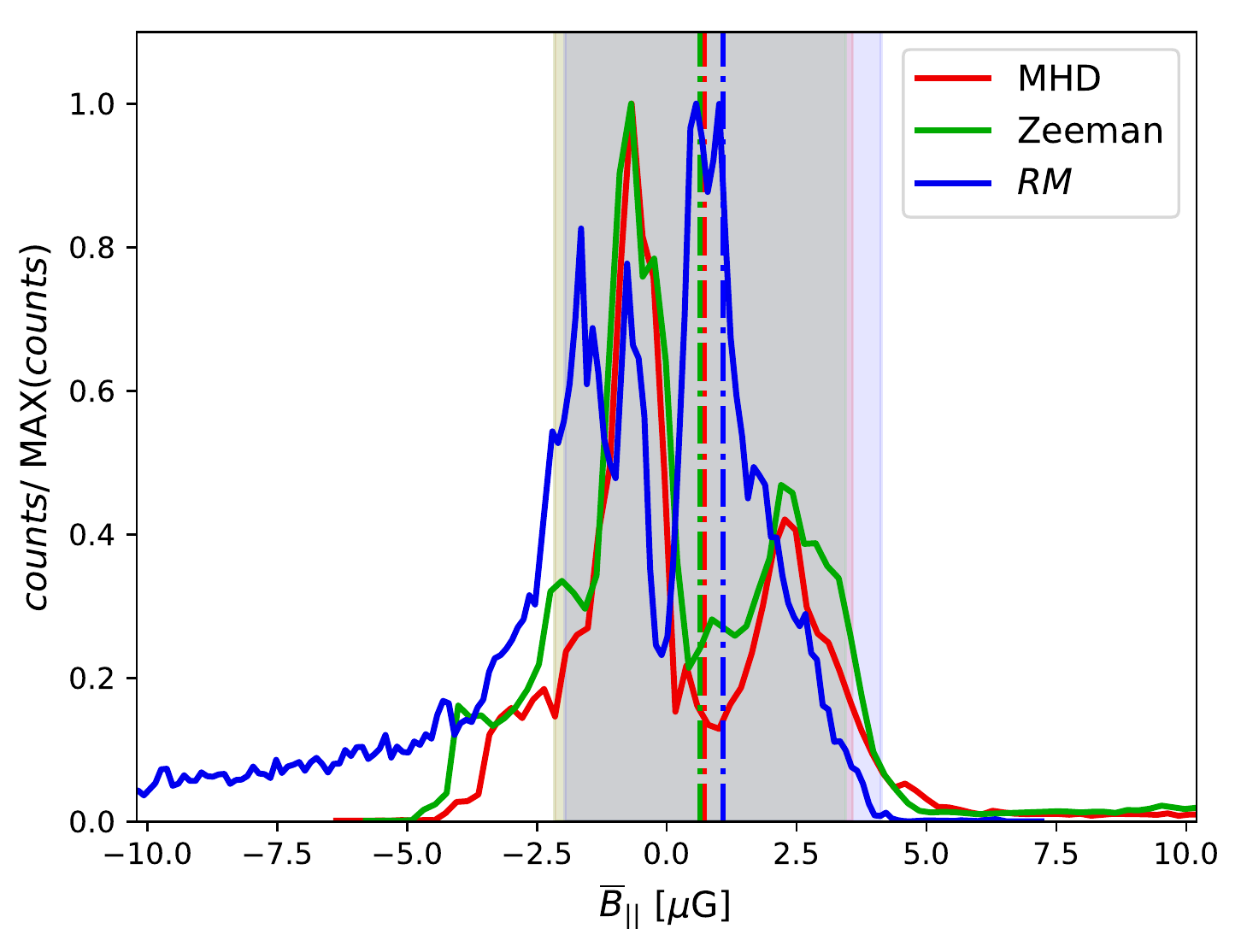}
\end{minipage}
\caption{Histogram of the average LOS magnetic field strength $\overline{B}_{||}$ of the SILCC-Zoom MHD field and the ones derived by Zeeman observations and $RM$, respectively. The histogram corresponds to the full set of pixels shown in Fig. \ref{fig:MagFields}. Vertical dash dotted lines are the average while the shaded areas represent the standard deviation.}
\label{fig:HistMag}
\end{figure}

\subsection{Zeeman effect and Faraday $RM$}
\label{sect:ZeemanAndRM}
In the left panel of Fig. \ref{fig:HIMapProf} we present the resulting Zeeman map of the total circular HI 21 cm line polarization $P_{\mathrm{c}}$ integrated over all velocity channels as well as the corresponding profiles perpendicular to the spine. We also compare the profiles of the analytical model with the ones of the SILCC-Zoom main filament in the right panel of the figure. 
Here, the model parameterization $heli_{\mathrm{30}}$ predicts a profile that is mostly positive with a central peak while model $flow$ is negative for $r<0\ \mathrm{pc}$ and positive for $r>0\ \mathrm{pc}$ with a discontinuity at $r=0\ \mathrm{pc}$ i.e. the spine of the analytical model. Along the entire spine we find no characteristic profiles indicating the presence of the kinked magnetic field morphology predicted from model $flow$.

For convenience, we focus our discussion again on the regions A, B, and C, respectively. On large scales, the Zeeman map recovers the magnetic field reversal along the spine of the main filament. However, neither of the profiles of the distinct regions A - C can be clearly assigned to any of the magnetic field parameterizations $flow$  nor $heli_{\mathrm{30}}$ considered in the analytical model. 

It is expected that for model $flow$ and the main filament that $P_{\mathrm{c}}=0\ \%$ exactly at the kink of the filament since the LOS and magnetic field are perpendicular (see \S~\ref{sect:ObsZeeman}). However, the spine of the main filament shows an offset with the contour of  $P_{\mathrm{c}}=0\ \%$ in Fig.~\ref{fig:HIMapProf}. In the cuts in Fig. \ref{fig:Midplanes}, the kink of the magnetic field follows the spine of the main filament. However, this kink is not only present close to the spine but also in the tail-like density structure along the negative $Y$ axis. 
This indicates that the change in sign of  $P_{\mathrm{c}}$ happens not exactly at the spine but most possibly also along the tail in the SILCC-Zoom cutout, i.e. along the LOS of the observer. Thus, the profiles of circular polarization of the regions A - C shown in Fig. \ref{fig:HIMapProf} are not comparable to each other because of the variations in the tail of the main filament. 

In Fig. \ref{fig:RMMapProf} we show the map and profiles for the Faraday $RM$. Yet again, for the discussion of the $RM$ map in Fig. \ref{fig:RMMapProf} we focus only on the regions A - C in our analysis. We note that the sign of the overall $RM$ map roughly coincides with the spine of the main filament. We find a particular spot within the diffuse cloud in the upper right corner (marked with a red arrow in Fig. \ref{fig:RMMapProf}). This spot represents a bubble of highly ionized gas surrounding one of the star-forming regions. This particular star-forming region harbors a massive star with $1.62\times 10^4\ \mathrm{L}_{\odot}$ (see also Fig. \ref{fig:3DMHD}). Hence, we find locally values for $RM$ up to $10^5\ \mathrm{rad\ m^{-2}}$ for that ionized bubble. 

The $RM$ profiles in Fig. \ref{fig:RMMapProf} of the regions A and C are roughly symmetric with the maximum close to the spine i.e. $r=0\ \mathrm{pc}$. The maximum of region B has a small offset between maximum and spine. Here, we note that the profile shape of region B would match better to the predictions of the model $heli_{\mathrm{30}}$ than to $flow$. This might be a result of a lower magnetic field strength for the side $X<0\ \mathrm{pc}$ of the SILCC-Zoom simulation. However, none of the profiles of the main filament posses the discontinuity at  $r=0\ \mathrm{pc}$ as predicted by the model $flow$.  Consequently, $RM$ observations seem not to be an unambiguous indicator of the kinked field morphology of the main filament.

Comparing the zero crossings with respect to the spine in the maps of both tracers (Zeeman and $RM$ maps) reveals some similarities in regions B and C. In both regions the zero crossing is somewhat parallel to the spine.  However, in regions A, the zero crossings of $RM$ and Zeeman do not coincide, where with the one of the $RM$ being more parallel to the $Z-$axis. Indeed, comparing the profiles of $RM$ and Zeeman of the SILCC-Zoom simulation reveals that both tracers also do not match concerning their profile shape.




\subsection{Quantifying the magnetic field strength}
\label{sect:MagFieldStrength}
In the previous sections we directly simulate and analyze the signal of different tracers in order to detect characteristic patterns of the underlying magnetic field morphology. However, the Zeeman effect and the $RM$ allow also to indirectly estimate the average magnetic field strength $\overline{B}_{\mathrm{||}}$ and its direction with respect to the LOS\footnote{We note that we use the notation $B_{\mathrm{||}}$ for the field strength parallel to the LOS at a distinct position while $\overline{B}_{\mathrm{||}}$ corresponds to the field strength averaged along the entire LOS. In the particular coordinate system of the SILCC-Zoom and the analytical model $B_{\mathrm{||}}$ is the same as the $B_{\mathrm{Y}}$ field component in each cell.}. In this section we scrutinise the magnetic field strength derived by different tracers for possible fingerprints of the underlying magnetic field morphology along the spine. In principle dust polarization may recover the plane-of-sky magnetic field strength $\overline{B}_{\mathrm{\bot}}$ by means of the Chandrasekhar-Fermi Method (CF) \citep[][]{ChandrasekharFermi1953}. However, this technique comes with numerous uncertainties \citep[e.g.][]{Ostriker2001,ChoYoo2016} that cannot be handled within the scope of this paper\footnote{Indeed, a separate publication dealing synthetic observations in the context of the CF Method is already in preparation.}.

Our approach is as follows: We calculate the average gas density weighted LOS field component $ { \overline{B}_{\mathrm{||}}=\int  n_{\mathrm{g}}(\ell) \times B_{\mathrm{||}}(\ell) \mathrm{d}\ell } $ of the SILCC-Zoom cutout. This map acts as a reference for comparison with the maps derived by Zeeman effect and $RM$. For the Zeeman effect we perform a least squares fit over all velocity channels of the RT simulated Stokes $V_\nu$ component to the intensity $I_\nu$ (compare Eq. \ref{eq:ZeemanI} and Eq. \ref{eq:ZeemanV}) for each pixel to get a map of $\overline{B}_{\mathrm{||}}$. 

The average magnetic field strength is derived from the $RM$ by assuming $RM = 0.81\int n_{\mathrm{el}}(\ell) B_{\mathrm{||}}(\ell)  \mathrm{d}\ell \approx 0.81 DM \times \int B_{\mathrm{||}}(\ell) \mathrm{d}\ell$. Here, the factor $0.81$ results from the natural constants in Eq. \ref{eq:DefinitionRM} given that $n_{\mathrm{el}}$ is in units of $\mathrm{cm^{-3}}$, $ B_{\mathrm{||}}$ is in $\mu\mathrm{G}$, and $\ell$ in $\mathrm{pc}$, respectively. The quantity ${ DM = \int n_{\mathrm{el}}(\ell) \mathrm{d}\ell }$ is the dispersion measure of the free electrons. In practise, the $DM$ may be derived by exploiting $\mathrm{H}_\alpha$ emission as a proxy for the free electrons \citep[e.g.][]{Schnitzeler2012,Betti2019}. However, modeling such an emission goes beyond the scope of this paper. For simplicity we assume a constant ratio between gas and electrons as derived in \S~\ref{sect:ToyModel} and apply the gas column density map calculated in \S~\ref{sect:ObsDust} to get $DM=8.1\times10^{-4} N_{\mathrm{H}}$. This gives the average LOS magnetic field strength deduced via ${ \overline{B}_{||} \approx RM /(6.4\times10^{-4}\times N_{\mathrm{H}}) }$ where the $RM$ corresponds to the map shown in Fig. \ref{fig:RMMapProf}.

Figure \ref{fig:MagFields} shows the resulting map of the SILCC-Zoom data and the maps derived by Zeeman and $RM$ mock observations. The Zeeman map of the average LOS magnetic field strength $\overline{B}_{\mathrm{||}}$ agrees with the SILCC-Zoom map with respect to magnitude and sign. The exception is the  region at $X>24\ \mathrm{pc}$ and $Z>6\ \mathrm{pc}$. In this region, the Zeeman derived $\overline{B}_{\mathrm{||}}$ appears to be much more pronounced than that of the SILCC-Zoom main filament. This generall good match is in contrast to the analytical models presented in \cite{Reissl2018}. One of their major findings was that the  Zeeman method underestimated the actual magnetic field strength throughout by a factor of 10 for different tracers. However, these models had much higher gas densities and temperatures and, consequently, a line broadening $\Delta \nu$ in the order of  the characteristic frequency shift $\Delta \nu_{\rm z}$ between  different Zeeman lines. Instead, the gas densities and temperatures in the SILCC-Zoom simulation as well as the analytical model are about a factor of $2-10$ smaller compared to  \cite{Reissl2018} with the consequence that $\Delta \nu \ll \Delta \nu_{\rm z}$ and the magnetic field strength can be better recovered by Zeeman observations, as demonstrated e.g. in the parameter study in \cite{Brauer2017B}. 

In contrast to the Zeeman derived $\overline{B}_{\mathrm{||}}$ map the $RM$ seems to underestimate the magnetic field strength close to the spine of the main filament while we see some overestimation up to $10\ \mu\mathrm{G}$ for the thin and diffuse gas. For the highly ionized bubble (marked by a red arrow in the figures \ref{fig:RMMapProf} and \ref{fig:MagFields}) we even report a field strength up to $\approx 10^4\ \mu\mathrm{G}$. This is because of the higher electron fraction within that bubble and not because an actually higher magnetic field strength.

Generally, all three maps (SILCC-Zoom,  Zeeman, $RM$) show the same trend for the zero crossing of $\overline{B}_{\mathrm{||}}$ along the spine of the main filament. Remarkably, this is despite the fact that $RM$ and Zeeman effect may in general not trace the same pieces of the main filament along the LOS.

In Fig. \ref{fig:ProfAllMag} we present the profiles of the original LOS magnetic field strength $\overline{B}_{\mathrm{||}}$ of the SILCC-Zoom and the $\overline{B}_{\mathrm{||}}$ indirectly derived by Zeeman and $RM$ mock observations. For simplicity, we focus only on the selected profiles for each of the zoom-in regions A - C. We create similar Zeeman and $RM$ mock observations for the parameterizations $heli_{\mathrm{30}}$ and $flow$ of the analytical model. 

The magnetic field profiles of the main filament derived from the Zeeman effect and $RM$ of the main filament somewhat agree with each other concerning the absolute magnitude as well as the zero crossings. As for the analytical model both parameterizations $heli_{\mathrm{30}}$ and $flow$ have a clear zero crossing from negative to positive in Zeeman and $RM$, respectively, at $r=0\ \mathrm{pc}$ i.e. the spine. The zero point coincides with the spine as it is expected by the underlying magnetic field morphology 
\citep[see also Figures 1 and 2 in][]{Reissl2018}. However, a similar trend concerning zero crossing cannot be found for any of the profiles in the regions A - C of the main filament. Indeed, the regions B and C show a clear zero crossing but with an offset with respect to the spine while the $\overline{B}_{\mathrm{||}}$ in region A is mostly positive. Hence, deriving the LOS magnetic field strength $\overline{B}_{\mathrm{||}}$ by means of Zeeman or $RM$ observations may detect field reversals along the LOS but seems not to allow to unambiguously identify the kinked field morphology in the main filament either.

We calculate the pixel per pixel ratio of the maps in Fig.~\ref{fig:MagFields}. Taking the average over all the ratios we see that the Zeeman technique may overestimate $\overline{B}_{\mathrm{||}}$ in the maps on average by a factor of about 2.1. However, we find also a factor of up to 7.5 for some pixel in the very close proximity of the star-forming regions. For the $RM$ map see that the $RM$ seems to overestimate the magnetic field strength throughout by a factor of 3.4 on average but we find values of about 0.4 close to the spine of the filament and peak values of up to 20.1 for the diffuse gas. Note that we masked the ionized bubble for calculating the later values. For the ionized bubble itself we report a peak value of 1400.


A histogram of the LOS magnetic field strength $\overline{B}_{\mathrm{||}}$ is shown in Fig. \ref{fig:HistMag}. It represents the pixel by pixel distribution of $\overline{B}_{\mathrm{||}}$ derived from the SILCC-Zoom simulations and by the Zeeman and $RM$ mock observations shown in Fig. \ref{fig:MagFields}. All three histograms have a mean close to  $\overline{B}_{\mathrm{||}}\approx 0\ \mu\mathrm{G}$ and a standard deviation of about $\pm 2\ \mu\mathrm{G}$. However, in this histogram we also see that the Zeeman observation and the $RM$ have the tendency to overestimate the LOS field strength $\overline{B}_{\mathrm{||}}$.

We note, that this does only hold for our SILCC-Zoom simulation and the particular set of parameters of the RT post-processing. The error might be vastly different for observations of other objects.

\section{The origin of the magnetic field tracer signals}
\label{sect:Origin}

\begin{figure*}
\centering
\begin{minipage}[c]{1.0\linewidth}
      \includegraphics[width=0.50\textwidth]{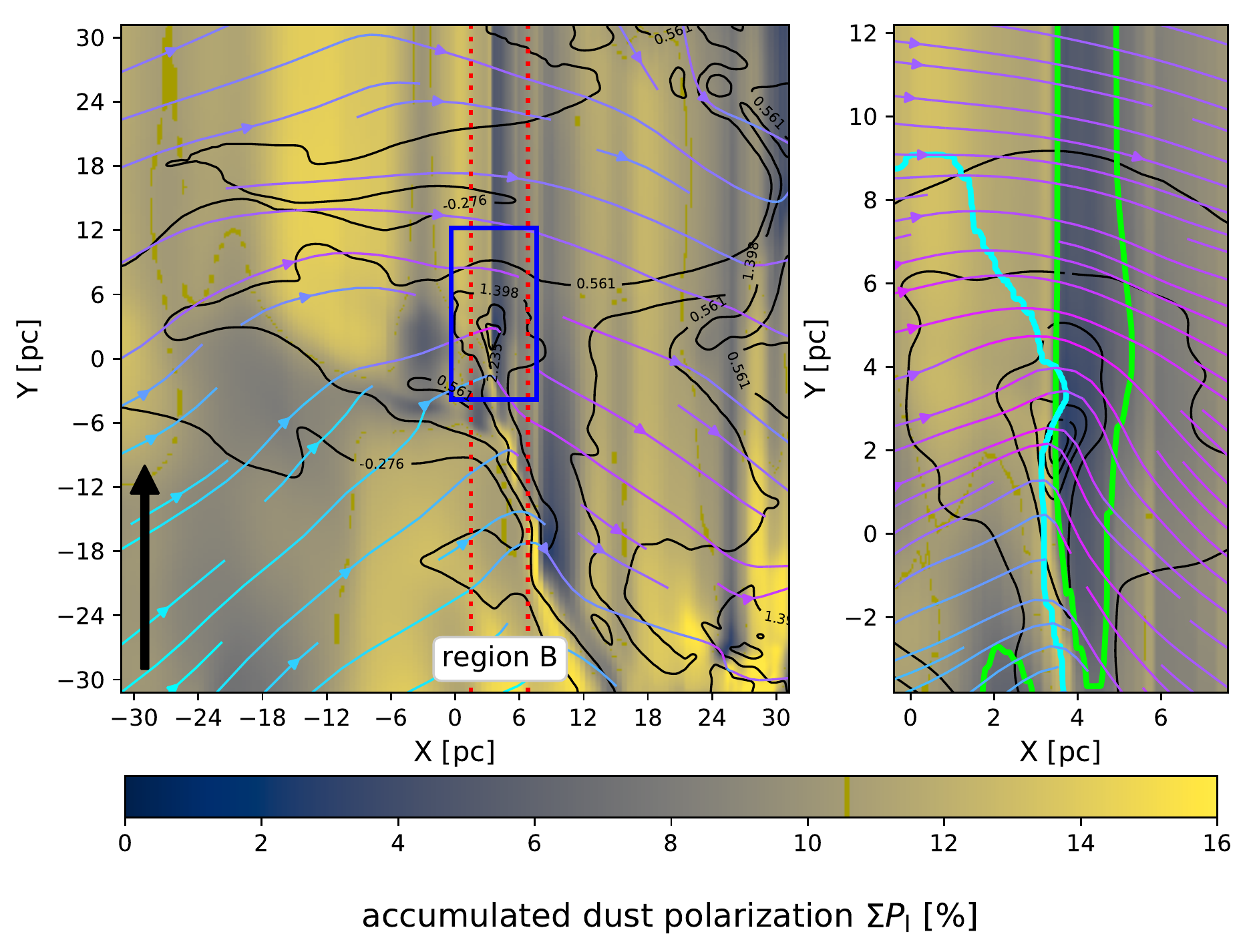}
      \includegraphics[width=0.50\textwidth]{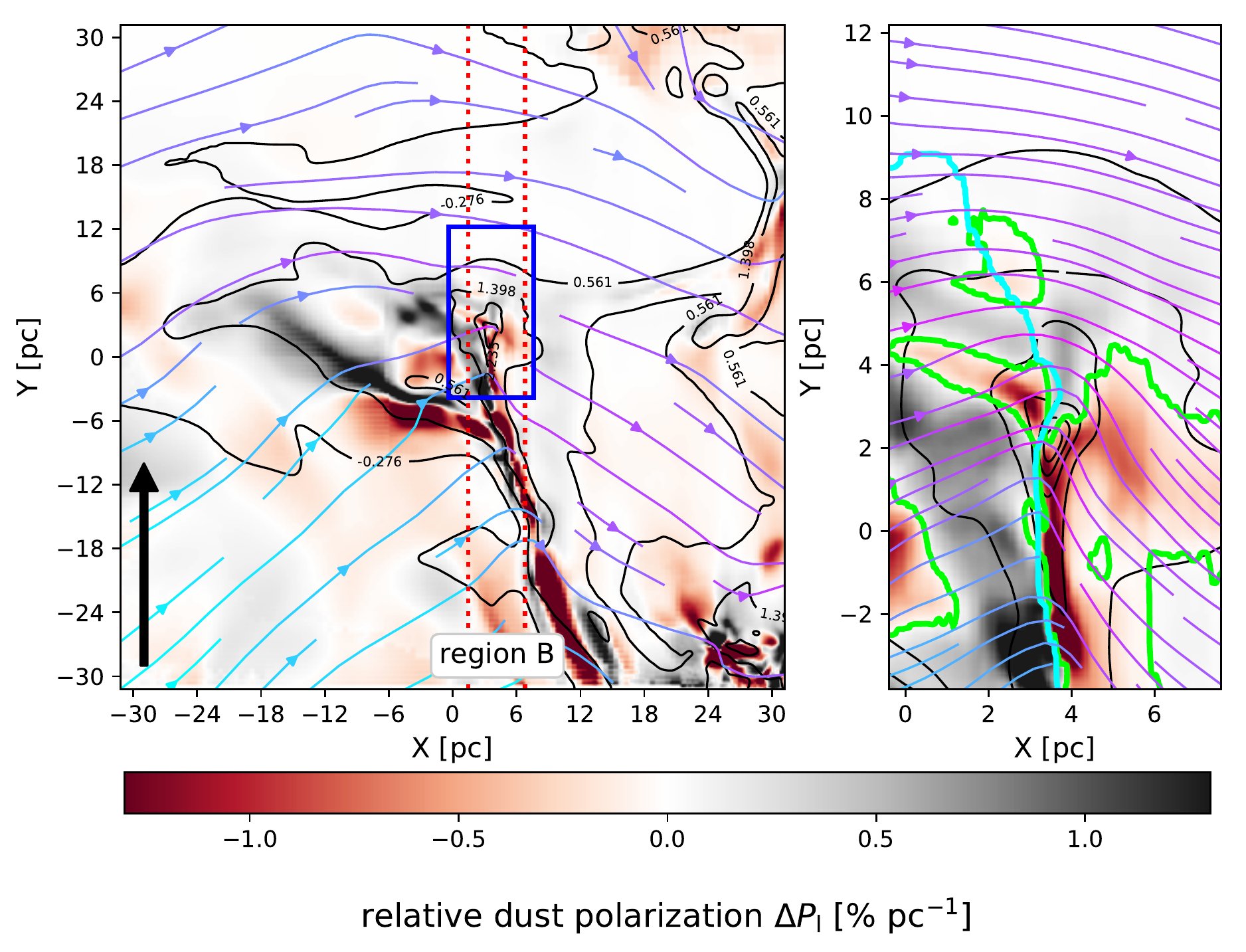}
\end{minipage}
\caption{This plot shows the change in linear dust polarization in the $XY-$plane at $Z=0\ \mathrm{pc}$. Note that the synthetic observations are in the $XZ-$plane and the LOS goes from $-31.21\ \mathrm{pc}$ to $31.21\ \mathrm{pc}$ as indicated by the black arrow.  
The planes show the cumulative increase of $\Sigma P_{\mathrm{l}}$ (left panels) and its relative change $\Delta P_{\mathrm{l}}$ (right panels) along the LOS. The smaller panels shows the zoom-in regions of the spine of the main filament marked by blues boxes in the larger panels. Black contour lines indicate the gas number density and the vector field represents the direction and magnitude of the magnetic field. Vertical red dotted lines depict the range of region B as shown in Fig. \ref{fig:DustEmission}. Green lines are the contours of $\Sigma P_{\mathrm{l}}=7\ \%$  (left panel) and $\Delta P_{\mathrm{l}}=0\ \%/\mathrm{pc}$ (right panel), respectively,  where as cyan lines represent the contour of LOS magnetic field component $B_{\mathrm{||}}=0\ \mu\mathrm{G}$, i.e. the points where $\vec{B}$ is perpendicular the LOS and the angular dependency of dust emission $\sin^2 \vartheta \approx 1$. We emphasise that the $\Sigma P_{\mathrm{l}}=7\ \%$ contours within region B is slightly to the right to the magnetic field reversal.}
\label{fig:AccDust}
\end{figure*}

\begin{figure*}
\centering
\begin{minipage}[c]{1.0\linewidth}
      \includegraphics[width=.50\textwidth]{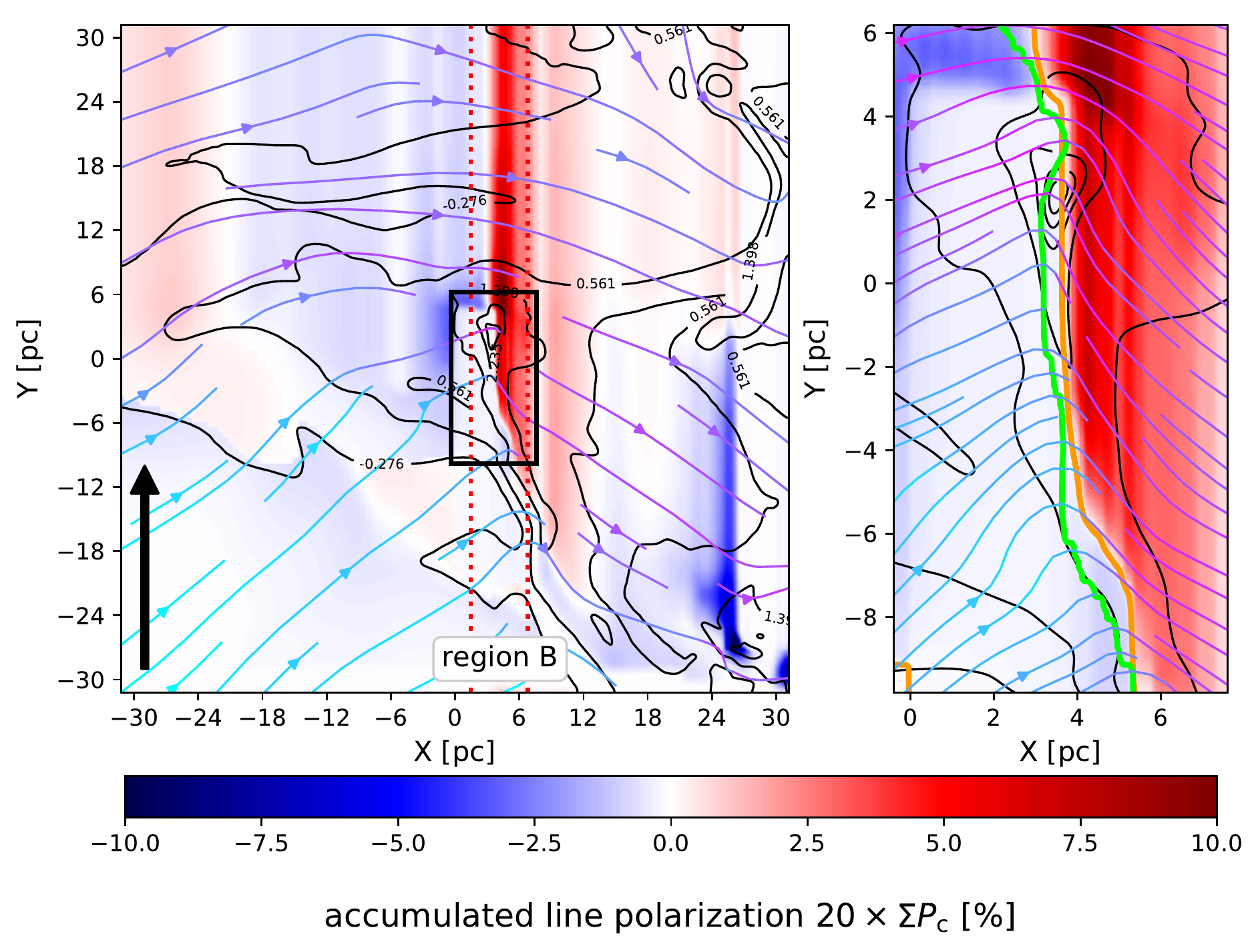}
      \includegraphics[width=.50\textwidth]{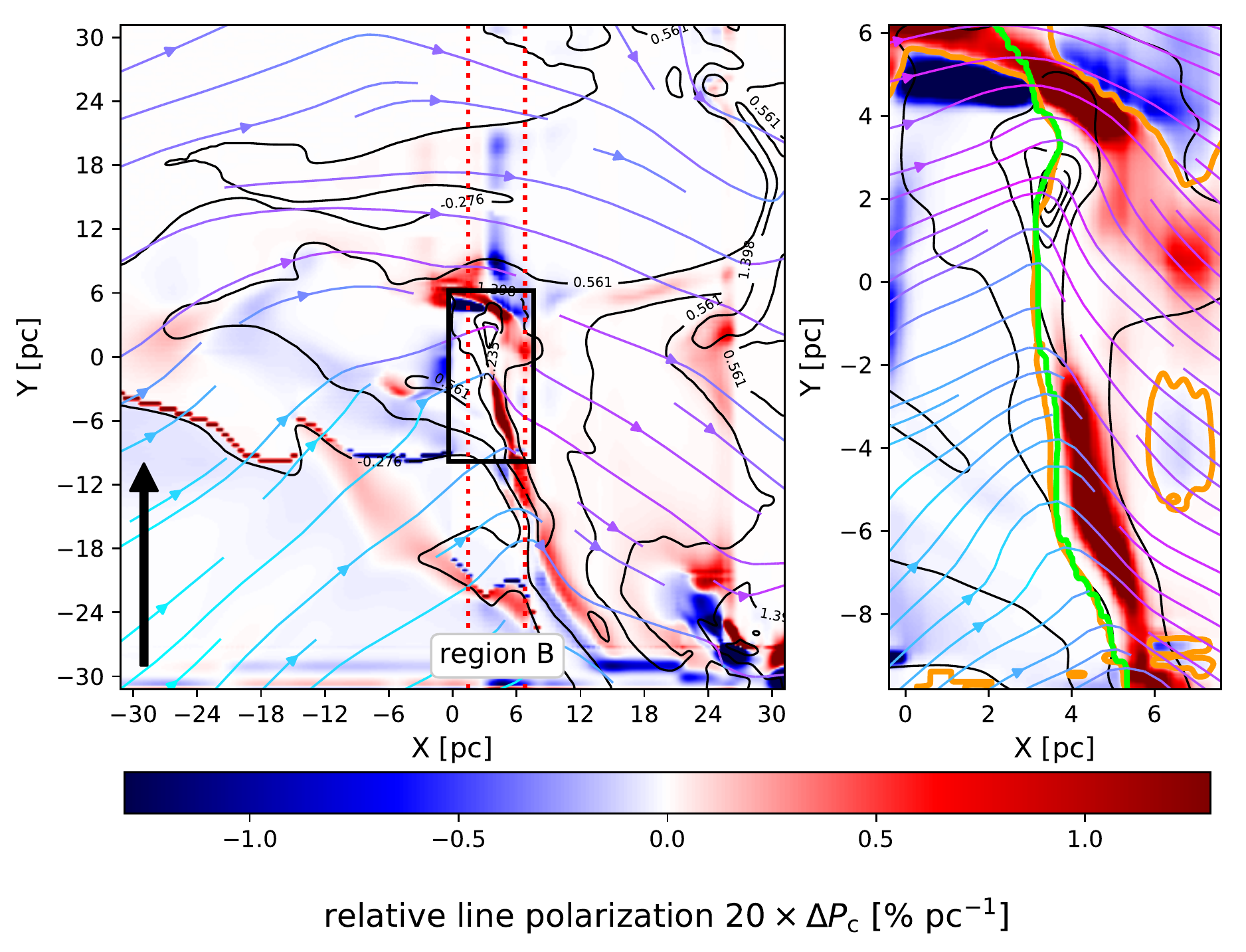}
\end{minipage}
\caption{{\bf The same as Fig. \ref{fig:AccDust} but for the circular polarization of the line RT with Zeeman effect. Green contours are $B_{\mathrm{||}}=0\ \mu\mathrm{G}$ whereas orange lines show the cumulative circular polarization $\Sigma P_{\mathrm{c}}=0\ \%$ (left zoom-in panel) and relative circular polarization $\Delta P_{\mathrm{c}}=0\ \%/\mathrm{pc}$ (right zoom-in panel), respectively. Here, the zero crossings for both $\Sigma P_{\mathrm{c}}$ and $\Delta P_{\mathrm{c}}$ coincide  roughly with the magnetic field reversal. }}
\label{fig:AccLine}
\end{figure*}

\begin{figure*}
\centering
\begin{minipage}[c]{1.0\linewidth}
      \includegraphics[width=0.50\textwidth]{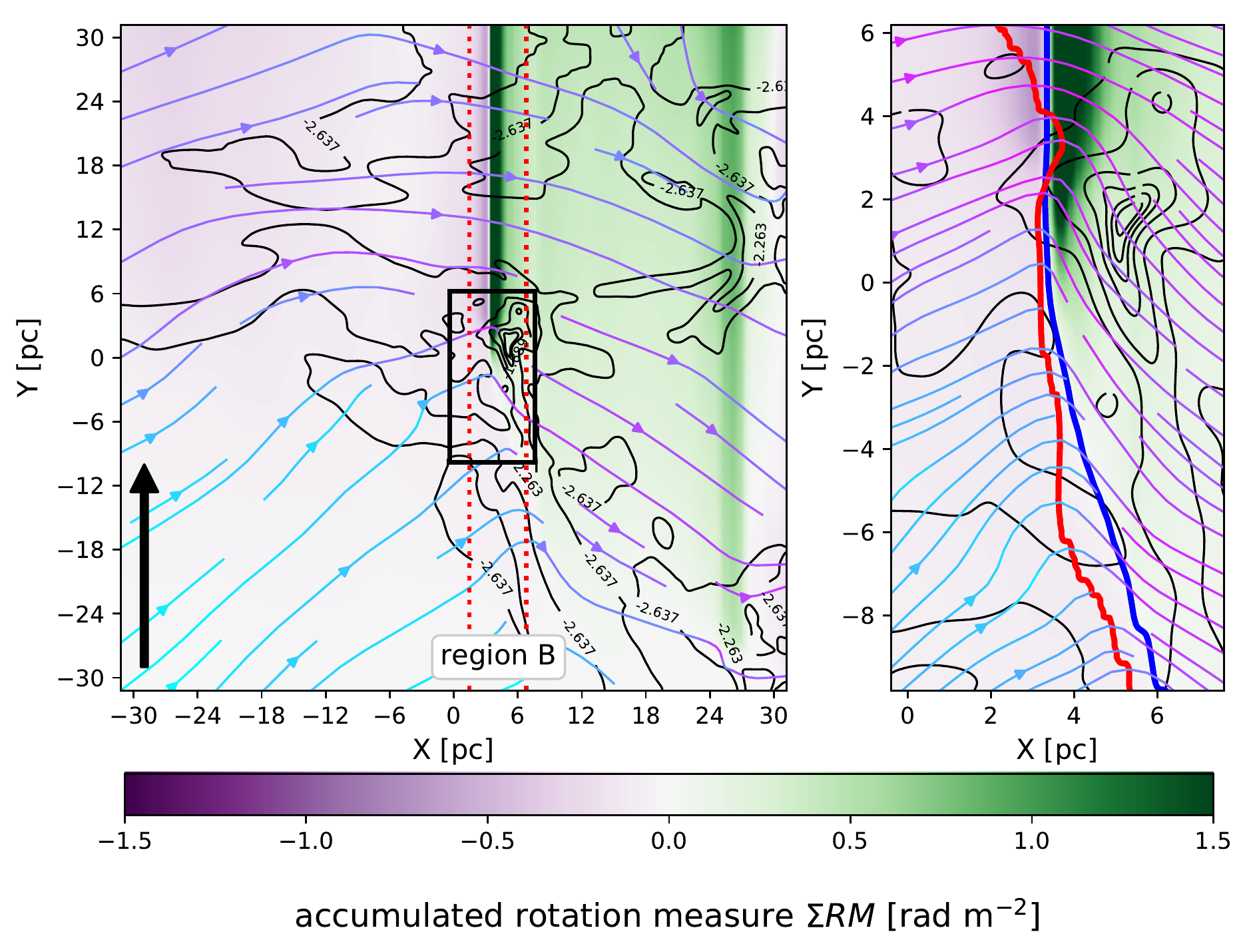}
      \includegraphics[width=0.50\textwidth]{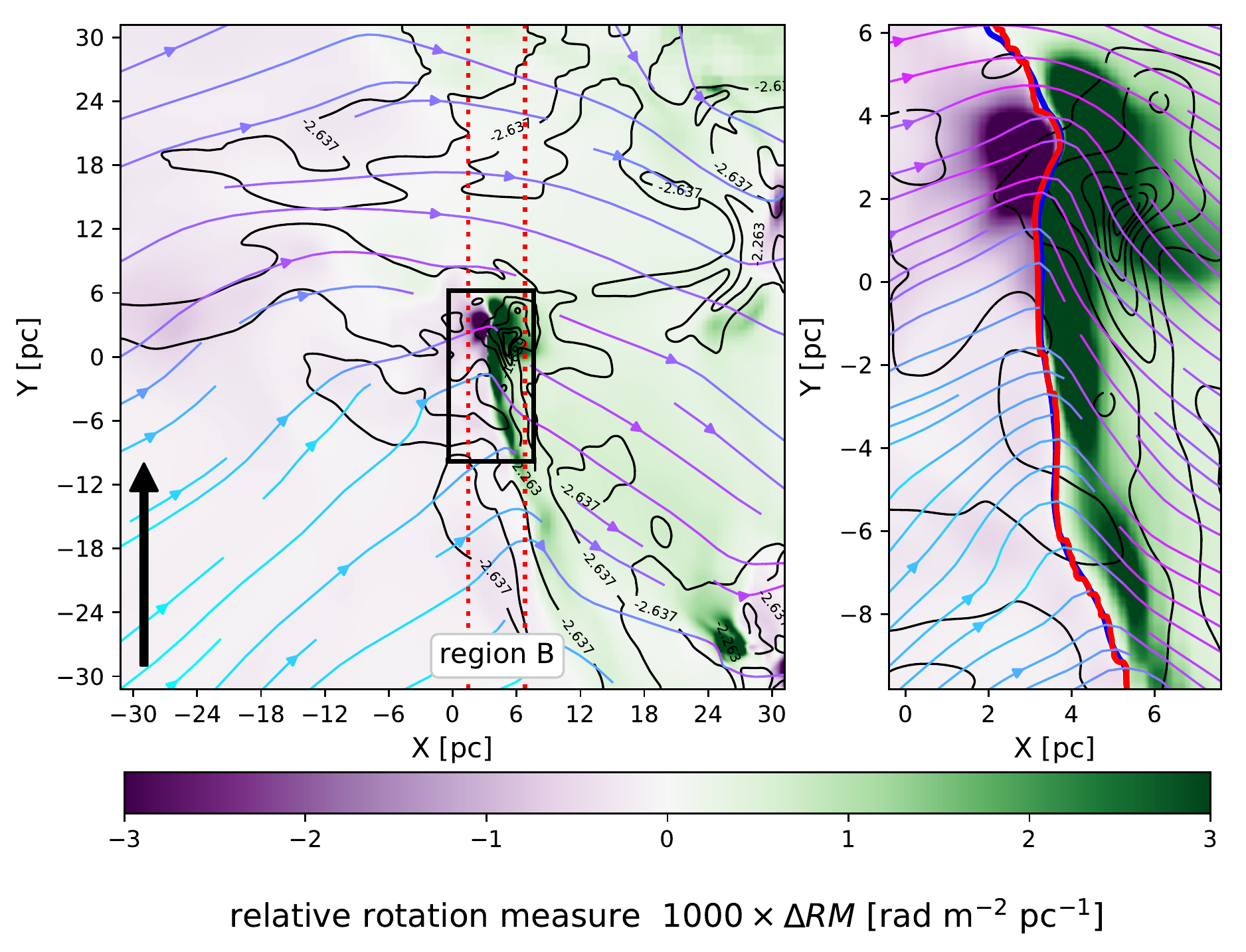}
\end{minipage}
\caption{The same as Fig. \ref{fig:AccDust} but for the Faraday $RM$. Here, the black contours show the electron density $n_{\mathrm{el}}$. Red contours in the zoom-in panels are $B_{\mathrm{||}}=0\ \mu\mathrm{G}$ whereas blue contours represent $\Sigma RM=0\ \mathrm{rad\ m^{-2}}$ and $\Delta RM=0\ \mathrm{rad\ m^{-2}\ pc}^{-1}$, respectively. The contours of $\Sigma RM=0\ \mathrm{rad\ m^{-2}}$ and $\Delta RM=0\ \%\ \mathrm{rad\ m^{-2}\ pc}^{-1}$ resemble the magnetic field reversals.}
\label{fig:AccRM}
\end{figure*}


We start the LOS analysis as outlined in \S~\ref{sect:LOSAnalysis} for the $850\ \mu\mathrm{m}$ dust polarization as it accumulates along the LOS towards the observer. From the change of $P_{\mathrm{l}}$ along the LOS (i.e. along the Y-axis) we derive the quantities of the cumulative value $\sum P_{\mathrm{l}}$ as well as its relative change $\Delta P_{\mathrm{l}}$ with the help of {\sc POLARIS}. First, we compare $\sum P_{\mathrm{l}}$ and $\Delta P_{\mathrm{l}}$, respectively, with gas pressure, the radiation field, angular dependency, and the magnetic field to quantify the particular impact of the field direction on the observed polarization signal. Later in this section, we present a similar LOS analysis for the cases of Zeeman observations and the $RM$ observations, respectively.

In Fig. \ref{fig:AccDust} we show the resulting data of the $XY-$plane at $Z=0\ \mathrm{pc}$ being identical with the planes presented in Fig. \ref{fig:Midplanes}. We emphasise that the region B introduced in \S~\ref{sect:DustAndExtionction} also intersects with this particular $XY-$plane. Similar plots for the regions A and C, respectively, are provided in Appendix \ref{app:RelChange}. On the left side of Fig. \ref{fig:AccDust} we present a panel  with $\Sigma P_{\mathrm{l}}$ as well as a zoom in panel of the region close to the spine. At $Y=-31.21\ \mathrm{pc}$ the dust polarization starts with $\Sigma P_{\mathrm{l}}= 0\ \%$. The spatial extension of the region B, as depicted e.g. in Fig. \ref{fig:DustEmission}, is indicated by vertical dotted red lines. We spot several regions of depolarization, particularly in close proximity to the spine of the main filament (compare Fig. \ref{fig:Midplanes}). Such depolarization regions are mostly an effect of decreasing grain alignment efficiency since the dust grain orientation becomes randomised by the increase in dust-gas collisions in high pressure regions  (see Eq. \ref{eq:OmegaRatio}). In the zoom-in panel we depict the central region around the spine of the main filament and the contour where $\Sigma P_{\mathrm{l}} = 7\ \%$ as well as the contour of $B_{\mathrm{||}}=0\ \mu\mathrm{G}$, i.e. the characteristic kink in the magnetic field direction. We emphasise that the contours do barely intersect. This may indicate that the magnetic field is not the most dominant factor contributing to $\Sigma P_{\mathrm{l}}$. We also see that the polarization signal does not originate at the spine of the main filament. This is consistent with the right panels of Fig. \ref{fig:AccDust}. In these panels, we show the relative change $\Delta P_{\mathrm{l}}$ of linear dust polarization along the LOS. Indeed, depolarization ($\Delta P_{\mathrm{l}} < 0\ \%/\mathrm{pc}$) takes place at the spine of the main filament but also seems to follows the tail of the main filament (compare Fig. \ref{fig:Midplanes}).

For the linear dust polarization, the inverse of the gas pressure ${ p_{\mathrm{g}}^{-1}=(n_{\mathrm{g}}T_{\mathrm{g}})^{-1} }$, the angular-dependency $\sin^2 \vartheta$ (compare Eq. \ref{eq:DCpol}) as well as the total radiation field are the most relevant parameters (see \S~\ref{sect:ObsDust}). For the radiation field we introduce the parameter 
\begin{equation}
\zeta =   \frac{1}{u_{\mathrm{rad}}} \int \lambda Q_{\mathrm{RAT}} \gamma_\lambda u_\lambda \mathrm{d}\lambda
\label{eq:Zeta}
\end{equation}
as an estimator where ${ u_{\mathrm{rad}} =  \int  u_\lambda \mathrm{d}\lambda }$ is the total radiation field. Hence, the parameter $\zeta$ represents the average dependency of the RAT alignment on the magnitude of the radiation field (see Eq. \ref{eq:OmegaRatio}). We quantify the correlation of $\Delta P_{\mathrm{l}}$ with these parameters by the Pearson correlation coefficient $\mathcal{R}$ (see \S~\ref{sect:LOSAnalysis}). The statistics is calculated for each path element $\mathrm{d}\ell$ along the LOS within the range of the region B (see red dotted line in Fig. \ref{fig:AccDust}). In Table \ref{tab:PearDust} we list the results of $\mathcal{R}$ of region B in comparison to that of the regions A and C, respectively. For all regions we find a moderate or strong negative correlation with $p_{\mathrm{g}}^{-1}$. The angular-dependency $\sin \vartheta$ is only the second most important quantity for the dust polarization signal in the main filament followed by the radiation field as the least important factor. This is consistent with the study of \cite{Reissl2020} where the inverse gas pressure is identified to be the most dominant factor for synthetic dust polarization in large-scale observations of the diffuse and translucent ISM. This is because of the strong influence of gas pressure on grain alignment (see Sect. \ref{sect:DustAndExtionction}).


In Fig. \ref{fig:AccLine} we present the LOS analysis of the accumulated circular polarization $\Sigma P_{\mathrm{c}}$ of the HI Zeeman split $1420\ \mathrm{MHz}$ line as well as its relative change $\Delta P_{\mathrm{c}}$. The definition of $\Sigma P_{\mathrm{c}}$ and $\Delta P_{\mathrm{c}}$ is again given in Sect.  \ref{sect:LOSAnalysis}. We note several local instances of a change in the sign of $\Sigma P_{\mathrm{c}}$. The largest change in $\Sigma P_{\mathrm{c}}$ in the maps of Fig. \ref{fig:AccLine} occurs in the lower right corner and close to the spine of the main filament. In the zoom-in panel we depict the surrounding of the spine as well as the contour where $\Sigma P_{\mathrm{c}}=0\ \%$ i.e. the zero crossing in comparison with contour of the magnetic field reversals. Both contours are roughly congruent and approximately parallel to the LOS. Hence, the circular polarization signal seems to be sensitive to the magnetic field direction. However, such a correlation cannot be seen along the entire filament where we find more deviations between the zero crossing of $\Sigma P_{\mathrm{c}}$ and the magnetic field reversal in the regions A and C (see Fig. \ref{fig:AccAppLine}). In contrast to linear dust polarization circular line polarization depends also directly on the magnetic field strength itself. We also note that most of the increase in circular polarization ($\Delta P_{\mathrm{c}}>0\ \%/\mathrm{pc}$) seems to be due to the fact that the magnetic field strength is generally higher for $X  \gtrsim 6\ \mathrm{pc}$ and not because of the field direction per se. In Table \ref{tab:PearLine} we present the Pearson $\mathcal{R}$ coefficients for $\Delta P_{\mathrm{c}}$ dependent on the HI density $n_{\mathrm{HI}}$, gas temperature $T_{\mathrm{g}}$, the total magnetic field strength $B$, and the angular-dependency $\cos \vartheta$ (compare Eq. \ref{eq:ZeemanV}), respectively, for the regions A - C. We report only a weak correlation of $\Delta P_{\mathrm{c}}$ with $n_{\mathrm{HI}}$, $T_{\mathrm{g}}$, and $B$. We emphasise that the impact of $\cos \vartheta$ is roughly of the same order as all the other parameters. Apparently, such a weak correlation seems to be enough to detect the reversals in the magnetic field direction in Zeeman observations (see \S~\ref{sect:MagFieldStrength}) but does not allow to deduce the underlying magnetic field morphology in the main filament.

Figure \ref{fig:AccRM} shows the LOS analysis for the accumulated $\Sigma RM$ in region B as well as the relative change $\Delta RM$. Here, the results are similar to that of the Zeeman measurements as far as most of the signal originates close to the denity peak of the main filament. In contrast to the Zeeman effect where $\Delta P_{\mathrm{c}}$ does not always coincide with the magnetic field reversal, i.e. $B_{\mathrm{||}}=0\ \mu\mathrm{G}$, the zero crossing of $\Delta RM$ coincides almost exactly with the magnetic field reversal along the entire filament (see also Fig. \ref{fig:AccAppRM}). However, the tail of the main filament seems only marginally to contribute to $\Delta RM$ compared to Zeeman and dust polarization.

We present the Pearson coefficient $\mathcal{R}$ for the correlation of $\Delta RM$ with electron density $n_{\mathrm{el}}$, the total field strength $B$, and the angular-dependency $\cos \vartheta$ (compare Eq. \ref{eq:DefinitionRM}) in Table  \ref{tab:PearRM}. The $\mathcal{R}$ reveals a clear trend for the regions A, B, and C, respectively. In the main filament the relative change $\Delta RM$ is mostly modulated by $n_{\mathrm{el}}$ followed by the total magnetic field strength $B$. {This may be because the electron fraction is connected to the turbulent gas component while the magnetic field seems rather well behaved in vicinity of the main filament compared to the more twisted field in the diffuse cloud (see figures \ref{fig:3DMHD} and \ref{fig:Midplanes}).  This finding that the $RM$ signal is dominated by the free electrons is also noted in \cite{ReisslRM2020} on Galactic scales for a Milky Way analogue. This tight interplay of the free electrons and the magnetic field strength on the resulting $RM$ becomes immediately obvious by comparing Eq. \ref{eq:DefinitionRM}. For the main filament the angular-dependency $\cos \vartheta$ seems to have the least impact on the absolute magnitude of $RM$ observed in Fig. \ref{fig:RMMapProf} though $\cos \vartheta$ is giving the correct sign of the field direction along the LOS (see Sect. \ref{sect:MagFieldStrength})}.



\begin{table}
\centering
   
\begin{tabular}[]{ |l | c |  c | c |}
   \hline
       region & $\log_{\mathrm{10}}\left(p_{\mathrm{g}}^{-1}\right)$ & $\log_{\mathrm{10}}\left( \zeta \right)$ & $\sin^2 \vartheta$ \\ 
   \hline  
   \hline  
      A & -0.87 & 0.282 & 0.283 \\ 
   \hline
     B & -0.26 & 0.200 & 0.426 \\
    \hline
     C & -0.32 & 0.221 & 0.391 \\ 
  \hline
  \end{tabular}
  
  \caption{Pearson correlation coefficient $\mathcal{R}$ for the relative change in $\Delta P_{\mathrm{l}}$ of the linear dust polarization with respect to the inverse pressure $p_{\mathrm{g}}^{-1}$, the radiation field $\zeta$, and the angular-dependency $\sin^2 \vartheta $ for the regions A, B, and C corresponding to the Figs. \ref{fig:AccDust} and \ref{fig:AccAppDust}.}
  \label{tab:PearDust}
\end{table}

\begin{table}
\centering
   \begin{tabular}[]{ |l | c |  c | c | c |}
   \hline
       region & $\log_{\mathrm{10}}(n_{\mathrm{HI}})$ & $\log_{\mathrm{10}}(T_{\mathrm{g}})$ & $B$ & $\cos \vartheta$\\ 
   \hline  
   \hline  
      A & 0.099 & -0.100 & -0.118 & 0.116 \\ 
   \hline
     B & 0.147 & -0.142 & 0.146 & 0.142 \\
    \hline
     C & -0.012 & -0.012 & -0.102 & -0.078  \\ 
  \hline
  \end{tabular}
  \caption{The same as table \ref{tab:PearDust} for the circular line polarization $\Delta P_{\mathrm{c}}$  with respect to HI density $n_{\mathrm{HI}}$, gas temperature $T_{\mathrm{g}}$, the total magnetic field strength B, and the angular-dependency $\cos \vartheta$ for the regions A, B, and C as corresponding to the Figs. \ref{fig:AccLine} and \ref{fig:AccAppLine}.}
  \label{tab:PearLine}
\end{table}

\begin{table}
\centering

   \begin{tabular}[]{ |l | c |  c | c |}
   \hline
       region & $n_{\mathrm{el}}$ & $B$ & $\cos \vartheta$\\
   \hline  
   \hline  
      A & 0.409 & 0.303 & 0.122 \\ 
   \hline
     B & 0.416 & 0.334 & 0.143 \\
    \hline
     C & 0.548 & 0.291 & 0.167 \\ 
  \hline
  \end{tabular}
  \caption{The same as table \ref{tab:PearDust} for $\Delta RM$ with respect to electron density $n_{\mathrm{el}}$, the total magnetic field strength $B$, and the angular-dependency $\cos \vartheta$ corresponding to the Figs. \ref{fig:AccRM} and Fig. \ref{fig:AccAppRM}.}
  \label{tab:PearRM}
\end{table}


\section{Discussion}
\label{sect:Discussion}
In this section, we discuss the implications and limits of our model as well as the observational possibilities of distinguishing between different magnetic field morphologies.

\subsection{Categorization of the main filament}
\label{sect:Categorization}
We categorise the SILCC-Zoom main filament with respect to the broader context of existing surveys of filamentary structures. We note that the M/L is usually given as a single number for the entire filament up to some radius $r$ while \cite{Stutz2016} and \cite{Alvarez2020} quantify their filaments as a radially dependent profile \mlr. 

For the main filament we find a line mass scaling factor for the profile of $K=63.24\ \mathrm{M_{\odot}\ pc^{-1}}$. This is lower than the values inferred for the Orion/ISF and the California L1482 filaments with $K=385\ \mathrm{M_{\odot}\ pc^{-1}}$ and $K=217\ \mathrm{M_{\odot}\ pc^{-1}}$, respectively. We note again that the SILCC-Zoom simulation analyzed here was not specifically designed to match these two high-mass filaments. Rather it is the only computational model available to us that provides sufficient physical detail and spatial resolution across the filament to allow for the current investigation, while consistently accounting for the larger-scale  dynamics of the surrounding interstellar medium. We point out that these numbers are well above the theoretical stability limit of self-gravitating unperturbed isothermal filaments of  $K_{\rm crit} = 16.0\ \mathrm{M_{\odot}\ pc^{-1}}$ \citep[][]{Inutsuka1997}. As additional physical processes can add to the stability against gravitational collapse, such as internal small-scale turbulent motions and magnetic pressure \citep{Seifried2015}, the presence of protostars or other signatures of star formation are typically associated with 'supercritical' filaments whose line mass lies significantly above $K_{\rm crit}$ \citep[e.g.][]{Marsh2016}. We also note 
that in the outer envelope of our simulated filament the slope $\gamma = 0.34$ of the \mlr profile approximately matches the one of the Orion/ISF $(\gamma = 0.38)$, while the inner portion more closely matches that of California/L1482. 

California has been previously designated a "sleeping giant" \citep[][]{Lada2017} based on the fact that it has a similar mass as Orion~A, but with a lower star-formation rate and its mass is preferentially spread over a lower column density $N_{\mathrm{H}}$ \citep[][]{Lada2009}. Later, \cite{Lada2012} proposed an evolutionary link between the two clouds, where Orion is the fully "matured" sibling, having achieved
star-formation conditions similar to the Orion Nebula Cluster \citep[see also][]{Stutz2016}. Hence, this SILCC-Zoom main filament with its sparsely distributed star-formation regions may better represent such a link between non-star-formaing and star-forming filaments but for lower mass systems like Musca or Taurus \citep[see][]{Andre2014,Cox2016,Arzoumanian2017}.

Recently, observations and simulations of interstellar filaments are sometimes interpreted as a universal characteristic width \citep[see e.g.][however, see  \citealt{Smith2014} and \citealt{Smith2016} for a critical discussion]{ArzoumanianA2011,JuvelaA2012,PalmeirimA2013,KainulainenA2016}. Having identified the SILCC-Zoom main filament as a "low-mass" system, it is supposed to have a constant width scattering around $0.1\ \mathrm{pc} \pm 0.15\mathrm{pc}$.  Meanwhile the main filament does not show a break in the \mlr profile on those scales, but instead has a very clear break further out at larger radii at about $r=1\ \mathrm{pc}$ (see Fig. \ref{fig:DiagML}). In conclusion, in terms of the \mlr profile and the star-formation the main filament looks like observations of low-mass systems like California, however, it fails to reproduce the corresponding characteristic line width. 

\subsection{Is there an impact of grain alignment physics?}
A concern in dust polarimetry is in accounting for the influence for the grain alignment mechanism. A variation in grain alignment efficiency may further modulate the polarization signal. In particular for RAT alignment one has to deal with two types of angular dependencies. The $\Psi$-dependency between the direction of the an-isotropic radiation field and the magnetic field lines as well as the $\vartheta$-dependency of dust polarization is a result the grain geometry of the dust grains (see \S~\ref{sect:ObsDust}). In order to measure the magnetic field morphology, both effects need to be quantitatively determined. 

Furthermore, the applied RAT grain alignment physics is also highly sensitive to gas density, gas temperature, and the magnitude of the radiation field  \citep[see e.g.][]{Lazarian2007}. Density and temperature have an inhibiting effect to the grain alignment efficiency since random gas collisions tend to kick the grains out of stable alignment and the grain alignment efficiency and subsequent dust polarization drops in the dense regions as well as close to the spine of the filament. 

The magnitude of the radiation field enhances the grain alignment according to the RAT mechanism \citep[][]{Lazarian2007}. A higher radiation field spins-up the dust grains more efficiently and, subsequently, the grains couple to the magnetic field by means of paramagnetic effects. Hence, one could expect a higher degree of polarization close to the three distinct star-forming regions of the SILCC-Zoom cutout. Indeed, in circumstellar disks and sub-parsec clouds, the $\Psi$-dependency of RATs may possibly become observable \citep[][]{Andersson2010,Reissl2016}. However, such specific RAT effects with respect to the radiation field are not observable on parsecs scales since gas pressure, i.e. the specific combination of density and temperature can become the dominant factors \cite{Reissl2020}. The mock observations at a wavelength of $850\ \mu\mathrm{m}$ presented in this paper show a similar trend. The degree of linear dust polarization $P_{\mathrm{l}}$ clearly correlates with gas properties of the main filament, leading to a depolarization toward the spine. However, we cannot spot any enhancement of $P_{\mathrm{l}}$ that may be particularly attributed to the star-forming regions where the radiation field is expected to be highest (see Fig. \ref{fig:DustEmission}). Consequently, the $\Psi$-dependency appears to be only a minor effect that does not manifest itself in the dust polarization signal of the main filament.

\subsection{Dissecting magnetic field morphologies}
The bending of field lines by star-forming filaments is predicted either along the filament \citep[][]{Gomez1018} or perpendicular to it \citep[][]{Seifried2017,LiKlein2019}. 
However, the exact field configuration is not well constrained by the available observational data. 
Some favor a helical field \citep[][]{Fiege2000A,Stutz2016,Schleicher2018}, while others interpret their observations as a magnetic field kinked around the filament or remain ambiguous in their conclusions \citep[][]{Heiles1997,Tahani2018}. 

A helical field in a filament may naturally arise by means of shear or collapse at one end of the filament. A subsequent torsion waves may then bend a poloidal field into a helical configuration as suggested in \cite{Fiege2000A}. However, no MHD setup to date seems to be capable of reproducing self-consistently such a helical magnetic field evolution. Hence, we can only rely on our analytical model for comparison.

For observed low-mass filamentary structures such as Musca or Taurus
the projected plane-of-the-sky magnetic field morphology is perpendicular to the spine at larger columns ($N_{\mathrm{H}}>10^{22}\ \mathrm{cm}^2$, \citealt{Planck2016}). Such a finding would be consistent with magnetic  field reversal associated with both a bent field as well as a helical field configuration. In contrast to a field being perpendicular to the filament, \cite{Pillai2015} and Pillai at al, sub. report a magnetic field in G0.253+0.016 that follows the spine. However, this may possibly be explained by a low inclination \citep[][]{Tomisaka2015,Reissl2018} or a helical field with a pitch angle of about $\alpha  \lesssim  45^{\circ}$ \citep[][]{Reissl2018} i.e. projection effects. 

Projection effects onto the plane of the sky impact the interpretation of observational data in an even more complex manner. Considering dust polarization, different magnetic field configurations may result in an analogous polarization pattern \citep{Reissl2014,Tomisaka2015,Reissl2018}. In order to break such degeneracies, \cite{Reissl2014} suggested a combination of linear and circular dust polarimetry. Later, \cite{Reissl2018} explored this idea in more detail by considering synthetic Zeeman observations in addition to dust polarization on the basis of analytical models of distinct density distributions and different magnetic field configurations. This study includes helical and kinked magnetic fields. 
Indeed, somewhat similar studies of predicting profiles by means of mimicking dust polarization for different magnetic field geometries are presented e.g. in \cite{Fiege2000A}, \cite{Padovani2012}, and \cite{Tomisaka2015}. However, these studies lack the proper treatment of grain alignment physics or turbulence. Naively, the radial profiles of linear and circular dust polarization of the SILCC-Zoom simulation are expected to follow the predictions of the analytical analytical model, particularly given that grain alignment seems to be not the dominant factor in large-scale observations \citep[][]{Seifried2019,Reissl2020}. A good part of the main filament does not reproduce  the characteristic feature in the dust polarization profile (see \S~\ref{sect:ObsDust}) predicted by the analytical model. In fact, only three distinct regions of the main filament resemble the linear dust polarization profiles of the kinked magnetic field. We find that the origin of the polarization signal is close to, but not identical, to the spine and the tail of the filament (see fig. \ref{fig:AccDust}). This is due to the fact that gas pressure ($p_{\mathrm{g}}$) and the overall radiation field ($\zeta$) contribute roughly of the same order as the angular-dependency along the LOS. Since the angular-dependency is not the  dominant factor, dust polarization cannot reliably trace the magnetic field morphology. 

Reversals in the magnetic field direction are observed by means of Zeeman measurements in {Orion A} by \cite{Heiles1997}. As a possible explanation a scenario of a shock running into a molecular cloud is discussed. Effectively, this scenario is similar to the "sushi" model suggested by \cite{Inoue2013} and \cite{Inoue2018} as well as the situation within the SILCC-Zoom main filament where the magnetic field lines are kinked in direction of the moving gas.  Although the LOS magnetic field strength derived by Zeeman measurements slightly overestimates the overall strength of the original MHD field on average by about a factor of 2, we detect the field reversal associated with the kinked field lines by the main filament. Signs of these field reversals are also present in our synthetic $RM$ observations but the $RM$ somewhat underestimates the overall field magnitude. 

A similar method to measure $B_{\mathrm{||}}$ on the basis of $RM$ is presented in the pioneering work of \cite{Tahani2018}. Additionally, they subtracted the $A_{\mathrm{V}}$ measured 'on' the filament and 'off' the filament in order to eliminate a possible $RM$ contamination of ionised material behind and in front of the object. We note that this step is not necessary for our modelling since we create mock observations for an isolated cube without background or foreground contamination. We note that many observed high-mass filaments, e.g.\ in Orion, have a higher field strength of up to $\sim 1\ \mathrm{mG}$, whereas we see values of $\sim 10\ \mu\mathrm{G}$ in the low-mass SILCC-Zoom simulation or $\sim 100\ \mu\mathrm{G}$ in the filament simulations of \cite{LiKlein2019}. However, the ratio of of electron density to gas density  $n_{\mathrm{el}}/n_{\mathrm{gas}}\approx 10^{-4}$ in the SILC-Zoom main filament and the analytical model
is of the same order as the one presented in \cite{Tahani2018}. Hence, the magnitude of the $RM$ observed in \cite{Tahani2018} is much larger than the one presented in Fig. \ref{fig:RMMapProf}. 

More recently, \cite{Tahani2019} determined that a kinked magnetic field may be the most likely configuration in Orion-A by means of a Monte Carlo analysis. However, in their study the 3D magnetic field is merely projected by integration along the LOS lacking the physics of our RT post-processing as outline in Sects. \ref{sect:ObsDust} - \ref{sect:ObsSync} as well as a turbulent gas component.

However, we demonstrate that the parameterizations $heli_{\mathrm{30}}$ as well as $flow$ predict identical field reversals along the spine of the filament of our analytical model. A further complication is that both the profile of $heli_{\mathrm{30}}$ and  $flow$ are very similar when probing them by Zeeman and $RM$ as tracer techniques. Furthermore, our LOS analysis of the SILCC-Zoom filament shows that the electron density seems to be a more dominant factor compared to the contribution of the magnetic field.

For the origin of the Zeeman signal in the SILCC-Zoom filament (see Fig. \ref{fig:AccLine}) we find that the correlations are more ambiguous, with density and temperature variations being as important as the angular-dependency, and so any fingerprint of a distinct magnetic field morphology becomes smeared out. 

\subsection{Limitations and outlook}
The LOS analysis discussed in \S~\ref{sect:Origin} strongly suggests that the magnetic field morphology may not unambiguously be detected by dust polarization, Zeeman measurements, or the Faraday $RM$. However, such a conclusion would only be valid within the limitations of your case study. In order to place it in the proper context, we measure the line mass profiles \mlr of the SILCC-Zoom main filament as outlined in \S~\ref{sect:Categorization}. We find the line mass $M/L$ of the SILCC-Zoom main filament to be in the low-mass regime (see Fig. \ref{fig:DiagML}). As already put forward by \cite{Stutz2016}, the lower mass systems appear to be turbulence-dominated, while the higher mass clouds are likely more complex, with turbulence becoming sub-dominant compared to gravity and the magnetic field  \citep[e.g.][]{GonzalezLobos2019}. These results indicate that mass and in particular $M/L$ are the most fundamental parameters to address. 

We speculate that for such high-mass filaments with their lower turbulence polarization measurements with different tracers may actually be a viable method for the determination of the magnetic field morphology. 
%
%
Hence, a systematic study of filaments with different $M/L$ would be a way forward, both in terms of numerical simulation where we have very few setups with multi-physics required for synthetic observations and in the observations where only very few objects have been studied so far.

Finally, a systematic analysis of position-velocity (PV) diagrams may be an alternative to Zeeman, dust, and $RM$ as tracer of the magnetic field morphology. For instance, it is noted in \citealt{Alvarez2020} that the gas in the filament L1482 follows a corkscrew like trajectory that cannot be accounted by gravity alone. The corresponding PV diagrams of L1482 of several lines indicate that the gas potentially moves along the field lines of a helical morphology. Consequently, the potential of PV diagrams to determine the magnetic field morphology will be explored in a followup study in great detail. 



\section{Summary and Conclusion}
\label{sect:Summary}
In this paper we examine the reliability of dust polarization, HI 21 cm line polarization including the Zeeman effect, and Faraday $RM$ as tracers of magnetic field properties in a low-mass filament. Especially, we analyze their potential to observationally infer the magnetic field strength and to distinguish between different field morphologies like a kink magnetic field and a helical field. Theory predicts that all of these tracers have an inherent angular-dependency between LOS and magnetic field direction. Hence, the signal of dust polarization, line polarization including the Zeeman effect, and Faraday $RM$ should be highly depend on the magnetic field morphology. 

To quantify this, we perform a case study and analyze mock observations of a cut-out of the SILCC-Zoom MHD simulation post-processed with the RT code {\sc POLARIS}. It contains three star-forming regions and a kinked magnetic field bent around a low-mass large-scale filament. For further comparison, we also create and study an analytical model of a filament with cylindrical symmetry and parameters similar to the SILCC-Zoom simulation. We superimpose a kinked magnetic field similar to the simulation and also study the effect of a simple helical field structure. This allows to interpret the synthetic observations and puts us into the position to qualitatively evaluate the influence of gas turbulence and variations in the magnetic field morphology and strength on the tracer signal. We derive profiles of the synthetic dust, Zeeman, and $RM$ observations perpendicular to the spine of the SILCC-Zoom filament and of the analytical model. The strength and direction along the LOS of the magnetic field is recovered based on the parameters of the synthetic $RM$ and Zeeman observations. Finally, we probe the signal of the three different tracers along the LOS and perform a statistical analysis in order to determine their origin with respect to the SILCC-Zoom filament.

The main results of our case study are summarised as follows:


\begin{itemize}
    \item  We note that the analytical model can make vastly different predictions concerning the synthetic observations of dust polarization, Zeeman effect, and $RM$ compared to the results coming from the SILCC-Zoom MHD filament. It seems that such simple analytical models are not viable to predict actual observations since they cannot accurately account for turbulence in density and the variations in the magnitude and direction of the magnetic field. 

    \item By evaluating radial profiles of linear dust polarization along the entire SILCC-Zoom filament we identify three distant regions. Within these regions the linear dust polarization profiles show the characteristic features of a kinked magnetic field morphology as predicted by the analytical model. However, such features cannot be detected along the entire SILCC-Zoom filament. For most of the filament the grain alignment and subsequently the dust polarization signal is modulated by the gas pressure as the dominant parameter and not the angular-dependency on the magnetic field. In general, linear dust polarization observations seem not to be a reliable means to distinguish between different magnetic field morphologies.
    
    \item The circular dust polarization profiles in turn cannot reproduce any of the characteristic profiles along the filament SILCC-Zoom predicted by the analytical model. On top of the grain alignment physics, non-parallel field lines along the LOS are a prerequisite to amplify circular dust polarization. Hence, circular dust polarization seems to be even more sensitive to gas turbulence and variations in the magnetic field as linear dust polarization. 
    
    \item Radial profiles of circular line polarization measurements including Zeeman effect do not match the predictions of the analytical model along the entire filament. Probing the origin of the line polarization signal within the SILCC-Zoom filament reveals no consistent correlation with the parameters of density, temperature, and the magnetic field. However, the Zeeman effect allows to recover the LOS magnetic field strength and its direction of the SILCC-Zoom filament with an average overprediction of about $2.1$. Comparing maps of the recovered magnetic field to that of the projected field of the SILCC-Zoom simulation reveals that the field reversals with respect to the spine in both maps do coincide. However, such reversals of the field orientation are common for many different field morphologies. By comparing the profiles of the recovered magnetic field strength resulting from the SILCC-Zoom filament and the analytical model we show that Zeeman observations remain highly ambiguous concerning the detection of any underlying magnetic field morphology.
    
    \item Synthetic $RM$ profiles of the SILCC-Zoom filament have barely any resemblance with the predictions of the analytical model. On the contrary, some of the synthetic $RM$ profiles show similarities with a helical field rather than with the actual kinked magnetic field morphology of the SILCC-Zoom filament. As a statistical analysis along the LOS reveals that the $RM$ is mostly modulated by electron density rather than the magnetic field. Indeed, synthetic $RM$ observations allow to recover a map of the LOS magnetic field strength of the SILCC-Zoom filament. The recovered field reversals agree with the kinked field morphology of the SILCC-Zoom simulation but we report an overestimation of the field strength by a factor of about $3.1$. Comparisons of profiles of the magnetic field strength recovered by the $RM$ with the profiles of the actual SILCC-Zoom field as well as the analytical model uncovers that the $RM$ remains inconclusive
    since it cannot distinguish between helical and kinked fields. Hence, the $RM$ seems to be an ambiguous and even misleading tracer for discriminating between different magnetic field morphologies.
\end{itemize}

We note again that these results are valid within the scope of our case study of a low-mass filament. A more systematic study of high-mass filaments may allow to distinguish between different magnetic field morphologies. We emphasise that we do not make any claim in this paper about the actual magnetic field morphology in any observations  of filamentary structures. Instead, we merely question the consensus that certain observational tracers may be particularly suitable to answer this outstanding problem reliably.  

In summary, the characteristic signals of dust polarization, the Zeeman effect, or the Faraday $RM$ of a low-mass system are modulated by the turbulence in density, temperature, and variations of the magnetic field strength. As a result of this, the characteristic angular-dependencies between the LOS and the magnetic field direction inherent in these tracers become convoluted and the observed polarization profiles do not agree with the profiles predicted from the analytical models. We conclude, that utilizing dust emission, the Zeeman effect, or $RM$ as a tracer allows to measure the magnetic field strength well, but it may not provide a reliable tool to unambiguously discriminate between different magnetic  field morphologies in the highly turbulent filamentary multi-phase interstellar medium.

\section*{Acknowledgements}
We thank our colleague Simon Glover for input and for many stimulating discussions. We are also grateful to the anonymous referee for constructive comments that helped to improve the analysis presented here. 

S.R. and R.S.K. acknowledge  support  from  the  Deutsche  Forschungsgemeinschaft in the Collaborative Research centre (SFB 881) ``The Milky Way System'' (subprojects B1, B2, and B8) and in the Priority Program SPP 1573 ``Physics of the Interstellar  Medium''  (grant  numbers  KL  1358/18.1,  KL  1358/19.2). The authors also acknowledge access to computing infrastructure support by the state of Baden-W\"urttemberg through bwHPC and the German Research Foundation (DFG) through grant INST 35/1134-1 FUGG.

A.M.S. gratefully acknowledges funding support through Fondecyt Regular (project code 1180350) and funding support from Chilean Centro de Excelencia en Astrof\'{i}sica y Tecnologías Afines (CATA) BASAL grant AFB-170002.

D.S. and S.W. acknowledge the support of the Bonn-Cologne Graduate School, which is funded through the German Excellence Strategy. D.S. and S.W.  also acknowledge funding by the Deutsche Forschungsgemeinschaft (DFG) via the Collaborative Research Center SFB 956 ``Conditions and Impact of Star Formation'' (subprojects C5 and C6). S.W. acknowledges support by the ERC Staring Grant RADFEEDBACK (grant. no. 679852). 

The {\sc FLASH} code used in this work was partly developed by the Flash Center for Computational Science at the University of Chicago. The authors acknowledge the Leibniz- Rechenzentrum Garching for providing computing time on SuperMUC via the project ``pr94du'' as well as the Gauss Centre for Supercomputing e.V. (www.gauss-centre.eu).






\bibliographystyle{mnras}
\bibliography{bibtex}


\appendix

\section{Relative change of the polarization signal}
\label{app:RelChange}

\begin{figure*}
\centering
\begin{minipage}[c]{1.0\linewidth}
      \includegraphics[width=0.50\textwidth]{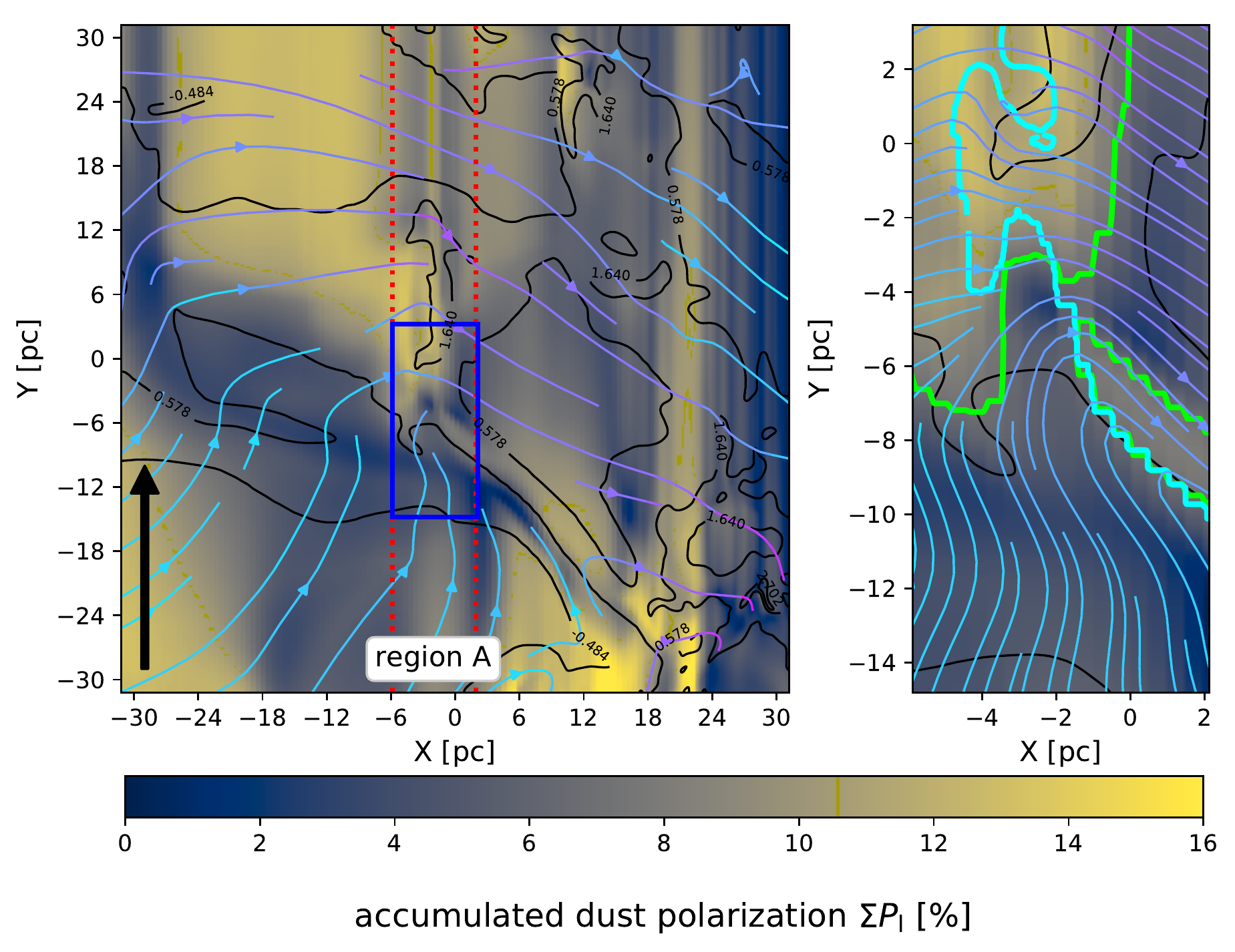}
      \includegraphics[width=0.50\textwidth]{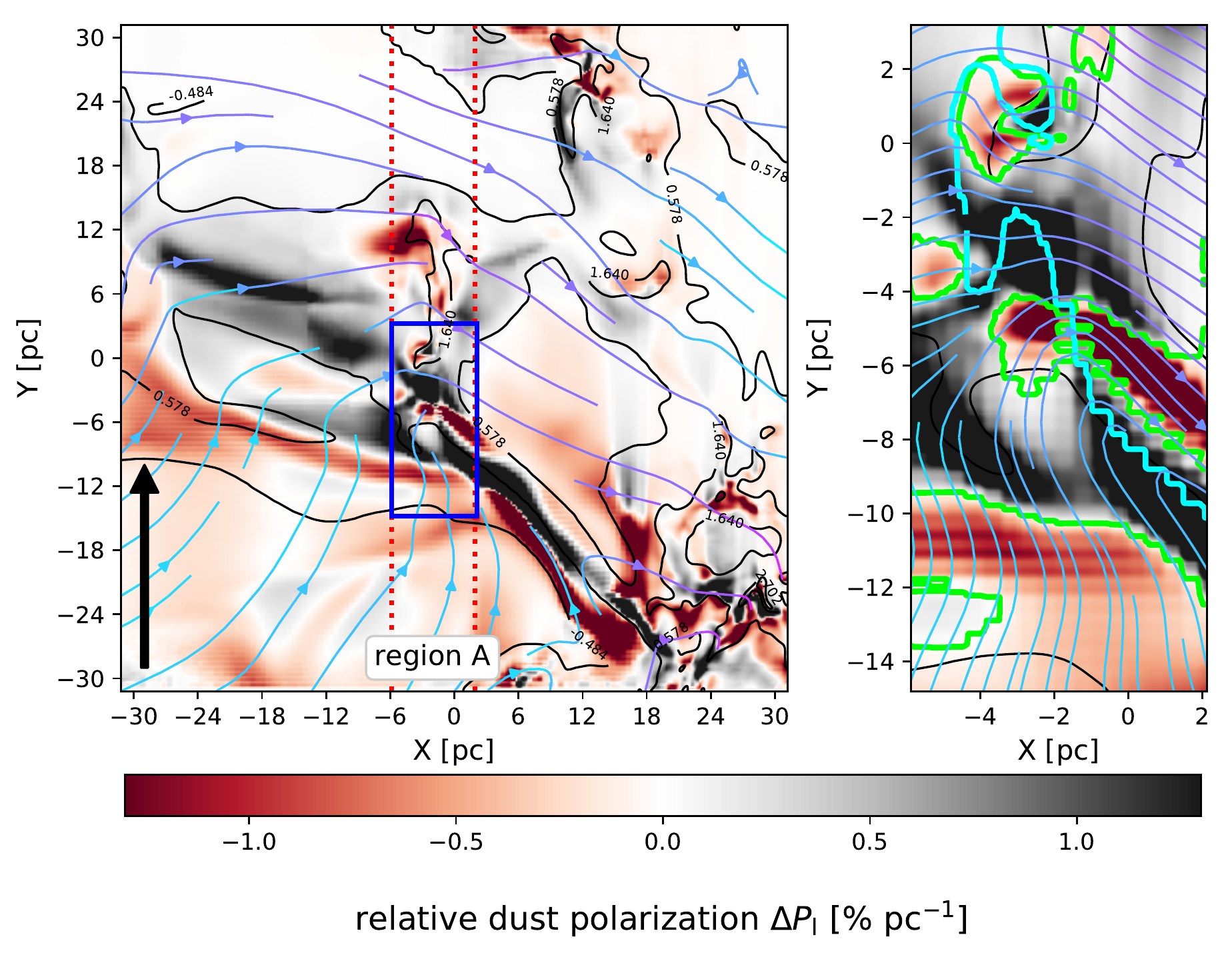}
\end{minipage}
\begin{minipage}[c]{1.0\linewidth}
      \includegraphics[width=0.50\textwidth]{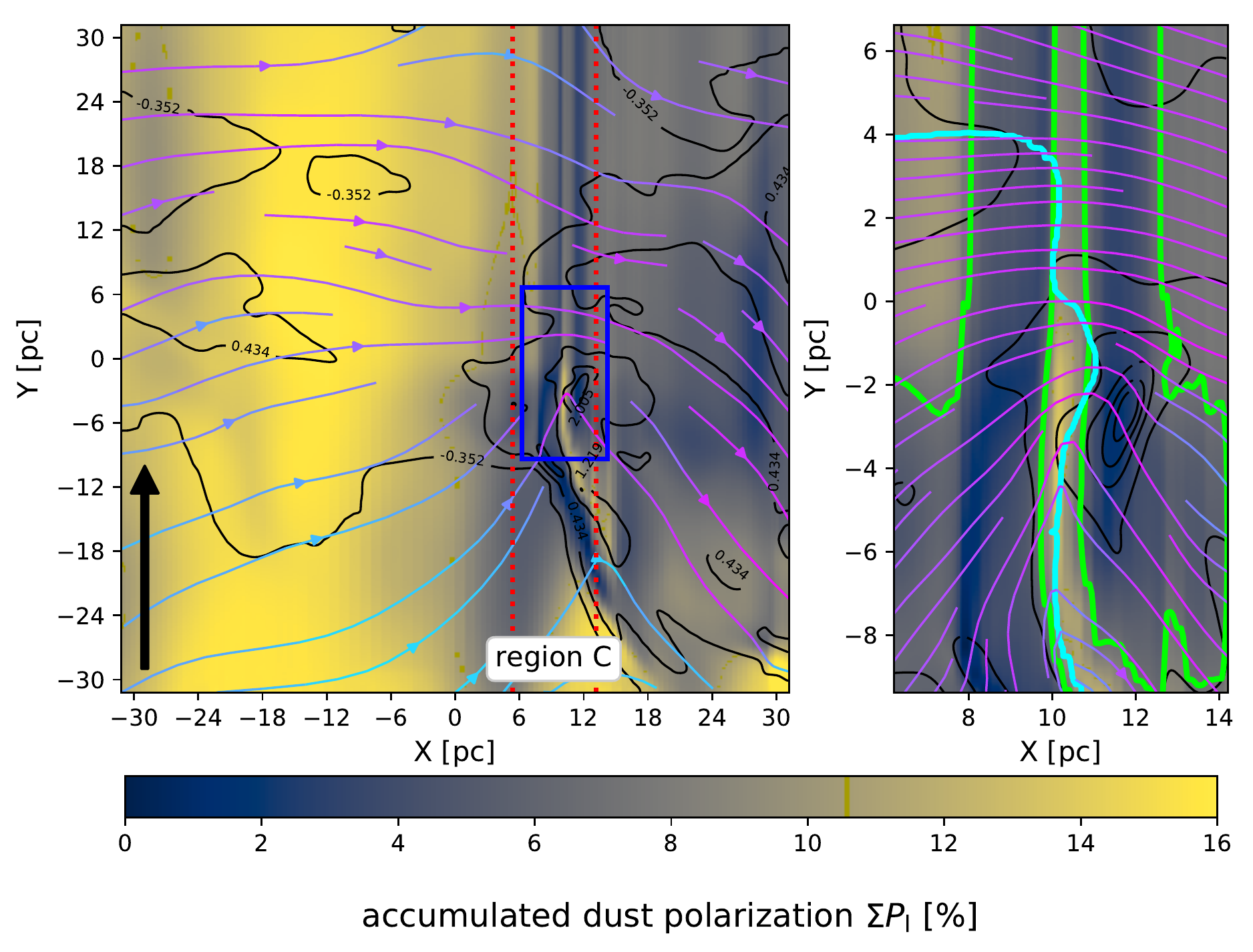}
      \includegraphics[width=0.50\textwidth]{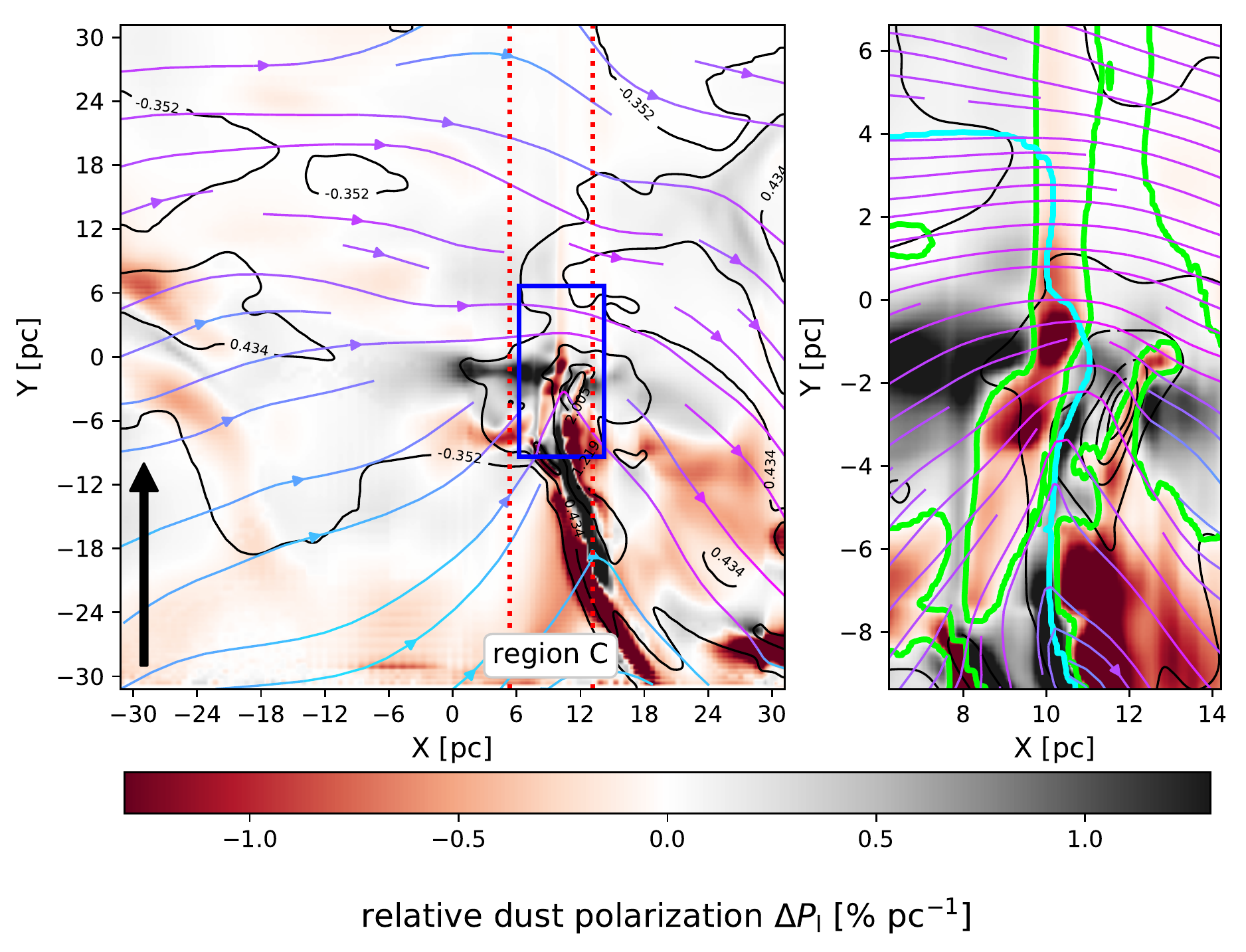}
\end{minipage}
\caption{The same as the linear dust polarization in Fig. \ref{fig:AccDust} but for region A within the $XY-$plane at $Z=22.6\ \mathrm{pc}$ (top) and for region C within the $XY-$plane at $Z=-15.5\ \mathrm{pc}$ (bottom), respectively.}
\label{fig:AccAppDust}
\end{figure*}

\begin{figure*}
\centering
\begin{minipage}[c]{1.0\linewidth}
      \includegraphics[width=0.50\textwidth]{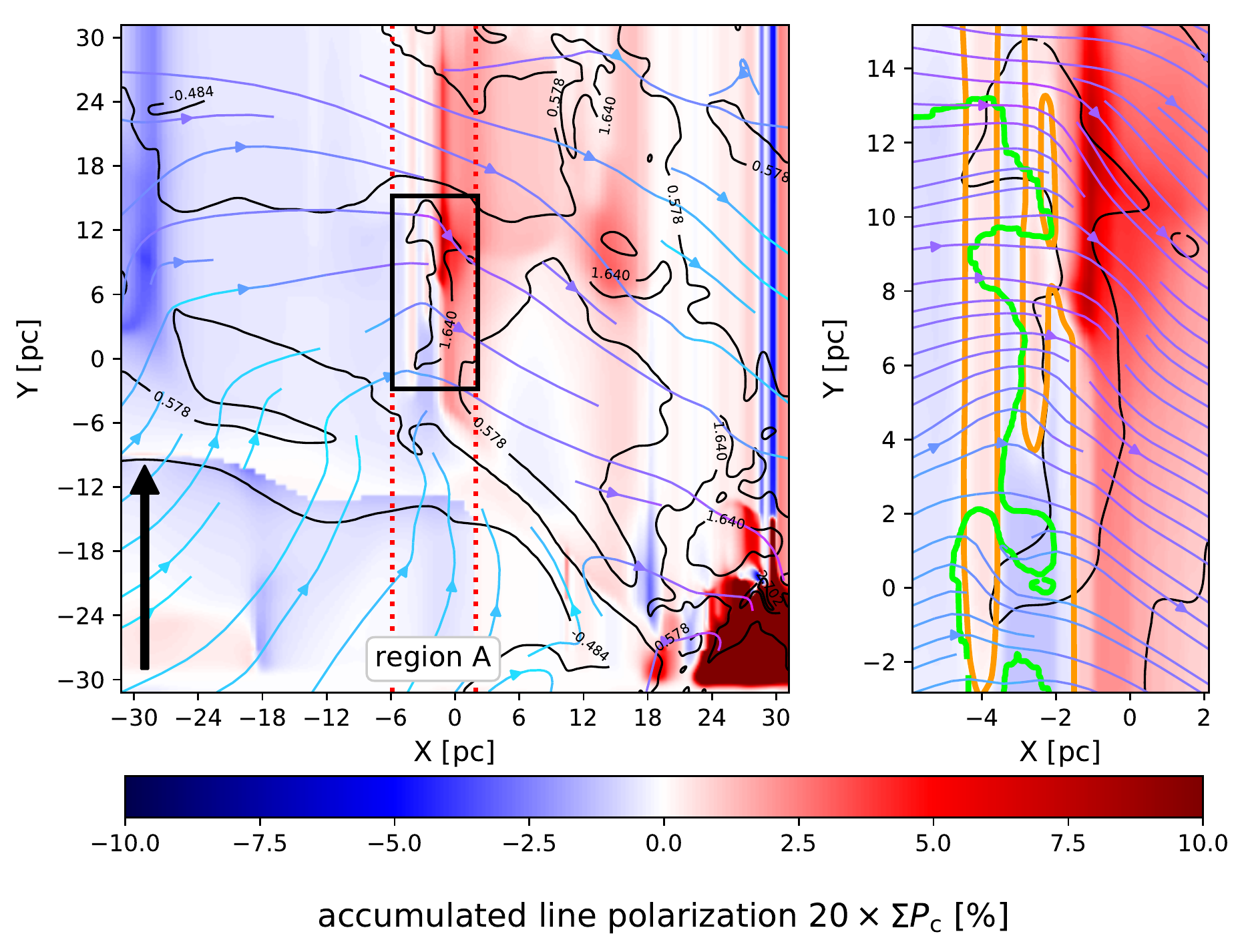}
      \includegraphics[width=0.50\textwidth]{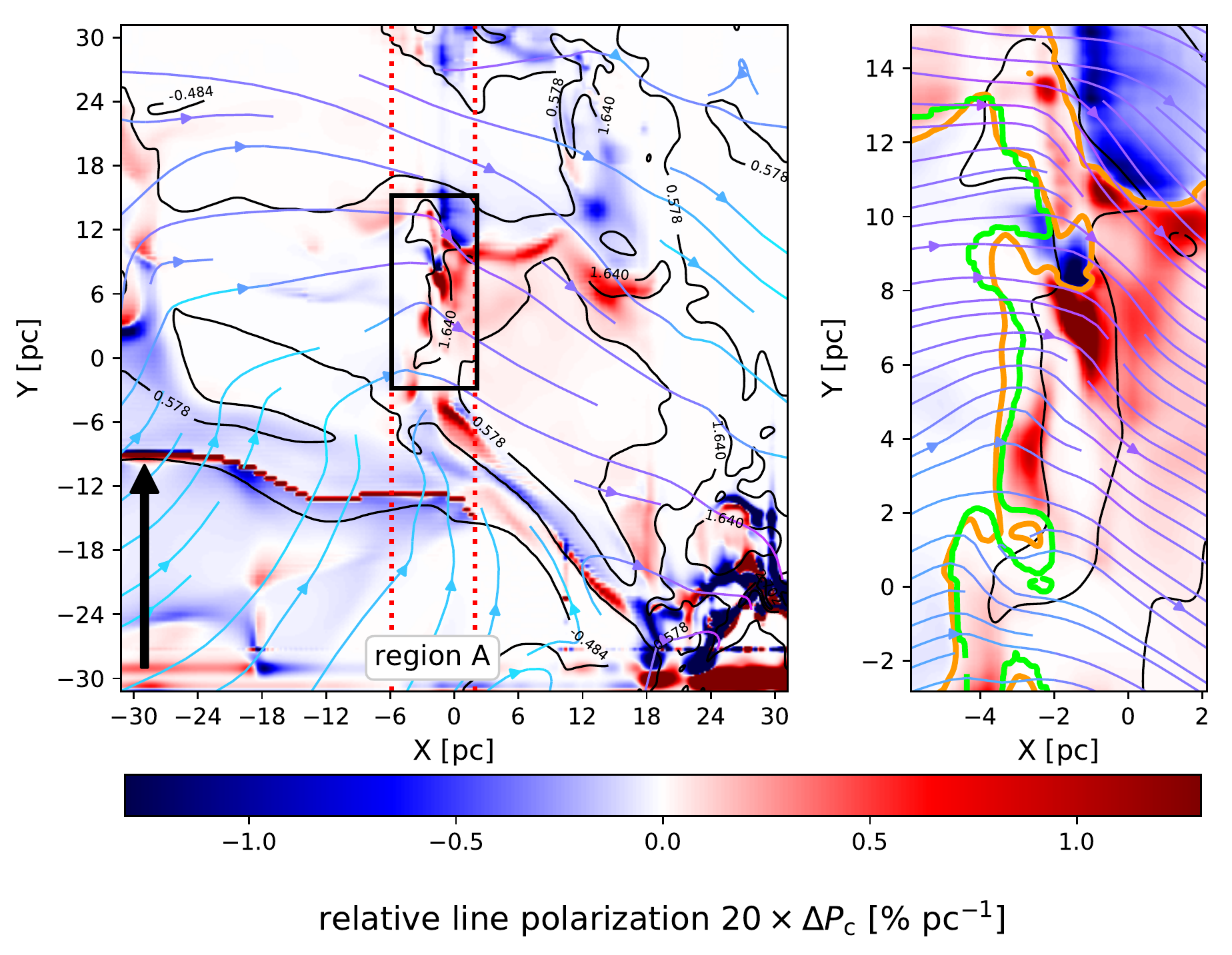}
\end{minipage}
\begin{minipage}[c]{1.0\linewidth}
      \includegraphics[width=0.50\textwidth]{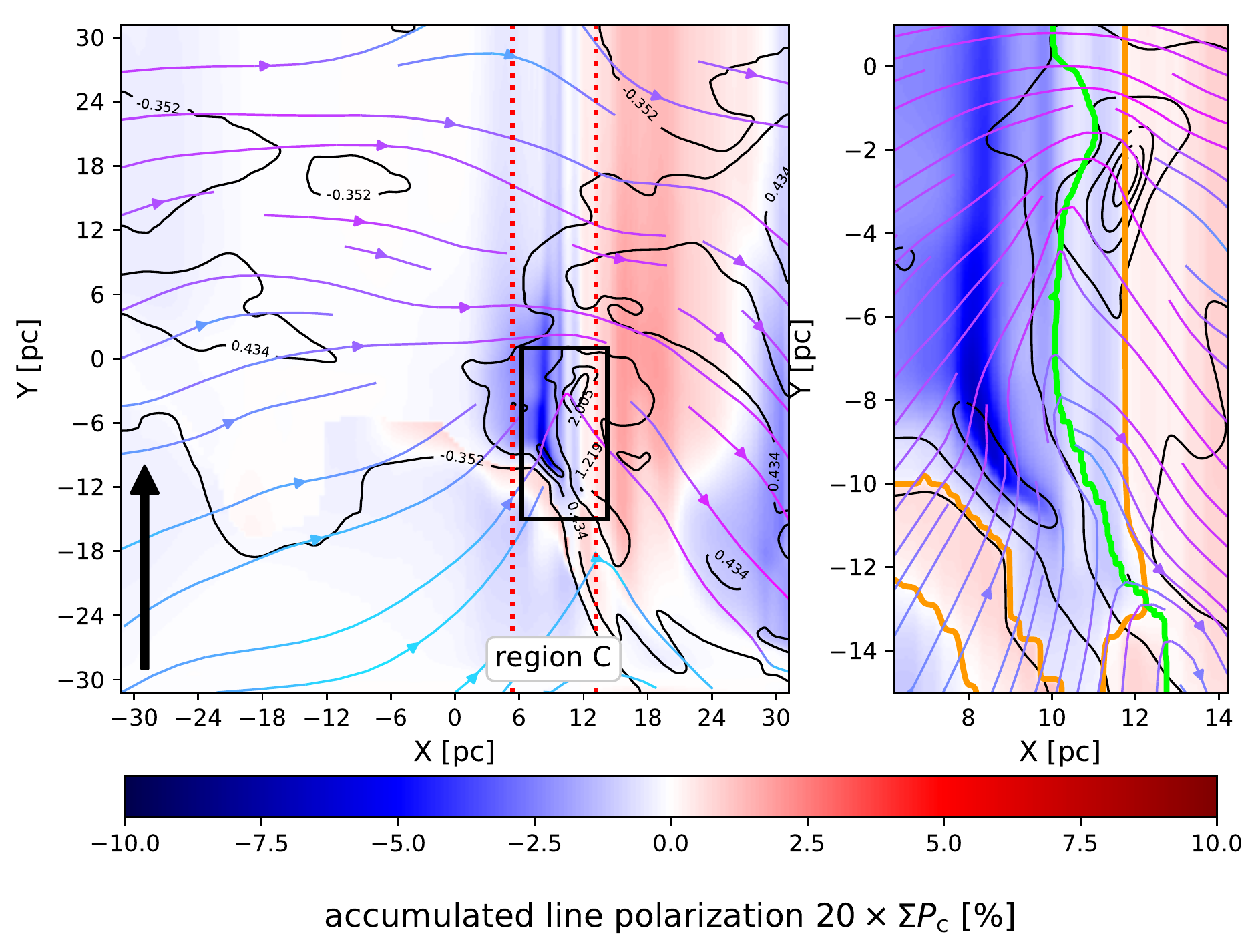}
      \includegraphics[width=0.50\textwidth]{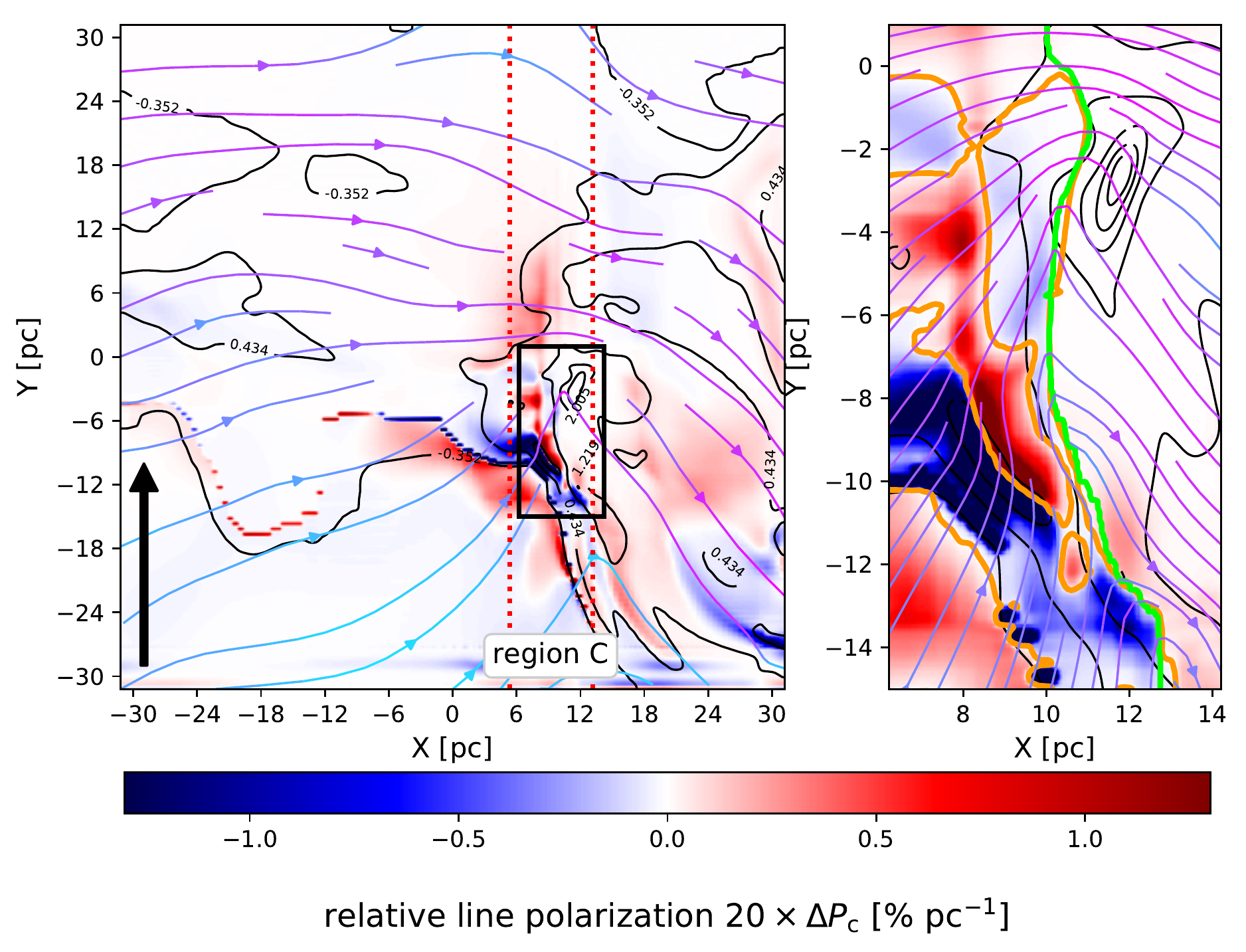}
\end{minipage}
\caption{The same as the circular line polarization with Zeeman effect in Fig. \ref{fig:AccLine} but for region A within the $XY-$plane at $Z=22.6\ \mathrm{pc}$ (top) and for region C within the $XY-$plane at $Z=-15.5\ \mathrm{pc}$ (bottom), respectively.}
\label{fig:AccAppLine}
\end{figure*}

\begin{figure*}
\centering
\begin{minipage}[c]{1.0\linewidth}
      \includegraphics[width=0.50\textwidth]{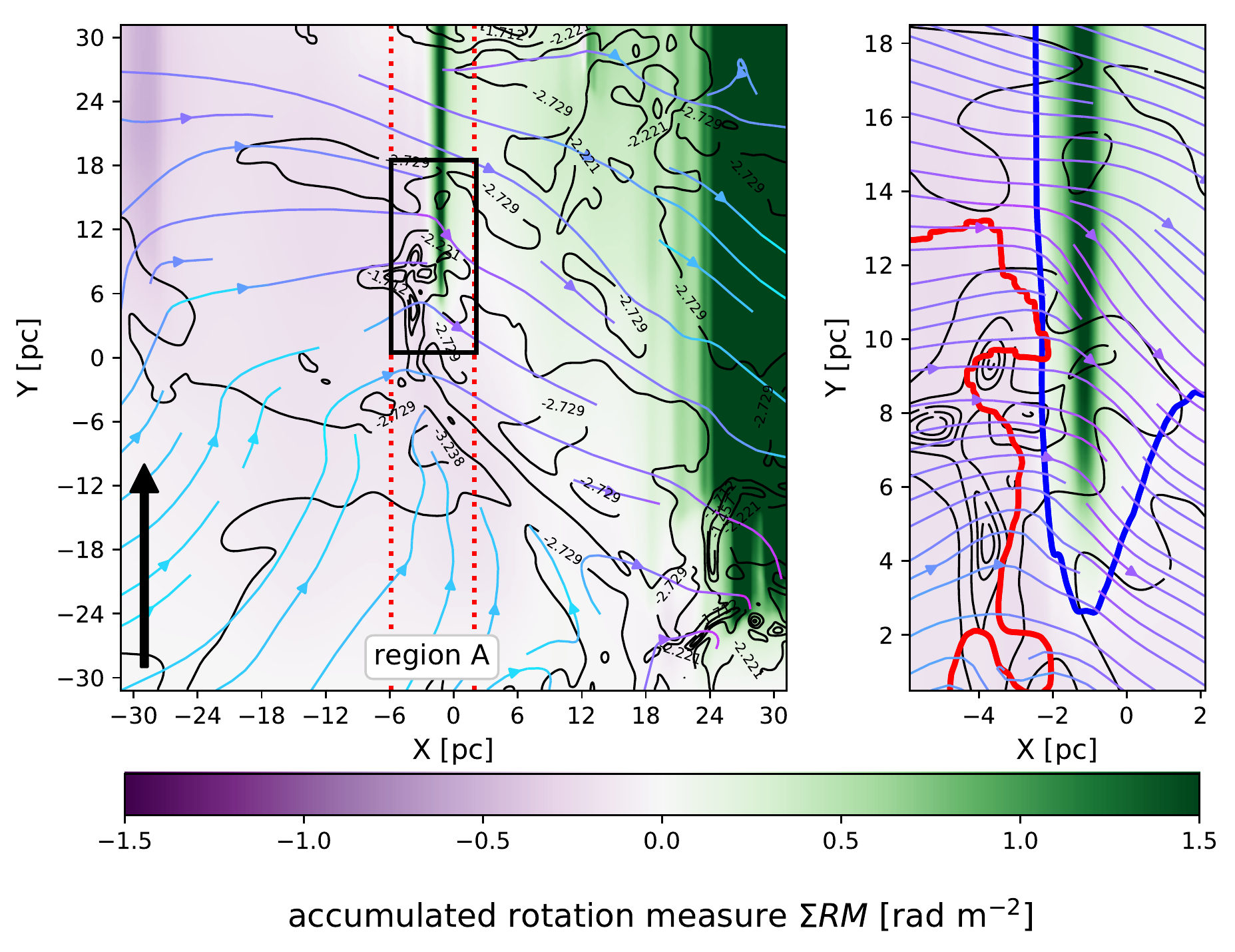}
      \includegraphics[width=0.50\textwidth]{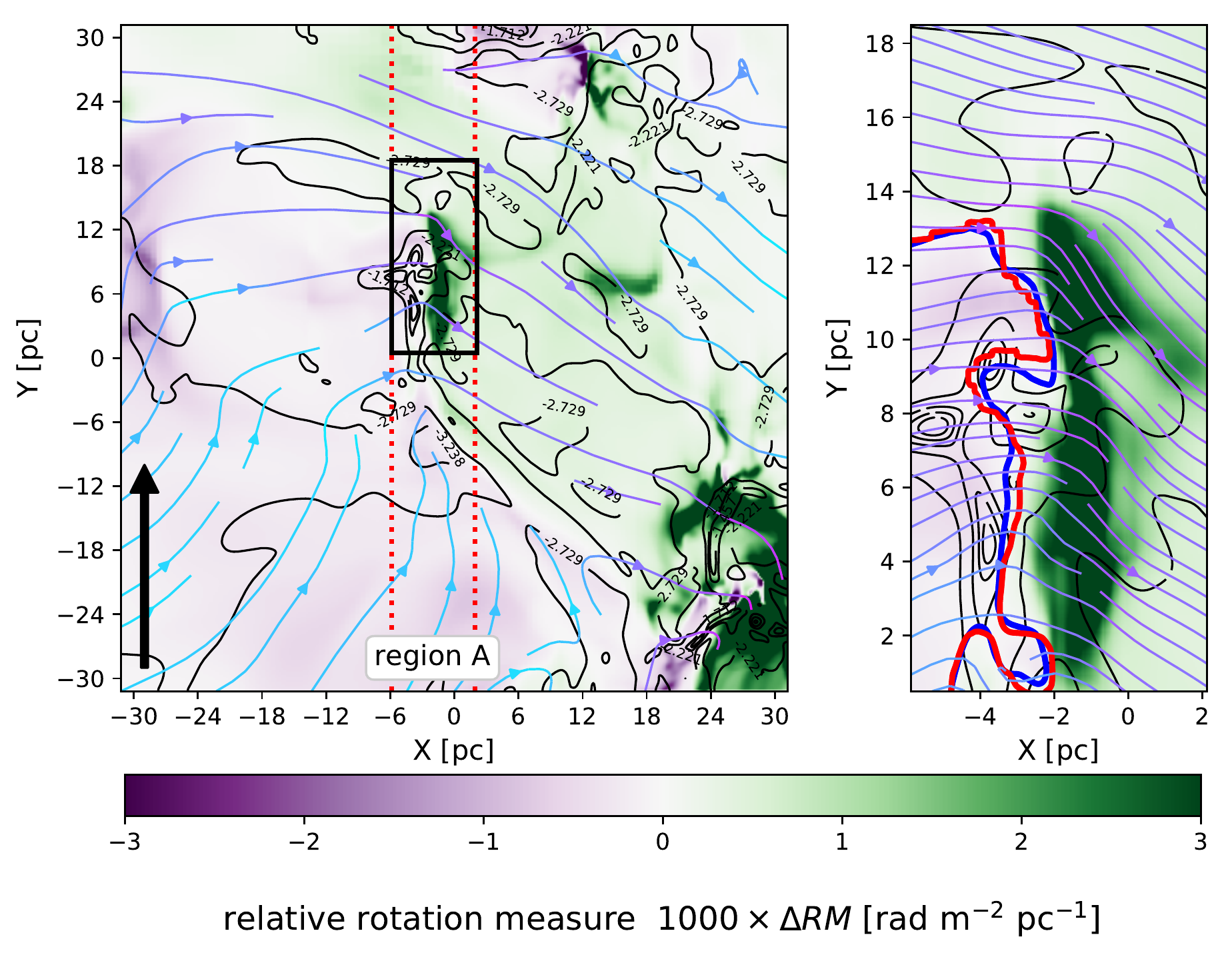}
\end{minipage}
\begin{minipage}[c]{1.0\linewidth}
      \includegraphics[width=0.50\textwidth]{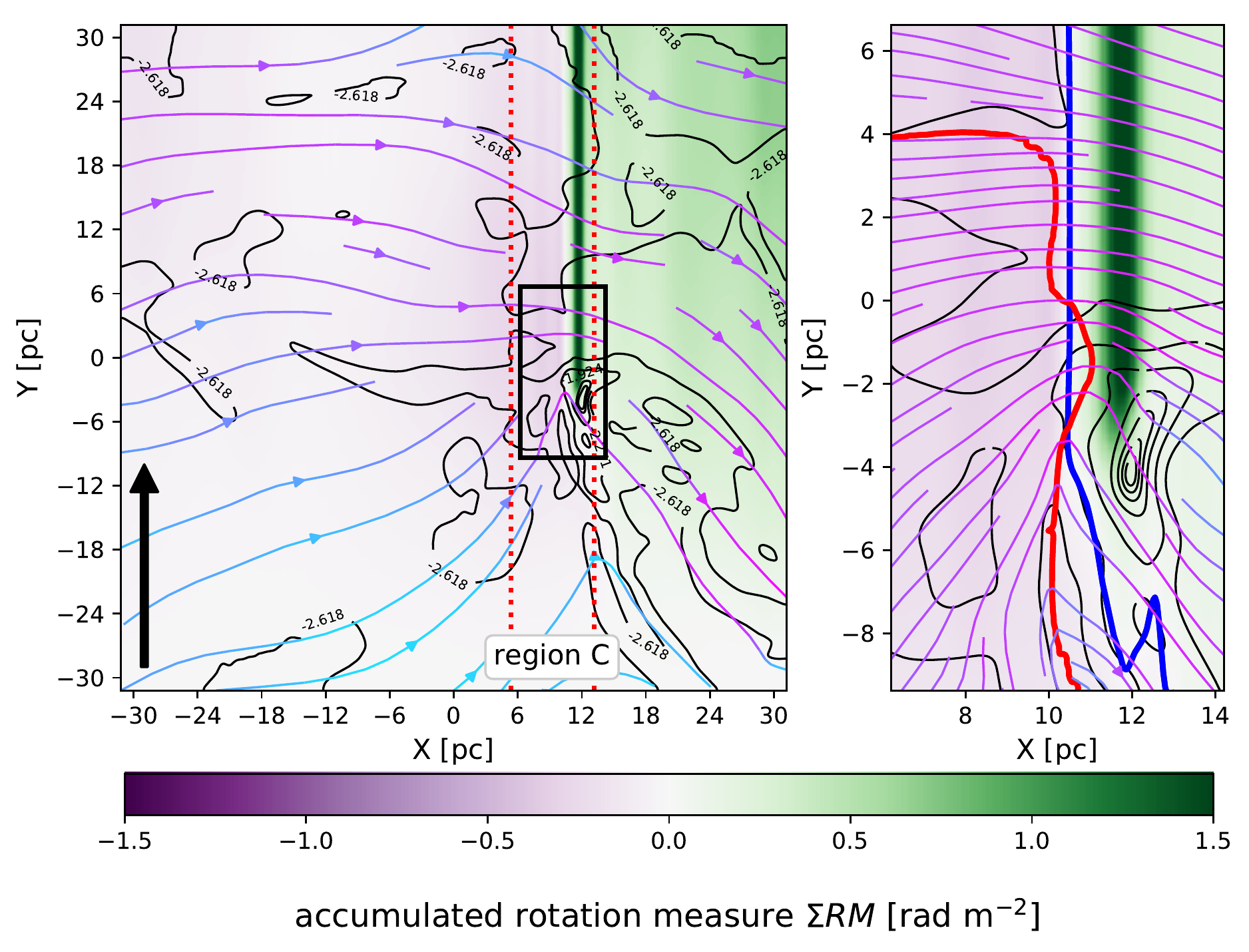}
      \includegraphics[width=0.50\textwidth]{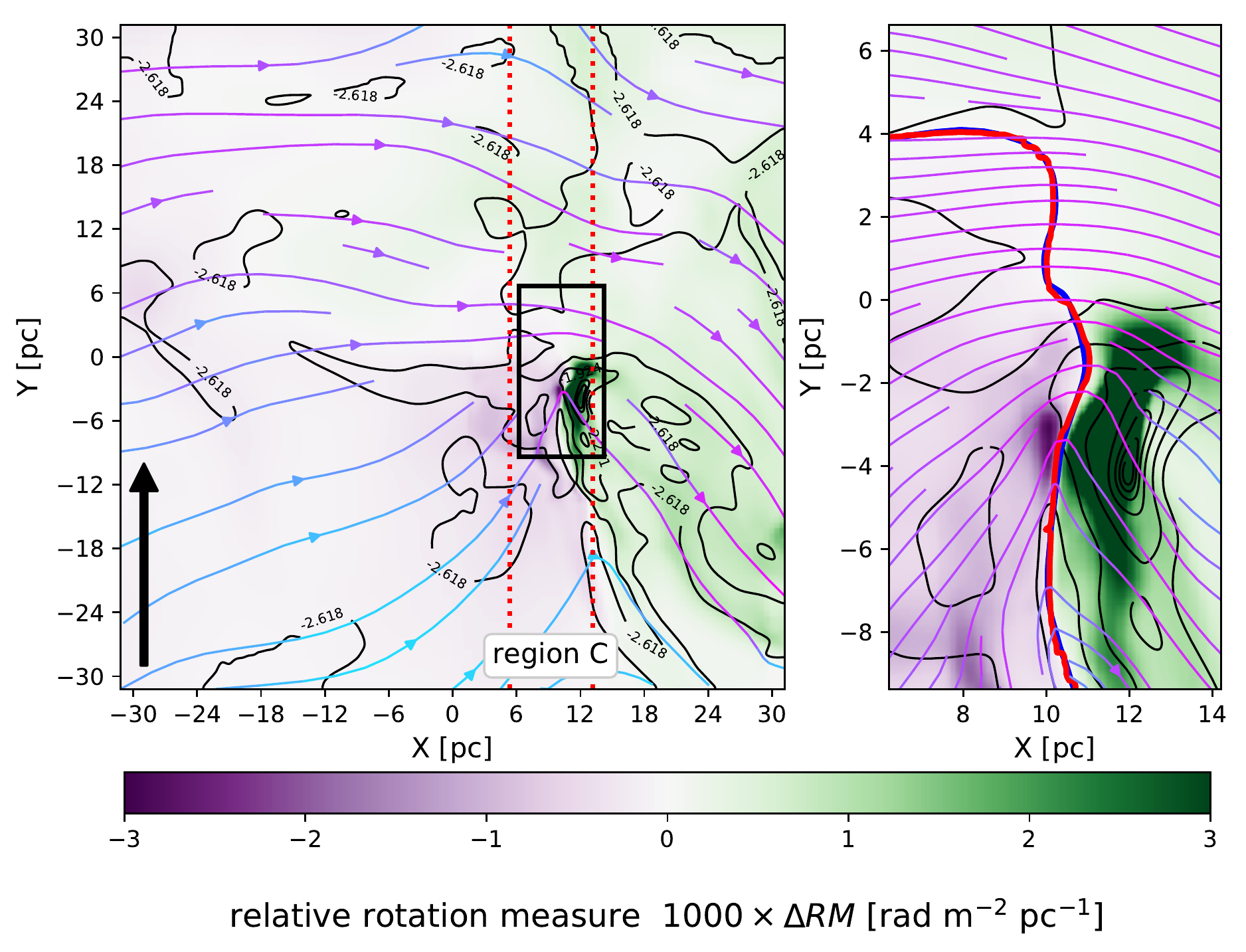}
\end{minipage}
\caption{The same as Faraday $RM$ in Fig. \ref{fig:AccRM} but for region A within the $XY-$plane at $Z=22.6\ \mathrm{pc}$ (top) and for region C within the $XY-$plane at $Z=-15.5\ \mathrm{pc}$ (bottom), respectively.}
\label{fig:AccAppRM}
\end{figure*}

In \S~\ref{sect:Origin} we showed the change of the polarization signal in a plane associated with the region B. In this section we present the maps for linear dust polarization $P_{\mathrm{l}}$ (Fig. \ref{fig:AccAppDust}), the circular line  polarization $P_{\mathrm{c}}$ (Fig. \ref{fig:AccAppLine}), and the $RM$ (Fig. \ref{fig:AccAppRM}) in the regions A and C for the sake of completeness.

\bsp	
\label{lastpage}
\end{document}